\documentclass[letterpaper,12pt]{article}
\usepackage[margin=22mm]{geometry}
\usepackage{amsmath}
\usepackage{amsfonts}
\usepackage{amssymb}
\usepackage{graphicx}
\usepackage{verbatim}
\usepackage{subfigure}
\usepackage{caption}
\usepackage{setspace}

\usepackage{color,bm}
\usepackage{graphicx}
\usepackage{array,multirow}
\usepackage{float}
\usepackage{afterpage}
\usepackage{arydshln}
\usepackage{titlesec}
\usepackage{blkarray}
\usepackage{mathtools} 
\usepackage{pgfplots}
\usepackage{multirow}
\usetikzlibrary{bayesnet}
\usepackage{blkarray}
\usepackage{multirow}
\usepackage[ruled,vlined]{algorithm2e}
\usepackage{booktabs}
\usepackage{enumitem}
\usepackage{xr}
\usepackage{hyperref} 
\hypersetup{
	colorlinks=true,       
	linkcolor=blue,        
	citecolor=blue,        
	filecolor=magenta,     
	urlcolor=blue         
}
\usepackage{setspace}

\newtheorem{theorem}{Theorem}
\newtheorem{assumption}{Assumption}

\newtheorem{definition}{Definition}
\newtheorem{example}{Example}

\newtheorem{remark}{Remark}

\usepackage{tikz}
\tikzset{
  LabelStyle/.style = { rectangle, rounded corners, draw,
                        minimum width = 2em, fill = yellow!50,
                        text = red, font = \bfseries },
  VertexStyle/.append style = { inner sep=2pt,
                                font = \bfseries},
  EdgeStyle/.append style = {->,  left} }

\immediate\write18{texcount -tex -sum  \jobname.tex > \jobname.wordcount.tex}

\usepackage[sort]{natbib}

\providecommand{\keywords}[1]
{
  \small	
  \textbf{\textit{Keywords---}} #1
}

\title{Learning Latent and Hierarchical Structures\\ in Cognitive Diagnosis Models
}
\author{Chenchen Ma, Jing Ouyang, and Gongjun Xu \\       Department of Statistics, University of Michigan }
\date{} 

\begin{document}
\maketitle

\pagestyle{plain}
\setcounter{page}{1}
\pagenumbering{arabic}

\vspace{-0.2in}
\begin{abstract}
\normalsize
Cognitive Diagnosis Models (CDMs) are a special family of discrete latent variable models that are widely used in educational and psychological measurement. A key component of CDMs is the $Q$-matrix characterizing the dependence structure between the items and the latent attributes. Additionally, researchers also assume in many applications certain hierarchical structures among the latent attributes to characterize their dependence. In most CDM applications, the attribute-attribute hierarchical structures, the item-attribute $Q$-matrix, the item-level diagnostic model, as well as  the number of latent attributes,  need to be fully or partially pre-specified, which however may be subjective and misspecified as noted by many recent studies. This paper considers the problem of jointly learning these latent and hierarchical structures in CDMs from observed data with minimal model assumptions. Specifically, a penalized likelihood approach is proposed to select the number of attributes and estimate the latent and hierarchical structures simultaneously. An expectation-maximization (EM) algorithm is developed for efficient computation, and statistical consistency theory is also established under mild conditions. The good performance of the proposed method is illustrated by simulation studies and   real data applications in educational assessment.
\end{abstract} 


\section{Introduction}
\normalsize
\label{sec-intro}

Cognitive Diagnosis Models (CDMs) are a popular family of discrete latent variable models that have been widely used in modern educational, psychological, social and biological sciences.
These models focus on multivariate discrete noisy observations, and assume the existence of discrete latent attributes which  explain or govern the observed variables. 
In practice, the discrete latent attributes often have special scientific interpretations.
For example, in educational assessments, the latent attributes are often assumed to be the mastery or deficiency of target skills 
\citep{junker2001cognitive, de2011generalized}; in psychiatric diagnosis, they may be modeled as the presence or absence of certain underlying mental disorders \citep{templin2006measurement, de2018analysis};
and in epidemiological and medical measurement studies, they can be interpreted as the existence or nonexistence of some disease pathogens \citep{wu2016nested, wu2016partially, o2019causes}.
A common feature of such latent variable models is that the probabilistic distribution of the observed responses is governed by the unobserved latent attributes.
Upon observing the responses, one can infer the underlying latent attributes.
As such, CDMs can provide fine-grained inference on subjects' latent attribute profiles based on their multivariate observed responses, and the corresponding latent subgroups of a population can also be detected based on the inferred latent attribute profiles.

In the CDM framework, the dependence structure between the observed variables and the latent attributes is encoded through a binary design matrix, the so-called $Q$-matrix \citep{tatsuoka1990toward}. 
Under different item models, the interactions between the observed variables and the latent attributes are also modeled differently.
Two basic models are the Deterministic Input Noisy Output “AND” gate \citep[DINA;][]{haertel1989} model and the Deterministic Input Noisy Output “OR” gate \citep[DINO;][]{templin2006measurement} model, where there are only two levels of item parameters for each item.
\cite{de2011generalized} proposed the the Generalized DINA (GDINA) model, where the interactions among all the latent attributes were included.
Other popularly used CDMs include the General Diagnostic Model \citep[GDM;][]{von2019general}, the reduced Reparameterized Unified Model \citep[reduced-RUM;][]{dibello1995unified}, and the Log-linear Cognitive Diagnosis Models \citep[LCDM;][]{henson2009defining}.

In many applications of CDMs, researchers are also interested in hierarchical structures among the latent attributes.
For example, in a learning context, the possession of lower level skills are often assumed to be the prerequisites for the possession of higher level skills in education \citep{dahlgren2006senior, jimoyiannis2001computer, simon2004explicating, wang2011using}.
Learning such latent hierarchical structures among the latent attributes is not only useful for educational research,
but can also be used to design learning materials and generate recommendations or remedy strategies based on the prerequisite relationships among the latent attributes.
\cite{leighton2004attribute} proposed the Attribute Hierarchy Model, a variation of Tatsuoka's rule-space approach \citep{tatsuoka1983rule}, which explicitly defined the hierarchical attribute structures through an adjacency matrix. 
Under the CDM framework, \cite{templin2014hierarchical} proposed the Hierarchical Cognitive Diagnosis Models (HCDMs), in which a Directed Acyclic Graph (DAG) is essentially used to impose hard constraints on possible latent attribute profiles under hierarchies.

To fit hierarchical CDMs, the $Q$-matrix,
the hierarchical structures among the attributes,  the item-level models, and the number of latent attributes all need to be pre-specified by domain experts, which however can be subjective and inaccurate. 
Moreover, in exploratory data analysis, these prior quantities may be even entirely unknown.
An important problem in cognitive diagnosis modeling then becomes how to efficiently and accurately learn such latent and hierarchical structures and   
model specifications from noisy observations with minimal prior knowledge and assumptions.

In the literature, many methods have been recently developed to learn the $Q$-matrix, including methods to directly estimate  the $Q$-matrix from the observational data, via either frequentist approaches \citep{liu2012data,chen2015statistical,xu2018identifying,li2022learning} or Bayesian approaches \citep{chung2018mcmc,chen2018bayesian,culpepper2019estimating}, 
 and methods to  validate the pre-specified Q-matrix \citep{de2008empirically, decarlo2012recognizing, chiu2013statistical, de2016general,gu2018hypothesis}.
Many of these $Q$-matrix learning or validation methods, however, do not take the hierarchical structures into consideration, or they implicitly assume the hierarchical structure is known; moreover, the number of attributes and the item-level diagnostic models are often assumed to be known.

In terms of learning attribute hierarchies from observational data, \cite{wang2021learning} recently studied two exploratory approaches including the latent variable selection \citep{xu2018identifying} approach and the regularized latent class modeling  \citep[regularized LCM,][]{chen2017regularized} approach. 
However, the latent variable selection approach in \cite{wang2021learning}  requires specification of the number of latent attributes and a known identity sub-matrix in the $Q$-matrix.
The regularized LCM approach may not require the number of latent attributes, but the number of latent classes needs to be selected.
Based on the simulation in \cite{wang2021learning}, the performance of the regularized LCM was less satisfactory -- the accuracy of selecting correct number of latent classes was often below 50\% and the accuracy of recovering latent hierarchy was almost 0 in some cases.
In \cite{gu2019identification}, the authors proposed a two-step algorithm for structure learning of hierarchical CDMs.
However, their algorithm also assumed that the number of latent attributes was known and they only considered the DINA and DINO models.

In this paper, to overcome the limitations of the aforementioned methods, we propose a regularized maximum likelihood approach 
with minimal model assumptions to achieve the following four goals simultaneously: 
(1) estimate the number of latent attributes; (2) learn the latent hierarchies among the attributes; (3) learn the $Q$-matrix; and (4) recover item-level diagnostic models.
Specifically, we employ two regularization terms: 
one penalty on the population proportion parameters to select significant latent classes, and the other on the differences of item parameters for each item to learn the structures of the item parameters. 
After the significant latent classes  and the structure of the item parameters are learned, a latent structure recovery algorithm is used to estimate the number of latent attributes, the latent hierarchies among the attributes, the $Q$-matrix, and the item-level models.
For the estimation, we develop an efficient Penalized EM algorithm using the Difference Convex (DC) programming and the Alternating Direction Method of Multipliers (ADMM) method. 
Consistent learning theory is established under  mild regularity conditions.
We also conduct simulation studies to show the good  performance of the proposed method.
Finally we demonstrate the application of our method to two real datasets and obtain  interpretable results which are consistent with the previous research.

The paper proceeds as follows: 
the model setup of hierarchical CDMs is provided in Section \ref{sec-model}.
The proposed penalized likelihood approach and its theoretical properties are presented in Section \ref{sec-RLAM}. 
An efficient algorithm is developed and related computational issues are discussed in Section \ref{sec-algo}.
Simulation studies are presented in Section \ref{sec-simu}. 
In Section \ref{sec-data}, the model is applied to two real data sets of international educational assessment. 
Finally, Section \ref{sec-discuss} concludes with some discussions. The proof of the main theorem and detailed derivations for the proposed algorithm  are presented in the supplementary material. 

\section{Model Setup}
\label{sec-model}

In this section, we introduce the general model setup of hierarchical CDMs   and illustrate the connections between hierarchical CDMs and restricted latent class models, which motivates the proposed approach in Section \ref{sec-RLAM}.
In the following, for an integer $K$, we use $[K]$ to denote the set $\{1,2,\dots,K\}$, and we use $|\cdot|$ to denote the cardinality of a set.

\subsection{Hierarchical Cognitive Diagnosis Models}
\label{sec-hcdm}

In a CDM with $J$ items which depend on $K$ latent attributes of interest, 
two types of subject-specific variables are considered, including the   responses $\bm{R} = (R_1,\dots,R_J)$, and the latent binary attribute profile $\bm{\alpha} = (\alpha_1,\dots,\alpha_K)$.
In this paper, both the   responses $\bm{R}$ and the latent attribute profile $\bm{\alpha}$ are assumed to be binary.
The $J$-dimensional vector $\bm{R}\in\{0,1\}^J$ denotes the   binary responses to the set of $J$ items, and the $K$-dimensional vector $\bm{\alpha}\in\{0,1\}^K$ denotes a profile of possession of $K$ latent attributes of interest.
Since each latent attribute $\alpha_k$ is binary, the total number of possible latent attribute profiles $\bm{\alpha} = (\alpha_1,\dots,\alpha_K)$ is $2^K$.
For each latent attribute profile, we use $\pi_{\bm{\alpha}}$ to denote its proportion parameter, and the latent attribute profile $\bm{\alpha}$ is modeled to follow a categorical distribution with the proportion parameter vector $\bm{\pi}=(\pi_{\bm{\alpha}}:\bm{\alpha}\in\{0,1\}^K)$.
The proportion parameter vector lies in the $(2^K-1)$-simplex and  satisfies $\pi_{\bm{\alpha}}\in [0,1]$ and $\sum_{\bm{\alpha}\in\{0,1\}^K} \pi_{\bm{\alpha}}=1$. 

A key feature of hierarchical CDMs is that there exist certain hierarchical structures among the latent attributes. 
For example, in cognitive diagnosis modeling, the possession of lower-level skills are often regarded as the prerequisites for the possession of higher-level skills \citep{leighton2004attribute, templin2014hierarchical}.
With such an attribute hierarchy, any latent attribute profile $\bm{\alpha}$ that does not respect the hierarchy will not exist and have population proportion $\pi_{\bm{\alpha}}=0$. 
For $1\leq k \neq l \leq K$, we use $\alpha_k \xrightarrow{} \alpha_l$ (or $k\xrightarrow{}l$) to denote the hierarchy that attribute $\alpha_k$ is a prerequisite of attribute $\alpha_l$.
We assume such hierarchy $\alpha_k \xrightarrow{} \alpha_l$ (or $k\xrightarrow{}l$) exists if and only if there are no latent attribute profiles such that $\alpha_l=1$ but $\alpha_k=0$,
or equivalently,   we have $\pi_{\bm{\alpha}}=0$ if $\alpha_l=1$ but $\alpha_k=0$. 
We denote an attribute hierarchy by a set of prerequisite relations $\mathcal{E} = \{k\xrightarrow{}l: \text{ attribute }k \text{ is a prerequisite for attribute } l, \ 1\leq k \neq l \leq K\}$, 
and denote the induced set of existent latent attribute profiles by $\mathcal{A}=\{\bm{\alpha}\in\{0,1\}^K:\pi_{\bm{\alpha}}\neq 0 \text{ under }\mathcal{E}\}$. 
One can see that an attribute hierarchy results in the sparsity of the proportion parameter vector $\bm{\pi}$, which will significantly reduce the number of model parameters especially when $K$ is large.
Example hierarchical structures and the corresponding induced sets of existent attribute profiles are shown in Figure \ref{fig:hier}.

\tikzstyle{place}=[circle,draw=black,fill=white,inner sep=0pt,minimum size=6mm]
\tikzstyle{group1}=[rounded corners,fill=blue!5,inner sep=1ex]
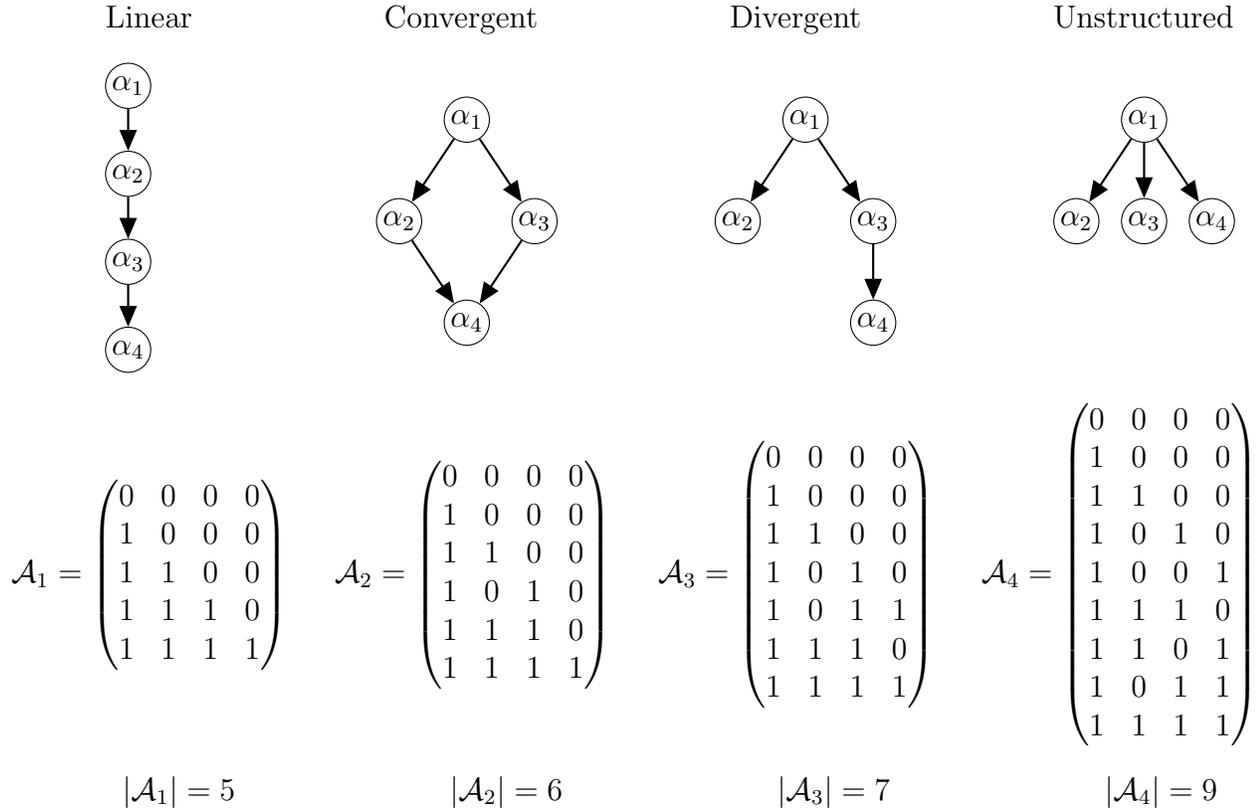
\begin{figure}[h!] 
    \centering
    \subfigure{
    Linear
    }
    \hspace{2cm}
    \subfigure{ 
    Convergent
    }
    \hspace{2cm}
    \subfigure{ 
    Divergent
    }
    \hspace{2cm}
    \subfigure{ 
    Unstructured
    }
    \\
    \subfigure{
    \begin{tikzpicture}[scale=0.9]
    \path
    node at (0, -1.5) [place] (1) {$\alpha_1$}
    node at (0, -2.8) [place] (2) {$\alpha_2$}
    node at (0, -4.1) [place] (3) {$\alpha_3$}
    node at (0, -5.4) [place] (4) {$\alpha_4$};
    \draw [->, thick] (1) to (2);
    \draw [->, thick] (2) to (3);
    \draw [->, thick] (3) to (4);
    
    \path
    node at (5, -2) [place] (1) {$\alpha_1$}
    node at (4, -3.5) [place] (2) {$\alpha_2$}
    node at (6, -3.5) [place] (3) {$\alpha_3$}
    node at (5, -5.0) [place] (4) {$\alpha_4$};
    \draw [->, thick] (1) to (2);
    \draw [->, thick] (1) to (3);
    \draw [->, thick] (2) to (4);
    \draw [->, thick] (3) to (4);

    \path
    node at (10, -2) [place] (1) {$\alpha_1$}
    node at (9, -3.5) [place] (2) {$\alpha_2$}
    node at (11, -3.5) [place] (3) {$\alpha_3$}
    node at (11, -5.0) [place] (4) {$\alpha_4$};
    \draw [->, thick] (1) to (2);
    \draw [->, thick] (1) to (3);
    \draw [->, thick] (3) to (4);
    
    \path
    node at (15, -2) [place] (1) {$\alpha_1$}
    node at (14, -3.5) [place] (2) {$\alpha_2$}
    node at (15, -3.5) [place] (3) {$\alpha_3$}
    node at (16, -3.5) [place] (4) {$\alpha_4$};
    \draw [->, thick] (1) to (2);
    \draw [->, thick] (1) to (3);
    \draw [->, thick] (1) to (4);
    \end{tikzpicture}
    }
    \\
    
   	\hspace{-1cm}
    \subfigure{
    $
    \mathcal{A}_1 = 
    \begin{pmatrix}
    0 & 0 & 0 & 0\\
    1 & 0 & 0 & 0\\
    1 & 1 & 0 & 0\\
    1 & 1 & 1 & 0\\
    1 & 1 & 1 & 1\\ 
    \end{pmatrix}{}
    $
    }
    \hspace{0cm}
    \subfigure{
    $
    \mathcal{A}_2 = 
    \begin{pmatrix}
    0 & 0 & 0 & 0\\
    1 & 0 & 0 & 0\\
    1 & 1 & 0 & 0\\
    1 & 0 & 1 & 0\\
    1 & 1 & 1 & 0\\
    1 & 1 & 1 & 1\\ 
    \end{pmatrix}{}
    $
    }
    \hspace{0cm}
    \subfigure{
    $
    \mathcal{A}_3 = 
    \begin{pmatrix}
    0 & 0 & 0 & 0\\
    1 & 0 & 0 & 0\\
    1 & 1 & 0 & 0\\
    1 & 0 & 1 & 0\\
    1 & 0 & 1 & 1\\
    1 & 1 & 1 & 0\\
    1 & 1 & 1 & 1\\ 
    \end{pmatrix}{}
    $
    }
    \hspace{0cm}
    \subfigure{
    $
    \mathcal{A}_4 = 
    \begin{pmatrix}
    0 & 0 & 0 & 0\\
    1 & 0 & 0 & 0\\
    1 & 1 & 0 & 0\\
    1 & 0 & 1 & 0\\
    1 & 0 & 0 & 1\\
    1 & 1 & 1 & 0\\
    1 & 1 & 0 & 1\\
    1 & 0 & 1 & 1\\
    1 & 1 & 1 & 1\\ 
    \end{pmatrix}{}
    $
    }
    \\
    \subfigure{
    $|\mathcal{A}_1| = 5$
    }
    \hspace{2.3cm}
    \subfigure{
    $|\mathcal{A}_2| = 6$
    }
    \hspace{2.3cm}
    \subfigure{
    $|\mathcal{A}_3| = 7$
    }
    \hspace{2.3cm}
    \subfigure{
    $|\mathcal{A}_4| = 9$
    }
    \caption{Examples of hierarchical structures of latent attributes. For $i=1, \dots, 4$, each $\mathcal{A}_i$ represents the induced set of existent attribute profiles under the hierarchical structure above it, where each row in $\mathcal{A}_i$ represents an attribute profile $\bm{\alpha}$ with $\pi_{\bm{\alpha}}\neq 0$.}
    \label{fig:hier}
\end{figure}

In CDMs, the structural matrix  $\bm{Q}=(q_{j,k})\in\{0,1\}^{J\times K}$ is an important component which imposes constraints on items to reflect the dependence between the items and the latent attributes. 
To be specific, $q_{j,k}=1$ if item $j$ requires (or depends on) attribute $k$. 
Then the $j$th row vector of $\bm{Q}$ denoted by $\bm{q}_j$ describes the full dependence of item $j$ on $K$ latent attributes.
In many applications, the matrix $\bm{Q}$ is pre-specified by domain experts \citep{george2015cognitive, junker2001cognitive, von2005general} to reflect some scientific assumptions.
See Figure \ref{fig:Q} for the illustration of the $Q$-matrix and the corresponding graphical representation.

\begin{figure}[ht]
    \centering
    \subfigure{
    $\bm{Q} = \begin{pmatrix}
    	1 & 1 & 0 & 0\\
    	1 & 0 & 1 & 0 \\
    	0 & 1 & 0 & 0 \\
    	1 & 1 & 0 & 1 \\
    	0 & 0 & 1 & 1 \\
    \end{pmatrix}$}
    \hspace{1in}
    \subfigure{
    \begin{tikzpicture}[baseline= -13ex][scale=0.8]
    \draw node at (-2.6, -1.5)[place] (1) {$\alpha_1$}
    node at (-1.3, -1.5) [place] (2) {$\alpha_2$}
    node at (0, -1.5) [place] (3) {$\alpha_3$}
    node at (1.3, -1.5) [place] (4) {$\alpha_4$}
    node at (-3.5, -3) [place,fill={rgb:black,1;white,2},] (11) {$R_1$}
    node at (-2, -3) [place,fill={rgb:black,1;white,2},] (22) {$R_2$}
    node at (-0.5, -3) [place,fill={rgb:black,1;white,2},] (33) {$R_3$}
    node at (1, -3) [place,fill={rgb:black,1;white,2},] (44) {$R_4$}
    node at (2.5, -3) [place,fill={rgb:black,1;white,2},] (55) {$R_5$};
  \draw [->] (1) to (11);
  \draw [->] (1) to (22);
  \draw [->] (1) to (44);
  \draw [->] (2) to (11);
  \draw [->] (2) to (33);
  \draw [->] (2) to (44);
  \draw [->] (3) to (22);
  \draw [->] (3) to (55);
  \draw [->] (4) to (44);
  \draw [->] (4) to (55);

	\end{tikzpicture}
	}
    \caption{Illustration of Q-matrix}
    \label{fig:Q}
\end{figure}
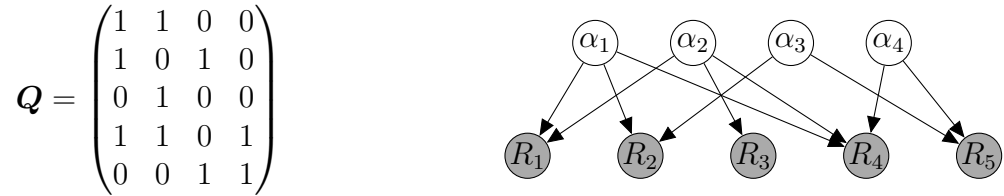{}

As in classical latent class analysis, given a subject's latent attribute profile $\bm{\alpha}$, the responses to $J$ items are assumed to be independent, which is known as the local independence assumption, and follow Bernoulli distributions with parameters $\theta_{1,\bm{\alpha}},\dots,\theta_{J,\bm{\alpha}}$, which are called item parameters.
Specifically, we have $\theta_{j,\bm{\alpha}}:=\mathbb{P}(R_j=1\mid \bm{\alpha})$.
Under the local independence assumption, the probability mass function of a subject's response vector $\bm{R} = (R_1,\dots,R_J)\in\{0,1\}^J$ can be written as 
\begin{equation}
    \mathbb{P}(\bm{R}\mid \bm{\Theta},\bm{\pi}) = \sum_{\bm{\alpha}\in\{0,1\}^K} \pi_{\bm{\alpha}}\prod_{j=1}^J \theta_{j,\bm{\alpha}}^{R_j}(1-\theta_{j,\bm{\alpha}})^{1-R_j}.
\end{equation}

So far we have a latent attribute profile $\bm{\alpha}$ for a subject to indicate the subject's possession of $K$ attributes, and a structural vector $\bm{q}$ for an item to reflect the item's dependence on $K$ latent attributes.
Moreover, the structural matrix $\bm{Q}$ puts constraints on item parameters to reflect the diagnostic model assumptions.  
One important and common assumption is that the item parameters $\theta_{j,\bm{\alpha}}$ only depends on whether the latent attribute profile $\bm{\alpha}$ contains the required attributes by item $j$, that is, the attributes in the set $\mathcal{K}_j = \{k\in[K]:q_{j,k}=1\}$, which is   the set of the  required attributes of item $j$. 
Therefore, for item $j$, the latent attribute profiles which are only different in the attributes outside of $\mathcal{K}_j$  have the same item parameters. 
In this way, the structural matrix $\bm{Q}$ forces some entries in the item parameter matrix $\bm{\Theta}$ to be the same.
The dependence of item parameters on the required attributes are modeled differently in different CDMs, 
 as shown in Example \ref{ex-two} and Example \ref{ex-multi}.

\begin{example}[DINA and DINO Models]
We first consider the  Deterministic Input Noisy output ``And" (DINA)   \citep{junker2001cognitive} and the Deterministic Input Noisy output ``Or" (DINO)   \citep{templin2006measurement}  models, where there are only two levels of item parameters for each item.
Specifically, we use $\theta_j^+$ and $\theta_j^-$ to denote the two levels for item $j$. 
We introduce a binary indicator matrix $\bm{\Gamma} = (\Gamma_{j,\bm{\alpha}}:j\in [J], \bm{\alpha}\in\{0,1\}^K)\in\{0,1\}^{J\times 2^K}$, which corresponds to the ideal responses under the DINA and DINO models.
Under the DINA model, which assumes a conjunctive ``And" relationship among the binary attributes,   the indicator matrix is defined as
\begin{equation}
    \Gamma^{\text{DINA}}_{j,\bm{\alpha}}=\prod_k^K \alpha_k^{q_{j,k}} = \prod_{k \in \mathcal{K}_j}\alpha_k.
\end{equation}
Under the DINO model assuming a conjunctive ``Or" relationship among the latent attributes, we have
\begin{equation}
    \Gamma_{j,\bm{\alpha}}^{\text{DINO}}=\mathbb{I}\big(\exists \ k \in [K], q_{j,k}=\alpha_k=1\big). 
\end{equation}
The indicator $\Gamma_{j,\alpha}$ in the DINA model indicates whether a subject possesses all the required attributes of item $j$, while that in the DINO model indicates whether a subject possesses any of the required attributes of item $j$.
In both models, the item parameters only depend on the set of the required attributes of an item $\mathcal{K}_j$, and the item parameters are defined as:
$$
\theta_j^+ = \mathbb{P}(R_j=1\mid \Gamma_{j,\bm{\alpha}}=1),\quad \theta_j^-=\mathbb{P}(R_j=1\mid \Gamma_{j,\bm{\alpha}}=0),
$$ where $\theta_j^-$ is also called the guessing parameter and $1-\theta_j^+ $ the slipping parameter. 
\label{ex-two}
\end{example}

\begin{example}[GDINA model]

The Generalized DINA model \citep[GDINA,][]{de2011generalized} is a more general model where all the interactions among the latent attributes are considered.
The item parameters for the GDINA model are written as
\begin{align}
\centering
	\theta_{j,\bm{\alpha}}^{\text{GDINA}} & = \beta_{j,0} + \sum_{k=1}^K \beta_{j,k} \alpha_k q_{j,k} + \sum_{k=1}^K \sum_{k' = k+1}^K \beta_{j,k,k'}\alpha_k \alpha_{k'}q_{j,k}q_{j,k'} +\cdots +\beta_{j,1,2,\dots,K} \prod_{k=1}^K \alpha_k q_{j,k}\\
		& = \beta_{j,0} + \sum_{k\in\mathcal{K}_j} \beta_{j,k} \alpha_k + \sum_{k, k'\in \mathcal{K}_j, k\neq k'} \beta_{j,k,k'}\alpha_k \alpha_{k'} + \cdots +\beta_{j,\mathcal{K}_j} \prod_{k \in \mathcal{K}_j} \alpha_k.
\end{align}
The coefficients in the GDINA model can be interpreted as following: $\beta_{j,0}$ is the probability of a positive response for the most incapable subjects with none required attributes present; 
$\beta_{j,k}$ is the increase in the probability due to the main effect of latent attribute $\alpha_{k}$;
$\beta_{j,1,2,\dots,K}$ is the change in the positive probability due to the interaction of all the latent attributes.
In the GDINA model, the intercept and main effects are typically assumed to be nonnegative to satisfy the monotonicity assumption, while the interactions may take negative values.
By incorporating all the interactions among the required attributes, the GDINA model is one of the most general CDMs.

\label{ex-multi}
\end{example}{}

\subsection{CDMs as Restricted Latent Class Models}
\label{sec-RLCM}

CDMs in fact can also be viewed as Restricted Latent Class Models \citep[RLCM,][]{xu2017identifiability}, a special family of more general Latent Class Models (LCMs).
We first give a brief description of the general model setup of LCMs \citep{goodman1974exploratory}.
In an LCM, we assume that each subject belongs to one of $M$ latent classes, $m=1,\cdots, M$.
For each latent class, we use $\pi_m$ to denote its proportion parameter for $m\in[M]$.
The latent classes follow a categorical distribution with the proportion parameter vector $\bm{\pi} = (\pi_m: m\in [M], \pi_m \geq 0, \sum_{m=1}^M \pi_m = 1)$.
As in classical finite mixture models, responses to items are assumed to be independent of each other given the latent class membership,
and we use $\bm{\Theta}=(\theta_{jk})\in [0,1]^{J\times M}$ to denote the item parameter matrix.
To be specific, for a subject's response $\bm{R} = (R_1, R_2, \dots, R_J)$,
we have $\theta_{jm}=\mathbb{P}(R_{j}=1\mid m)$. 
Then the probability mass function of an LCM can be written as 
\begin{equation}
    \mathbb{P}(\bm{R}\mid \bm{\Theta},\bm{\pi}) = \sum_{m=1}^M \pi_m\prod_{j=1}^J \theta_{jm}^{R_{j}}(1-\theta_{jm})^{1-R_{j}}.
\end{equation}
This unrestricted LCM is saturate in the sense that no constrains are imposed on the latent classes' response distributions. 

CDMs can be viewed as special cases of LCMs with $M=2^K$ latent classes and  additional constrains  imposed on the components' distributions.
Recall that in CDMs with $K$ latent attributes, each latent attribute profile $\boldsymbol{\alpha}$ is a $K$-dimensional binary vector and has a proportion parameter $\pi_{\boldsymbol{\alpha}}$.
We can use a one to one correspondence from $\{\boldsymbol{\alpha}: \boldsymbol{\alpha}\in\{0,1\}^K\}$ to $\{m: m=1,\ldots, 2^K\}$ such as $m = \sum_{k=1}^K \alpha_k \cdot 2^{k-1}+1$. Then in a CDM with $K$ latent attributes and no hierarchical structure, we have $M = 2^K$ latent classes. 
In hierarchical CDMs, the number of allowed latent attribute profiles is smaller than $2^K$ and we have $M = |\cal A|$, as discussed in Section \ref{sec-hcdm}.
Moreover, in CDMs, there are additional restrictions on the item parameter matrix $\bm{\Theta}$ through the $Q$-matrix.
Under these restrictions, for each item, certain subsets of item-level response probabilities will be constrained to be the same.
Thus, a CDM with or without any hierarchical structure can be viewed as a submodel of a saturated LCM.

\section{Regularized Estimation Method} 
\label{sec-RLAM}

\subsection{Motivation and Proposed Method}
\label{sec-Motivation}

To fit hierarchical CDMs, the $Q$-matrix,
the hierarchical structures among the attributes,  the item-level models, and the number of latent attributes are often needed to be pre-specified by domain experts, which however can be subjective and inaccurate.  
An important problem in cognitive diagnosis modeling then becomes how to efficiently and accurately learn these quantities from noisy observations.

In this section, we propose a unified modeling and inference approach   to learn the latent structures, including the number of latent attributes $K$, the attribute-attribute hierarchical structure, the item-attribute $Q$-matrix, and the item-level diagnostic models.
In particular, based on the observation in Section \ref{sec-RLCM}, we propose  to learn a hierarchical CDM with minimal model assumptions starting with an unrestricted LCM.
 We use the following discussion and examples to first illustrate the key idea. 

\begin{itemize}
	\item  As discussed in Section \ref{sec-hcdm}, when there exist hierarchical structures among the latent attributes, the number of truly existing latent attribute profiles is smaller than $2^K$.
For example, in Figure \ref{fig:hier}, when $K=4$, the total number of possible attribute profiles without any hierarchical structure is $2^K = 16$.
Under different hierarchical structures as shown in Figure \ref{fig:hier}, the numbers of   existing attribute profiles are all smaller than 16.
Therefore, to learn a hierarchical cognitive diagnosis model, we need first select significant latent attribute profiles that truly exist in the population.

\item Furthermore, to reconstruct the $Q$-matrix and item models in hierarchical cognitive diagnosis models, it is also essential to examine the inner structure of the item parameter matrix.
One key challenge here is that under certain model assumptions, there may exist some equivalent $Q$-matrices.
Here we say two $Q$-matrices are equivalent under certain hierarchical structure $\mathcal{E}$, denoted by $\mathbf{Q}_1 \overset{\mathcal{E}}{\sim} \mathbf{Q}_2$, if they give the same item parameter matrices, that is, $\boldsymbol{\Theta}(\mathbf{Q}_1,\mathcal{A}_{\mathcal{E}}) = \boldsymbol{\Theta}(\mathbf{Q}_2,\mathcal{A}_{\mathcal{E}})$, where $\mathcal{A}_{\mathcal{E}}$ is the induced latent attribute profile set under hierarchy $\mathcal{E}$.
As we introduced in Example \ref{ex-two}, for the DINA model, 
the item parameters only depend on the highest interactions among the required latent attributes.
For such models, we have equivalent $Q$-matrices under hierarchical structures. 
For example, consider three latent attributes with linear hierarchy, that is, $\mathcal{E} = \big\{1\rightarrow 2 \rightarrow 3 \big\}$.
We have 
\begin{equation}
	\mathbf{Q}^{(1)} = \begin{pmatrix}
		1 & 0 & 0 \\
		0 & 1 & 0 \\
		0 & 0 & 1 \\
	\end{pmatrix} \overset{\mathcal{E}}{\sim}
	\mathbf{Q}^{(2)} = \begin{pmatrix}
		1 & 0 & 0 \\
		1 & 1 & 0 \\
		1 & 1 & 1 \\
	\end{pmatrix} \overset{\mathcal{E}}{\sim}
	\mathbf{Q}^{(*)} = \begin{pmatrix}
		1 & 0 & 0 \\
		* & 1 & 0 \\
		* & * & 1 \\
	\end{pmatrix}, 
	\label{fig:equivalent-Q}
\end{equation}
where ``$*$" can be either 0 or 1.
However, when the underlying model is the GDINA model,
since all the interactions among the latent attributes are considered, 
there would not exist such equivalent $Q$-matrices.
For example, consider the second item of the $Q$-matrices in (\ref{fig:equivalent-Q}), for $\bm{Q}^{(1)}$, $\mathbf{q}^{(1)}_2 = (0, 1, 0)$ and the corresponding item parameter vector under the GDINA model is $\boldsymbol{\theta}_2^{(1)} = (\beta_0,\ \beta_0,\ \beta_0 + \beta_2, \ \beta_0 + \beta_2)$.
For $\bm{Q}^{(2)}$, $\mathbf{q}^{(2)}_2 = (1, 1, 0)$ and the corresponding item parameter vector under the GDINA model is $\boldsymbol{\theta}_2^{(2)} = (\ \beta_0,\ \beta_0+\beta_1,\ \beta_0 + \beta_1 + \beta_2+\beta_{1,2}, \ \beta_0 + \beta_1 + \beta_2+\beta_{1,2})$, which is different from that of $\bm{Q}^{(1)}$, and thus, the equivalence no longer holds under the GDINA model.
Therefore, to learn the $Q$-matrix and infer the item models in hierarchical CDMs, it is also necessary to learn the item parameter matrix and investigate the inner constraint structure of it.
Moreover, after learning the item parameter matrix, we can get the partial orders among the selected latent classes, which  in turn enable us to recover the latent hierarchies, the $Q$-matrix and item models.
We leave the details of the reconstruction of these quantites in Section \ref{sec-hier}.

\end{itemize}

Motivated by the above discussions and the fact that CDMs are a special family of  LCMs  with additional restrictions, we propose to start with an  unrestricted latent class model and then put additional regularization terms, to select significant latent classes and learn the item parameter matrix simultaneously.
Specifically, we start with a latent class model with $M$ latent classes, where $M$ is a large number, serving as an upper bound for the true number of latent classes.
If the true number of latent classes is smaller than $M$, some of the proportion parameters will be zeros.
Let  $\bm{\pi} = (\pi_1, \dots, \pi_M)$ be the proportion parameter vector, and $\bm{\Theta} = (\theta_{jk})\in(0,1)^{J\times M}$ be the item parameter matrix of the LCM, with $\bm{\theta}_j = (\theta_{jk}, k=1,\ldots, M)$ being the $j$th item's item parameter vector. 
Then for a response data matrix $\mathcal{R} = (R_{ij}: i \in [N], j \in [J]) \in \{0,1\}^{N\times J}$, where $N$ is the sample size, the likelihood can be written as 
\begin{equation}
    L_N(\bm{\pi},\bm{\Theta};\mathcal{R}) = \prod_{i=1}^N \Big( \sum_{k=1}^M\big(\pi_k\prod_{j=1}^J \theta_{jk}^{R_{ij}}(1-\theta_{jk})^{1-R_{ij}}\big)\Big).
\end{equation}
And the log-likelihood is
\begin{equation}
    l_N(\bm{\pi},\bm{\Theta};\mathcal{R}) = \sum_{i=1}^N \log \Big( \sum_{k=1}^M\big(\pi_k\prod_{j=1}^J \theta_{jk}^{R_{ij}}(1-\theta_{jk})^{1-R_{ij}}\big)\Big)
    \label{eq:log-like}
\end{equation}

We consider the following objective function with two additional penalty terms: 
\begin{equation}
 l_N(\bm{\pi},\bm{\Theta};\mathcal{R}) - \lambda_1 \sum_{k=1}^M \log_{[\rho_N]}\pi_k - \lambda_2 \sum_{j=1}^J \mathcal{J}(\bm{\theta}_j),   
\label{eq-objective}
\end{equation}
where   $\lambda_1$ and $\lambda_2$ are two nonnegative tuning parameters.
The terms $\log_{[\rho_N]}\pi_k $ and $\mathcal{J}(\bm{\theta}_j)$ are two penalties and we discuss them one by one as follows. 

\begin{figure}[ht]
    \centering
    \subfigure[]{\includegraphics[width=2.5in]{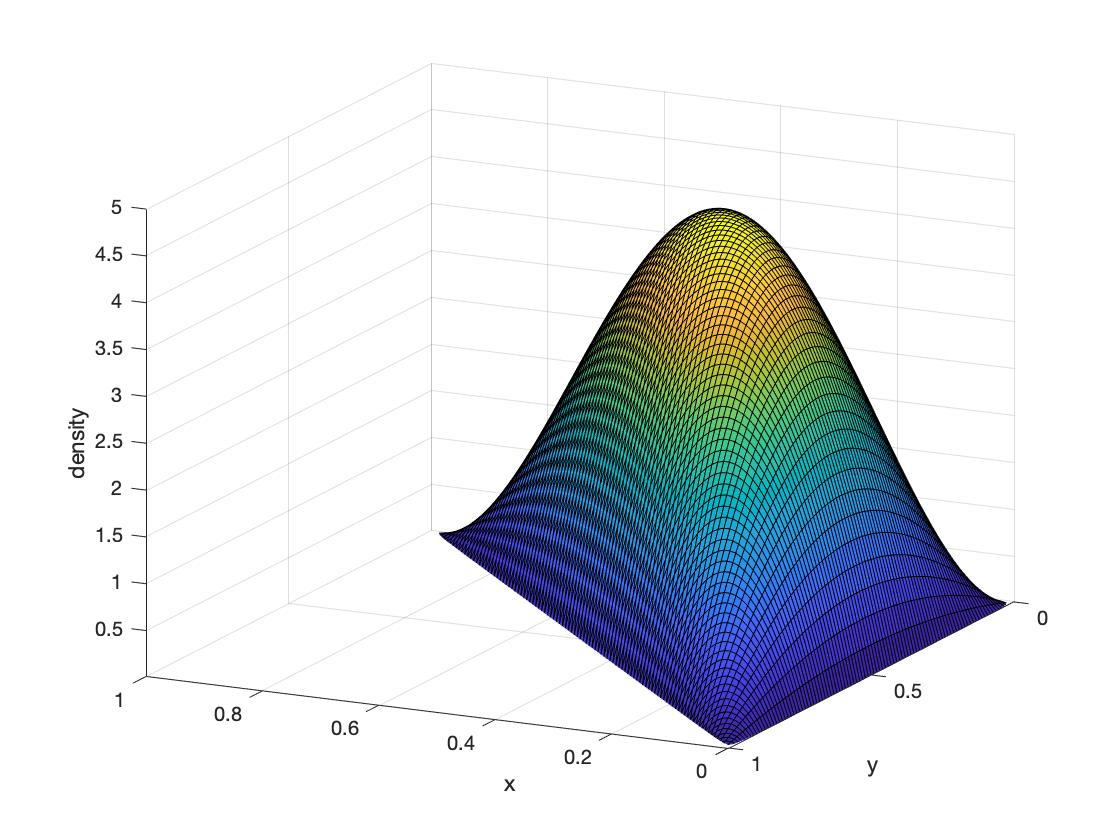}}
    \hspace{0.5in}
    \subfigure[]{
    \includegraphics[width=2.5in]{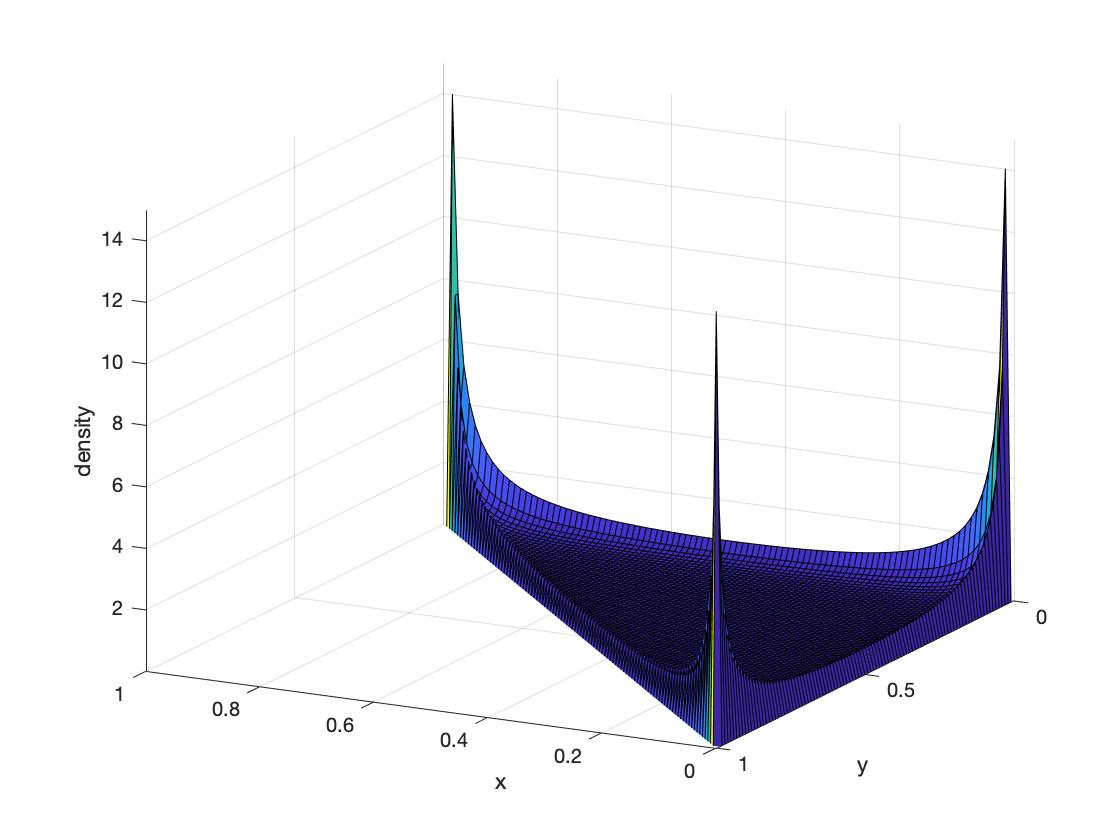}}
      
    \caption{(a): probability density function of 3-dimensional Dirichlet distribution with parameters all equal to $1 - \lambda_1 = 1$; (b): (improper) probability density function of 3-dimensional Dirichlet distribution with parameters all equal to $1 - \lambda_1 = -0.9$. }
    \label{fig:Dirichlet}
\end{figure}

The term $\log_{[\rho_N]}\pi_k = \log\pi_k\cdot\mathbb{I}\big(\pi_k>\rho_N\big) + \log{\rho_N}\cdot\mathbb{I}\big(\pi_k\leq \rho_N\big)$, is a log-type penalty \citep{gu2019learning}  on the proportion parameters, where $\rho_N$ is a small threshold to circumvent the singularity of the log function at zero. 
 Following \cite{gu2019learning}, we can take $\rho_N$ to be a small value, such as $N^{-d}$ for some $d \geq 1$.
The log penalty is imposed on the proportion parameters, which forces small values in the proportion parameters to be zero.
This log-type penalty also makes computation efficient in the E-step, as shown in our EM algorithm in Section \ref{em}.
We can also interpret this log penalty from a Bayesian perspective, where we use a Dirichlet prior with parameter $ 1 - \lambda_1$ for the proportions.
When $1 - \lambda_1 < 0$, it's not a proper Dirichlet distribution.
But allowing $1 - \lambda_1 < 0$ would help us select significant proportion parameters more efficiently compared to the traditional proper Dirichlet priors. 
As shown in Figure \ref{fig:Dirichlet}, when $\lambda_1 < 1$, the density concentrates more in the interior of the parameter space, while with $\lambda_1 > 1$ the density concentrates more on the boundary encouraging sparsity of the proportion parameter vector.
Therefore, it is essential to allow $\lambda_1$ to be nonnegative and even larger than 1 for the purpose of selecting the significant latent classes.

\begin{figure}[ht]
    \centering
    \subfigure[]{     \includegraphics[width=2in]{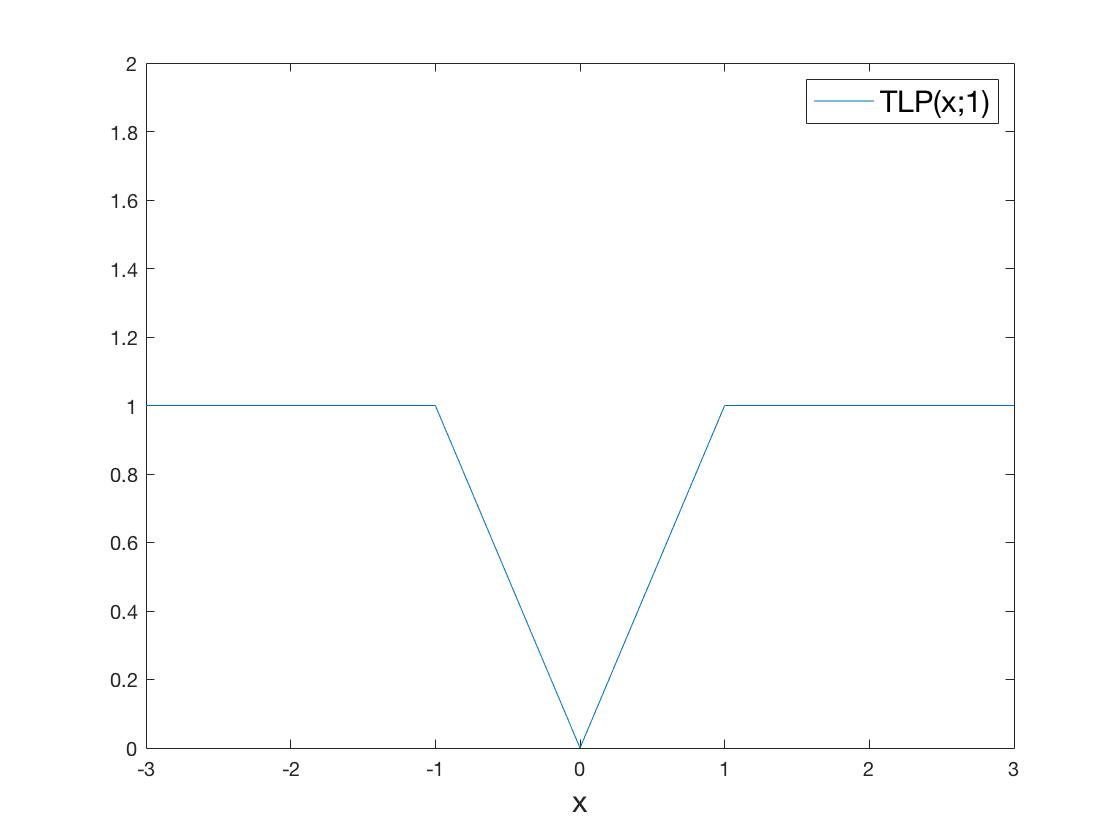} \label{fig:tlp}}
    \subfigure[]{
    \includegraphics[width=2in]{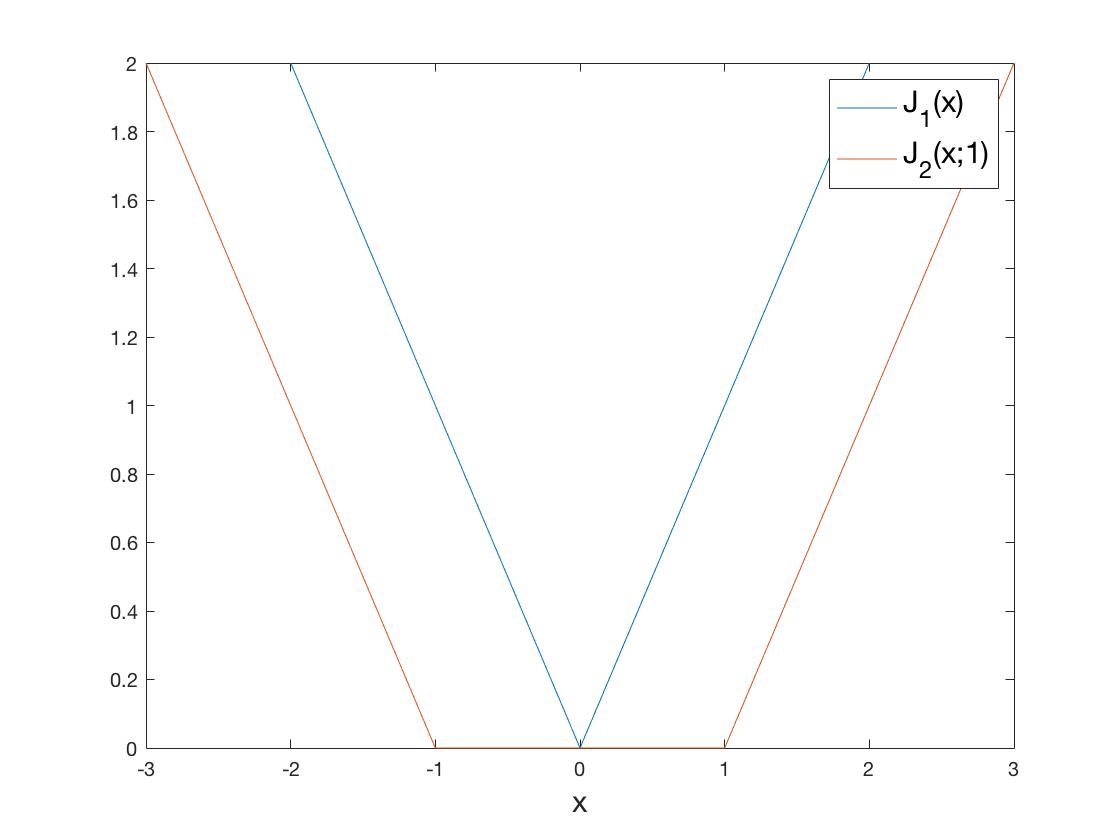} \label{fig:dc}}
    \subfigure[]{     \includegraphics[width=2in]{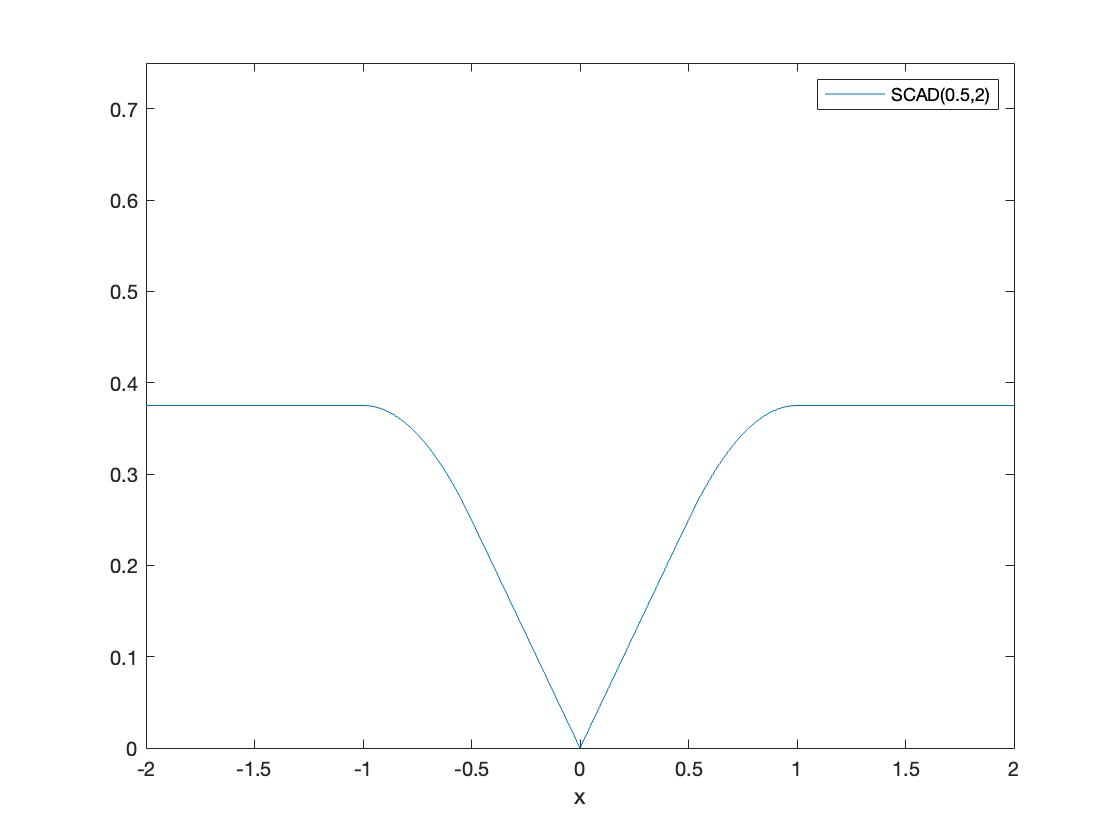} \label{fig:SCAD}}
    \caption{ (a): truncated Lasso function $\text{TLP}(x;\tau)$ with $\tau = 1$; (b): the DC decomposition into a difference of two convex functions $J_1(x)$ and $J_2(x;1)$; (c): SCAD penalty with $\lambda = 0.5$ and $a = 2$. }
\end{figure}

The penalty $\mathcal{J}(\bm{\theta}_j)$ is enforced on the item parameters for different latent classes item-wisely, which aims to learn the inner structure of the item parameter matrix. 
In particular, as  discussed in Section \ref{sec-hcdm}, due to the restrictions of the $Q$-matrix and item model assumptions, for each item, some subset  of latent attribute profiles have the same item parameters; and such constraint structure of the item parameter matrix can be used to further estimate the hierarchical structure and Q-matrix.
Therefore, in order to learn the set of latent classes that share the same item parameters we put the penalty function $\mathcal{J}(\cdot)$  on the differences among the item parameters for each item. 
A popular choice for shrinkage estimation is the Lasso penalty, which however is known to  produce biased estimation results.
To overcome this issue, we propose to use the grouped truncated Lasso penalty \citep{shen2012likelihood}, 
$$\mathcal{J}_{\tau}(\bm{\theta}_j)=\sum_{1\leq k < l \leq M} \text{TLP}(|\theta_{jk} - \theta_{jl}|;\tau),$$
where $\text{TLP}(x;\tau)=\min(|x|,\tau)$, and $\tau$ here is a positive tuning parameter.
Figure \ref{fig:tlp} provides an example for the TLP with $\tau = 1$.
Moreover, since we only focus on the item parameters for significant latent classes, we use $$\mathcal{J}_{\tau, \rho_N}(\bm{\theta}_j)= \underset{\substack{1\leq k < l \leq M,\\ \pi_l > \rho_N, \pi_k > \rho_N}}{\sum}
\text{TLP}(|\theta_{jk} - \theta_{jl}|;\tau).$$
A key feature for the truncated Lasso penalty is that it can be regarded as $L_1$ penalty for a small $|x|\leq \tau$, while it does not put further penalization for a large $|x|>\tau$. 
In this  way, it corrects the Lasso bias through adaptive shrinkage combining   with thresholding.
It discriminates small from large differences through thresholding and consequently is capable of handling low-resolution difference through tuning $\tau$.

s
In \cite{chen2017regularized}, the authors used the Smoothly Clipped Absolute Deviation (SCAD) penalty \citep{fan2001variable}, which can also be used to merge similar item parameters.
The SCAD penalty is similar to the TLP, while there is a quadratic spline function between the $L_1$ penalty for small values and the constant penalty for large values.
Specifically, the SCAD penalty is expressed as below
\begin{equation*}
	p_{\lambda, a}^{\text{SCAD}}(x) = 
		\begin{cases} 
      		\lambda |x| & \text{if } |x| \leq \lambda, \\
      		-\Big(\frac{|x|^2 - 2a\lambda |x| + \lambda^2}{2(a-1)} \Big) & \text{if } \lambda < x < a\lambda, \\
      		\frac{(a+1)\lambda^2}{2} & \text{if } |x| > a\lambda.
   \end{cases}
\end{equation*}
\noindent 
Figure \ref{fig:SCAD} plots the SCAD penalty with $\lambda=0.5$ and $a=2$.
Here, we want to mention several additional advantages for using the truncated Lasso penalty.  
First, it performs the model selection task of the $L_0$ function by providing a computationally efficient surrogate. 
When $\tau$ is sufficiently small, the truncated Lasso penalty has a good approximation to the $L_0$ penalty. 
Moreover, although it is not a convex function, it is piecewise linear and can be decomposed into a difference of two convex functions as illustrated in Figure \ref{fig:tlp} and Figure \ref{fig:dc}, which allows us to use Difference Convex (DC) programming \citep{tuy1995dc}, gaining computational advantages. 
 TLP also has nice likelihood oracle properties studied in previous literatures \citep{shen2012likelihood}.

\begin{remark}
	Our approach shares some similarities with the regularized LCM approach in \cite{chen2017regularized} that both use exploratory LCM to estimate the latent structures.
	However, in \cite{chen2017regularized}, 
	 the number of latent classes is pre-specified or selected in a way that all the possible values should be considered.
	This could require significantly more computational efforts when the number of latent attributes $K$ is large since there will be $2^K$ possible latent classes. For instance $K=10$ would lead to $2^K=1024$ possible candidate $M$ values.
	On the contrary, in our method, we only need an upper bound for the number of latent classes, and our model would perform the selection of significant latent classes more efficiently through the added log penalty.  Moreover, we also develop in Section \ref{sec-algo} a novel estimation algorithm to recover the number of latent attributes, the hierarchical structures among the attributes, and the $Q$-matrix, based on the proposed regularization estimation results. 
\end{remark}

\begin{remark}
	In \cite{wang2021learning}, the authors also studied the latent variable selection approach, which, however, required a pre-specified number of latent attributes and a known identity sub-matrix in the $Q$-matrix.
	Moreover, a hard cutoff for proportion parameters was used to select significant latent classes.
	For example, they chose 0.05 as the cutoff when $K=3$ and 0.025 when $K=4$. 
	This hard cutoff in fact plays a decisive role in determining the significant latent classes.
	However, there is neither a systematical way nor theoretical guarantee to select this cutoff, making it less practical in real applications.
\end{remark}

\subsection{Theoretical Proporties}
\label{sec-theory}

In this section, we present statistical properties of the regularized estimator obtained from (\ref{eq-objective}). 
We first present some identifiability results of hierarchical cognitive diagnosis model from \cite{gu2019learning}.
Then we will show that, under suitable conditions, the regularized estimator is consistent for model selection.
In the following, for two vectors $\bm{a}$ and $\bm{b}$ of dimension $n$, we say $\bm{a} \preceq \bm{b}$ if $a_i \leq b_i$ for $i = 1, \dots, n$, and $\bm{a} \succeq \bm{b}$ if $a_i \geq b_i$ for $i = 1, \dots, n$.

The identifiability of the model parameters depends on the restrictions of the item parameter matrix.
To characterize the identifiability conditions, we introduce an indicator matrix of most capable classes as $\bm{\Gamma}:= \big(\mathbb{I}\{\theta_{j,m}=\max_{m'\in [M]}\theta_{j,m'}\}, j \in [J], m \in [M] \big)\in\{0,1\}^{J\times M}$, indicating whether the latent classes possess the highest level of each item's parameters. Let $\boldsymbol{\Gamma}_{\cdot,m} $ denote the $m$th column vector of the $\bm{\Gamma}$ matrix.
Based on the indicator matrix, we can define a partial order among latent classes.
For $1\leq m_1\neq m_2\leq M$, we say latent class $m_1$ is of a larger order than latent class $m_2$ under $\boldsymbol{\Gamma}$ if $\boldsymbol{\Gamma}_{\cdot,m_1} \succeq \boldsymbol{\Gamma}_{\cdot,m_2}$. 
See Figure \ref{fig-gamma} for an illustrative example, where we use a Directed Acyclic Graph (DAG) to represent partial orders, where
$\boldsymbol{\Gamma}_{\cdot,m_1}$ points to $\boldsymbol{\Gamma}_{\cdot,m_2}$ if $\boldsymbol{\Gamma}_{\cdot,m_1}\preceq \boldsymbol{\Gamma}_{\cdot,m_2}$.

\begin{figure}[ht]
	\centering
	\subfigure{
	$\bm{\Theta} = \begin{pmatrix}
		0.2 & 0.8 & 0.8 \\
		0.2 & 0.2 & 0.8 \\
		0.2 & 0.2 & 0.8
	\end{pmatrix}$,
		}
	\hspace{0.2in}	
	\subfigure{
	$\bm{\Gamma} = \begin{pmatrix}
		0 & 1 & 1 \\
		0 & 0 & 1 \\
		0 & 0 & 1
	\end{pmatrix}$,
	}
	\hspace{0.2in}
	\subfigure{
	\begin{tikzpicture}[scale=0.9]
    \path
    node at (-1.5, -2.5) [place] (1) {$\Gamma_{\cdot, 1}$}
    node at (0, -2.5) [place] (2) {$\Gamma_{\cdot, 2}$}
    node at (1.5, -2.5) [place] (3) {$\Gamma_{\cdot, 3}$};
    \draw [->, thick] (1) to (2);
    \draw [->, thick] (2) to (3);
	\end{tikzpicture}
	}
	\caption{Indicator matrix and partial orders}
\label{fig-gamma}	
\end{figure}

As in \cite{gu2019learning}, for CDMs,
we define the indicator matrix for a set of latent attribute profiles  $\mathcal{A}$ as $\bm{\Gamma}^{\mathcal{A}}:= \big(\mathbb{I}\{\theta_{j,\boldsymbol{\alpha}}=\max_{\boldsymbol{\alpha}'\in\mathcal{A}}\theta_{j,\bm{\alpha}'}\}: \ j\in [J], \ \bm{\alpha} \in \mathcal{A}\big)\in\{0,1\}^{J\times|\mathcal{A}|}$.
Note that if we take the set of latent classes as the set of attribute profiles, the indicator matrix of an LCM is equivalent to that of a CDM.
Similarly, we define the proportion parameter vector and item parameter matrix for a set of latent attribute profiles $\mathcal{A}$ as $\bm{\pi}^{\mathcal{A}} = \big( \pi_{\bm{\alpha}} :\ \bm{\alpha} \in \mathcal{A} \big)$ and $\bm{\Theta}^{\mathcal{A}} = \big(\theta_{j,\bm{\alpha}}: \ j\in [J], \ \bm{\alpha} \in \mathcal{A}\big)$.
Following \cite{gu2019learning}, for any subset of items $S\subset [J]$, we define a partial order among the latent attribute profiles. 
For $\bm{\alpha}, \bm{\alpha}' \in \mathcal{A}$, we say $\bm{\alpha}\succeq_{S}\bm{\alpha}'$ under $\bm{\Gamma}^{\mathcal{A}}$ if $\Gamma_{j,\bm{\alpha}}^{\mathcal{A}} \geq \Gamma_{j,\bm{\alpha}'}^{\mathcal{A}}$ for $j\in S$. 
And for two item sets $S_1$ and $S_2$, we say ``$\succeq_{S_1}=\succeq_{S_2}$" if for any $\bm{\alpha}, \bm{\alpha}' \in \mathcal{A}$, we have $\bm{\alpha}\succeq_{S_1}\bm{\alpha}'$ if and only if  $\bm{\alpha}\succeq_{S_2}\bm{\alpha}'$.
Note that if we take the item set to be the set of all items, the definitions of indicator matrix and partial orders are the same as those in Section \ref{sec-Motivation}.
For a subset of items $S\subset [J]$ and a set of  attribute profiles $\mathcal{A}$, we define the corresponding indicator matrix $\Gamma^{(S,\ \mathcal{A})} = \big(\Gamma_{j, \bm{\alpha}}: \ j \in S, \ \bm{\alpha} \in \mathcal{A} \big)$.

We first state the definition of strict identifiability for latent hierarchy and model parameters.

\begin{definition}[strict identifiability, \cite{gu2019learning}]
\ \linebreak
Consider a CDM with a hierarchy $\mathcal{E}_0$ and the induced latent attribute profile set $\mathcal{A}_0$.
$\mathcal{A}_0$ is said to be (strictly) identifiable if for any indicator matrix $\bm{\Gamma}^{\mathcal{A}}$ of size $J \times |\mathcal{A}|$ with $|\mathcal{A}| \leq |\mathcal{A}_0|$, any proportion parameter vector $\bm{\pi}^{\mathcal{A}}$ and any valid item parameter matrix $\bm{\Theta}^{\mathcal{A}}$ respecting constraints given by $\bm{\Gamma}^{\mathcal{A}}$, the following equality
\begin{equation}
	\mathbb{P}(\bm{R}\mid \bm{\pi}^{\mathcal{A}}, \bm{\Theta}^{\mathcal{A}}) = \mathbb{P}(\bm{R}\mid \bm{\pi}^{\mathcal{A}_0}, \bm{\Theta}^{\mathcal{A}_0})
	\label{eq-identify}
\end{equation}
implies $\mathcal{A} = \mathcal{A}_0$. 
Moreover, if \eqref{eq-identify} implies $\big(\bm{\pi}^{\mathcal{A}}, \bm{\Theta}^{\mathcal{A}}\big) = \big(\bm{\pi}^{\mathcal{A}_0}, \bm{\Theta}^{\mathcal{A}_0}\big)$, then we say the model parameters $\big(\bm{\pi}^{\mathcal{A}_0}, \bm{\Theta}^{\mathcal{A}_0}\big)$ are (strictly) identifiable.
\end{definition}

The following theorem provides sufficient conditions for strict identifiability of latent hierarchies and model parameters.

\begin{theorem}[strict identifiability, \cite{gu2019learning}]
\ \linebreak
Consider a CDM with a hierarchy $\mathcal{E}_0$.
The hierarchy is identifiable if the following conditions of the indicator matrix $\bm{\Gamma}^{\mathcal{A}_0}$ corresponding to the induced latent attribute profile set $\mathcal{A}_0$ are satisfied:
\renewcommand\labelenumi{(\theenumi)}
\begin{enumerate}
    \item There exist two disjoint item sets $S_1$ and $S_2$, such that $\bm{\Gamma}^{(S_i,\ \mathcal{A}_0)}$ has distinct column vectors for $i=1,2$ and ``$\succeq_{S_1}$"=``$\succeq_{S_2}$" under $\bm{\Gamma}^{\mathcal{A}_0}$.
    \item For any $\bm{\alpha},\bm{\alpha}'\in \mathcal{A}_0$ where $\bm{\alpha}'\succeq_{S_i}\bm{\alpha}$ under $\bm{\Gamma}^{\mathcal{A}_0}$ for $i=1$ or 2, there exists some $j\in \big(S_1\cup S_2\big)^c$ such that $\Gamma_{j,\bm{\alpha}}^{\mathcal{A}_0}\neq \Gamma_{j,\bm{\alpha}'}^{\mathcal{A}_0}$.
    \item Any column vector of $\bm{\Gamma}^{\mathcal{A}_0}$ is different from any column vector of $\bm{\Gamma}^{\mathcal{A}_0^c}$, where $\mathcal{A}_0^c = \{0,1\}^K\setminus \mathcal{A}_0$.
\end{enumerate}{}
Moreover, under Conditions (1) - (3), the model parameters $\big(\bm{\pi}^{\mathcal{A}_0}, \bm{\Theta}^{\mathcal{A}_0}\big)$ associated with $\mathcal{A}_0$ are also identifiable.
\label{thm:identify}
\end{theorem}{}

Theorem \ref{thm:identify} provides conditions for strict identifiability of hierarchical structures and model parameters.
The strict identifiability can be relaxed to generic identifiability, where the hierarchy and model parameters can be identified expect for a zero measure set. 
The definition of generic identifiability is defined below.

\begin{definition}[generic identifiability, \cite{gu2019learning}]
\ \linebreak
Consider a CDM with a hierarchy $\mathcal{E}_0$ and the induced latent attribute profile set $\mathcal{A}_0$.
Denote the parameter space of $(\bm{\pi}^{\mathcal{A}_0}, \bm{\Theta}^{\mathcal{A}_0})$ constrained by $\bm{\Gamma}^{\mathcal{A}_0}$ by $\Omega$.
We say $\mathcal{A}_0$ is generically identifiable, if there exists a subset $\mathcal{V}\subset \Omega$  that has a Lebesgue measure zero, such that for any $(\bm{\pi}^{\mathcal{A}_0}, \bm{\Theta}^{\mathcal{A}_0}) \in \Omega \setminus \mathcal{V}$, Equation \eqref{eq-identify} implies $\mathcal{A} = \mathcal{A}_0$.
Moreover, for any $(\bm{\pi}^{\mathcal{A}_0}, \bm{\Theta}^{\mathcal{A}_0}) \in \Omega \setminus \mathcal{V}$, if\eqref{eq-identify} implies $\big(\bm{\pi}^{\mathcal{A}}, \bm{\Theta}^{\mathcal{A}}\big) = \big(\bm{\pi}^{\mathcal{A}_0}, \bm{\Theta}^{\mathcal{A}_0}\big)$, then we say  $\big(\bm{\pi}^{\mathcal{A}_0}, \bm{\Theta}^{\mathcal{A}_0}\big)$ are generically identifiable.
\end{definition}
 
The next theorem presents the generic identifiability results of latent hierarchies and model parameters.

 \begin{theorem}[generic identifiability, \cite{gu2019learning}]
Consider a CDM with a hierarchy $\mathcal{E}_0$.
The hierarchy is generically identifiable, if the following conditions of the indicator matrix $\bm{\Gamma}^{\mathcal{A}_0}$ corresponding to the induced latent attribute profile set $\mathcal{A}_0$ are satisfied:

\begin{enumerate}[label=(\Alph*)]
    \item There exist two disjoint item sets $S_1$ and $S_2$, such that altering some entries from 0 to 1 in $\bm{\Gamma}^{(S_1\cup S_2,\ \mathcal{A}_0)}$ yields a $\Tilde{\bm{\Gamma}}^{(S_1\cup S_2,\ \mathcal{A}_0)}$ satisfying that $\Tilde{\bm{\Gamma}}^{(S_i,\ \mathcal{A}_0)}$ has distinct column vectors for $i=1,2$ and ``$\succeq_{S_1}$"=``$\succeq_{S_2}$" under $\Tilde{\bm{\Gamma}}^{\mathcal{A}_0}$.
    \item For any $\bm{\alpha},\bm{\alpha}'\in \mathcal{A}_0$ where $\bm{\alpha}'\succeq_{S_i}\bm{\alpha}$ under $\bm{\Gamma}^{\mathcal{A}_0}$ for $i=1$ or 2, there exists some $j\in \big(S_1\cup S_2\big)^c$ such that $\Tilde{\Gamma}_{j,\bm{\alpha}}^{\mathcal{A}_0}\neq \Tilde{\Gamma}_{j,\bm{\alpha}'}^{\mathcal{A}_0}$.
    \item Any column vector of $\bm{\Gamma}^{\mathcal{A}_0}$ is different from any column vector of $\bm{\Gamma}^{\mathcal{A}_0^c}$, where $\mathcal{A}_0^c = \{0,1\}^K\setminus \mathcal{A}_0$.
\end{enumerate}{}
Moreover, under conditions (A) - (C), the model parameters $\big(\bm{\pi}^{\mathcal{A}_0}, \bm{\Theta}^{\mathcal{A}_0}\big)$ associated with $\mathcal{A}_0$ are also generically identifiable.
\label{thm:generic}
\end{theorem}

Theorem \ref{thm:generic} establishes generic identifiability conditions where the hierarchical structure and model parameters can be identified except for a zero measure set of parameters.
To establish consistency results, we need to make the following assumption.

\begin{assumption}
$\big[l_N(\hat{\bm{\pi}}^*, \hat{\bm{\Theta}}^*) - l_N(\hat{\bm{\pi}}_0, \hat{\bm{\Theta}}_0 )\big] / N =O_p(N^{-\delta}),$ for some $1/2 < \delta \leq 1$,
 where $(\hat{\bm{\pi}}^*, \hat{\bm{\Theta}}^*)$ is the maximum likelihood estimator (MLE) directly obtained from \eqref{eq:log-like}, and $(\hat{\bm{\pi}}_0, \hat{\bm{\Theta}}_0)$ is the Oracle MLE obtained under the condition that the number of latent attributes, the hierarchical structure, the $Q$-matrix and item-level diagnostic models are known.
\label{assum-mle}
\end{assumption}
When $\delta = 1$,  Assumption 1 corresponds to the usual root-N convergence rate of the estimators, while $1/2  <  \delta \leq 1$ corresponds to a slower convergence rate. 
Here we make a general assumption to cover different situations.  
In \cite{gu2019learning}, the authors   made a similar assumption about the convergence rate of the likelihood.

We use $(\bm{\pi}^0, \bm{\Theta}^0)$ to denote the true model parameter and $M_0 := |\mathcal{A}_0|$ to denote the true number of latent classes, where $\mathcal{A}_0$ is the reduced latent attribute profile set under the true hierarchical structure $\mathcal{E}_0$.
For $(\hat{\bm{\pi}}, \hat{\bm{\Theta}})$ obtained from optimizing \eqref{eq-objective}, we define the selected latent classes as $\{m: \hat{\pi}_m > \rho_N, \ m\in [M] \}$, and the number of selected latent classes $\hat{M} := \big|\{m: \hat{\pi}_m > \rho_N, \ m\in [M] \}\big|$.
For the true item parameter matrix $\bm{\Theta}^0$, we defined the set $S^0 = \big\{(j,k_1,k_2):\theta_{j,k_1}^0=\theta_{j,k_2}^0,1\leq k_1<k_2\leq M_0, 1 \leq j \leq J\big\}$ to indicate the constraint structure of the item parameter matrix.
Similarly, for $(\hat{\bm{\pi}}, \hat{\bm{\Theta}})$, we define $\hat{S} = \big\{(j,k_1,k_2): \hat{\theta}_{j,k_1}=\hat{\theta}_{j,k_2},\ 1\leq k_1<k_2\leq M, \ \hat{\pi}_{k_1}>\rho_N, \ \hat{\pi}_{k_2}>\rho_N\big\}$.
We say $\hat{S} \sim S^0$ if there exists a column permutation $\sigma$ of $\hat{\bm{\Theta}}$ such that $\hat{S}_{\sigma} = S^0$. 
Given the above assumptions, we have the following consistency results.

\begin{theorem}[consistency]\label{thm:consistency}
Suppose the identifiability conditions in Theorem \ref{thm:identify} are satisfied and Assumption \ref{assum-mle} holds.
For $\lambda_1$, $\lambda_2$, $\tau$ and $\rho_N$ satisfying
$N^{1-\delta}|\log \rho_N|^{-1}  = o(\lambda_1)$, $ \lambda_1 = o( N |\log \rho_N|^{-1})$  and 
$\lambda_2 \tau = o(\lambda_1 |\log \rho_N| )$, 
we can select the true number of latent classes consistently, that is, 
\begin{equation}
    \mathbb{P} \big(\hat{M} \neq M_0 \big)\xrightarrow{} 0, \text{ as } N\xrightarrow{}\infty.
    \label{eq:consistency-pi}
\end{equation}
Moreover, the estimated parameter $(\hat{\bm{\pi}}, \hat{\bm{\Theta}})$ is also consistent of $(\bm{\pi}^0, \bm{\Theta}^0)$.
If we further assume $\lambda_1 = o(N^{1/2})$, $\lambda_2 \tau = o(N^{1/2})$, $\lambda_2 N^{-1/2}\rightarrow\infty$ and $\tau N^{1/2}\rightarrow \infty$,
up to a column permutation,
the identical item parameter pair set $S^0$ is also consistently estimated,
\begin{equation}
	\mathbb{P}\big(\hat{S} \nsim S^0  \big) \rightarrow 0, \text{ as } N\rightarrow \infty.
\end{equation}

\end{theorem}
Theorem \ref{thm:consistency} implies that with suitable choices of hyperparameters, we can correctly select the number of latent classes and learn the inner structure of the item parameter matrix consistently as sample size $N$ goes to infinity.
For example, we can take $\rho_N \sim  N^{-d}$ for some $d \geq 1$, $\lambda_1 \sim N^{\frac{1}{2} - \epsilon_1}$, $\lambda_2 \sim N^{\frac{1}{2}+ \epsilon_2}$ and $\tau \sim N^{-\epsilon_3}$ for some small positive constants $\epsilon_1, \epsilon_2, \epsilon_3$ satisfying that $0 < \epsilon_1 < \delta - 1/2, \ 0 < \epsilon_2 < \epsilon_3 < 1/2$ and $\epsilon_3 - \epsilon_2 > \epsilon_1$.
Moreover, if the conditions in Theorem \ref{thm:generic} are satisfied, we can consistently estimate the true number of latent classes and inner structure of the item parameter matrix except for a zero measure set of model parameters.
In practice, we can use  information criteria, such as the Bayesian Information Criterion \citep[BIC,][]{schwarz1978estimating}, to help select the tuning parameters, which will be  further discussed in Section \ref{em}. 
The proof of the theorem is presented in the supplementary material.

Based on the learned latent classes and estimated item parameter matrix, we develop a latent structure recovery algorithm outlined in Algorithm~\ref{algo-binary} in Section \ref{sec-algo}. 
Specifically, we recover the number of latent attributes, the latent hierarchies, and the $Q$-matrix based on the partial orders among the selected latent classes.
Under the identifiability conditions, due to the consistency of the item parameter estimator $\hat{\Theta}$ and the inner structure $\hat{S}$ established, we can also consistently recover the partial orders among the latent classes, which ultimately leads to the consistency of the estimated number of latent attributes, the hierarchical structures and the $Q$-matrix. 
For algorithm details, please see Section \ref{sec-algo}.

\section{Learning Algorithms}
\label{sec-algo}

\subsection{Penalized EM Algorithm}
\label{em}
In this section, we develop an efficient EM algorithm for the proposed model.
For an LCM, the complete data log-likelihood function can be written as 
\begin{equation}
    l_C(\mathcal{R},\bm{z};\bm{\pi},\bm{\Theta}) = \sum_{i=1}^N\sum_{k=1}^M z_{ik}\log \big(\pi_k \varphi(\bm{R}_i;\bm{\theta}_k)\big),
\end{equation}
where $\varphi(\bm{R}_i;\bm{\theta}_k) = \prod_{j=1}^J \theta_{jk}^{R_{ij}}(1-\theta_{jk})^{1-R_{ij}}$ and $\bm{z}\in\{0,1\}^{N\times M}$ is the latent variable in which $z_{ik}$ indicates whether the $i$th subject belongs to the $k$th latent class. 
Then in an EM algorithm without additional penalty, we maximize the following objective function at the $
(c+1)$th iteration:
\begin{equation}
\underset{\boldsymbol{\pi}, \boldsymbol{\Theta}}{\max}\ 
    Q(\bm{\pi},\bm{\Theta}\mid \bm{\pi}^{(c)},\bm{\Theta}^{(c)}) = \sum_{i=1}^N \sum_{k=1}^M s_{ik}^{(c)}\big(\log \pi_k + \log \varphi_k (\bm{R}_i;\bm{\theta}_k)\big),
\end{equation}
where $$s_{ik}^{(c)}=\mathbb{E}_{\bm{\pi}^{(c)},\bm{\Theta}^{(c)}}[z_{ik}=1\mid \bm{R}_i]=\frac{\pi_k^{(c)}\varphi_k(\bm{R}_i;\bm{\theta}_k^{(c)})}{\sum_{k'}\pi_{k'}^{(c)}\varphi_{k'}^{(c)}(\bm{R}_i;\bm{\theta}_{k'}^{(c)})}.$$ 
With the additional penalty terms in (\ref{eq-objective}),  the new objective function denoted as $G(\bm{\pi},\bm{\Theta}\mid \bm{\pi}^{(c)},\bm{\Theta}^{(c)})$ becomes:
\begin{equation}
   \underset{\boldsymbol{\pi}, \boldsymbol{\Theta}}{\min} \ G(\bm{\pi},\bm{\Theta}\mid \bm{\pi}^{(c)},\bm{\Theta}^{(c)})= -\frac{1}{N} Q(\bm{\pi},\bm{\Theta}\mid \bm{\pi}^{(c)},\bm{\Theta}^{(c)}) + \tilde{\lambda}_1 \sum_{k=1}^M \log_{[\rho_N]} \pi_k + \tilde{\lambda}_2 \sum_{j=1}^{J}\mathcal{J}_{\tau, \rho_N}(\bm{\theta}_j),
\end{equation}
where $\tilde{\lambda}_1 = \lambda_1 / N$ and $\tilde{\lambda}_2 = \lambda_2 / N$.

As we mentioned in Section \ref{sec-Motivation}, the truncated Lasso penalty can be decomposed into a difference of two convex functions.
Therefore we can utilize DC programming \citep{tuy1995dc} to optimize $G$. 
Moreover, we also exploit the Alternating Direction Method of Multipliers \citep[ADMM,][]{boyd2011distributed} method to facilitate solving the problem. 
There are several advantages of using ADMM to perform optimization here.
Updating the parameters in an alternating or sequential fashion takes advantage of the decomposability of dual ascent,
while using the method of multipliers enables superior convergence properties \citep{boyd2011distributed}.
In practice, we also observe that the ADMM algorithm converges within a few tens of iterations in our simulation and real data studies.
The algorithm is summarized in Algorithm \ref{algo-pem}
and the derivations of the algorithm is presented in the supplementary material.

\begin{algorithm}[h!]
\caption{PEM: Penalized EM with $\log$-penalty and truncated Lasso penalty}
\label{algo-pem}
\SetKwInOut{Input}{Input}
\SetKwInOut{Output}{Output}

\KwData{Binary response matrix $\mathcal{R}=(R_{i,j})_{N\times J}$.}
Set hyperparameters $\tilde{\lambda}_1,\ \tilde{\lambda}_2, \ \tau$, $\gamma$ and $\rho$.

Set an upper bound for the number of latent classes $M$.

Initialize parameters $\bm{\pi}$, $\bm{\Theta}$, and the conditional expectations $\bm{s}$.

\While{not converged}{

In the $(c+1)th$ iteration,

\For{$(i,k)\in[N]\times[M]$}{
$s_{ik}^{(c+1)}=\frac{\pi_k^{(c)}\varphi_k(\bm{R}_i;\bm{\theta}_k^{(c)})}{\sum\pi_{k'}^{(c)}\varphi_{k'}(\bm{R}_i;\bm{\theta}_{k'}^{(c)})}$,  where $\varphi(\bm{R}_i;\bm{\theta}_k) = \prod_{j=1}^J \theta_{jk}^{R_{ij}}(1-\theta_{jk})^{1-R_{ij}}$.
}

\For{$k\in[M]$ and $\pi_k^{(c)} > \rho$}{
$\pi_k^{(c+1)} = \frac{\sum_{i=1}^N s_{ik}^{(c+1)}/N - \tilde{\lambda}_1}{1-M\tilde{\lambda}_1}$.
}

\For{$(j,k)\in[J]\times[M]$ and $\pi_k  ^{(c+1)} > \rho $}{
\begin{align*}
        \theta_{jk}^{(c+1)} = \underset{\theta_{jk}}{\mathrm{argmin}}&
        \Big\{-\frac{\sum_{i=1}^N s_{ik}^{(c)}R_{ij}}{N}\log(\theta_{jk})
        - \frac{\sum_{i=1}^N s_{ik}^{(c)}(1-R_{ij})}{N} \log (1-\theta_{jk})\\
        &+ \frac{\gamma}{2}\sum_{l>k}\big(\hat{d}_{jkl}^{(c)}-(
        \theta_{jk}-\hat{\theta}_{jl}^{
        (c)})+\hat{\mu}_{jkl}^{(c)}\big)^2
        \\
        &+\frac{\gamma}{2}\sum_{l<k}\big(\hat{d}_{jlk}^{(c)}-(
        \hat{\theta}_{jl}^{(c+1)}-\theta_{jk})+\hat{\mu}_{jlk}^{(c)}\big)^2
        \Big\}
    \end{align*}
}

\For{$j\in[J]$, $1\leq k < l \leq M$ and $\pi_k^{(c+1)}  >\rho $, $\pi_l^{(c+1)}  >\rho $}{
\begin{align*}
    \hat{d}_{jkl}^{(c+1)}= & \left\{
        \begin{aligned}
        \big( &\hat{\theta}_{jk}^{(c+1)} - \hat{\theta}_{jl}^{(c+1)} -\hat{\mu}_{jkl}^{(c)}\big)\cdot I\big( |\hat{d}_{jkl}^{(c)}| \geq \tau\big)\\
        &\text{ST}\big(\hat{\theta}_{jk}^{(c+1)} - \hat{\theta}_{jl}^{(c+1)} -\hat{\mu}_{jkl}^{(c)};\tilde{\lambda}_2/\gamma \big)\cdot I\big( |\hat{d}_{jkl}^{(c)}| < \tau\big),  
        \end{aligned}
    \right.\\
    & \quad \text{  where ST}(x;\gamma) = (|x|-\gamma)_+ x/|x|.\\
     \hat{\mu}_{jkl}^{(c+1)} = & \hat{\mu}_{jkl}^{(c)} + \hat{d}_{jkl}^{(c+1)} - \big(
    \hat{\theta}_{jk}^{(c+1)} - \hat{\theta}_{jl}^{(c+1)}\big)
\end{align*}{}
}
}
\Output{$\big\{\bm{\hat{\pi}},\ \bm{\hat{\Theta}}, \ \bm{\hat{s}}\big\}$}

\end{algorithm}{}

We want to note that our algorithm can naturally handle missing values. 
If $\mathcal{R} = (\mathcal{R}_{\text{obs}},\mathcal{R}_{\text{miss}})$ is the decomposition of the full data matrix into the observed part $\mathcal{R}_{
\text{obs}}$ and the missing part $\mathcal{R}_{\text{miss}}$, then after marginalization over the missing values, the initial likelihood $l(\mathcal{R};\bm{\pi},\bm{\Theta})$ is simplified to $l(\mathcal{R}_{\text{obs}};\bm{\pi},\bm{\Theta})$. 
Then a natural implementation could be based on indexing the inference procedure so that the posterior conditionals only involve sums over the observed values.
The detailed Penalized EM algorithm with missing values is summarized in the supplementary material.

To address the computational bottleneck when faced with large scale datasets, we can also use a stochastic version of the aforementioned EM algorithm. 
In each iteration, we randomly subsample a subset $S_r$ of rows (subjects), and a subset $S_c$ of columns (items).
Then update the conditional expectation $s_{i}^{(c)}$ for $i \in S_r$ with items in $S_c$. 
Updates in M-step remain the same, which will give us an intermediate model parameter $(\hat{\bm{\pi}}^{(c+1/2)}, \hat{\bm{\Theta}}^{(c+1/2)})$.
Then we use a weighted average of $(\hat{\bm{\pi}}^{(c)},\hat{\bm{\Theta}}^{(c)})$ and $(\hat{\bm{\pi}}^{(c+1/2)}, \hat{\bm{\Theta}}^{(c+1/2)})$ to update the model parameters.
Appropriate weights will provably lead to convergence to a local optimum \citep{delyon1999convergence}.

In terms of hyperparameter tuning, we use BIC defined as below:
\begin{equation}
	\text{BIC}(\bm{\pi},\bm{\Theta}) = -2 l_N (\bm{\pi},\bm{\Theta}) + \log N \big(M_{\rho_N} - 1 + \sum_{j=1}^J\text{dim}(\bm{\theta}_j) \big)
	\label{eq:bic}
\end{equation}
where $l_N$ is the log-likelihood, 
$M_{\rho_N} := \big| \{ m: \pi_m > \rho_N, m\in [M] \} \big|$ is the selected number of latent classes, 
and $\text{dim}(\bm{\theta}_j)$ is the number of distinct values in the set $\{\theta_{j,m}: \pi_m > \rho_N, \ m \in [M] \}$, that is, the number of distinct item parameters for item $j$ corresponding to the selected latent classes.
Our simulation results in Section \ref{sec-simu} show that BIC performed well.
We can also use other selection criteria such as EBIC \citep{chen2008extended} when the number of latent attributes $K$ is large.
From the matrix completion perspective, we may also perform cross validation to choose tuning parameters.

\subsection{Recover Latent Hierarchies and $Q$-matrix}
\label{sec-hier}

Once we fit the model and get the estimates of the model parameters including the number of significant latent classes $\hat{M}$, proportion parameters $\hat{\bm{\pi}}$ and item parameter matrix $\hat{\bm{\Theta}}$, 
our next goal is to recover the number of latent attributes, the latent hierarchical structure, the $Q$-matrix and item models. 

To this end, we develop an algorithm based on the indicator matrix $$\bm{\Gamma}:= \left(\mathbb{I}\{\hat{\theta}_{j,m}=\max_{l\in[\hat{M}]}\hat{\theta}_{j,l}\}: \ j \in [J], \ m \in [\hat{M}]\right)\in\{0,1\}^{J\times\hat{M}},$$ indicating whether a latent class possesses the highest level of an item's parameters.
One common assumption in CDMs is that more capable subjects have higher item parameters and thus larger indicator vectors, that is, $\boldsymbol{\Gamma}_{\cdot.\bm{\alpha}} \succeq \boldsymbol{\Gamma}_{\cdot,\bm{\alpha}^*}$, if $\bm{\alpha}\succeq \bm{\alpha}^*$.
Based on this assumption, we can get partial orders among the latent classes.
Then we can find the smallest integer $K$ such that some binary representations with $K$ digits satisfy these partial orders, and the binary representations can be treated as the  learned latent attribute profiles.
With these reconstructed latent attribute profiles, we can subsequently recover the hierarchical structures among the latent attributes and the $Q$-matrix.

Specifically, based on the indicator matrix, we get the partial orders among the latent classes.
We use a matrix $\boldsymbol{P}\in\{0,1\}^{\hat{M}\times \hat{M}}$ to represent the partial orders, where $P_{m_1, m_2} = 1$ indicates that $\boldsymbol{\Gamma}_{\cdot,m_1}\preceq \boldsymbol{\Gamma}_{\cdot,m_2}$.
Since we only want to include direct partial orders, for any $(m_1, m_2)$ such that $\big(\boldsymbol{P}^2\big)_{m_1, m_2} > 0$, we set $P_{m_1, m_2} = 0$.
For example, if $\boldsymbol{\Gamma}_{\cdot,m_1}\preceq \boldsymbol{\Gamma}_{\cdot,m_2}$, $\boldsymbol{\Gamma}_{\cdot,m_1}\preceq \boldsymbol{\Gamma}_{\cdot,m_3}$ and $\boldsymbol{\Gamma}_{\cdot,m_2}\preceq \boldsymbol{\Gamma}_{\cdot,m_3}$, since $m_2$ here is an intermediate latent class between $m_1$ and $m_3$, we will not include the partial order $\boldsymbol{\Gamma}_{\cdot,m_1}\preceq \boldsymbol{\Gamma}_{\cdot,m_3}$ in $\boldsymbol{P}$.
From $\boldsymbol{P}$, we can get a partial order set $\{m_1 \rightarrow m_2: P_{m_1,m_2}=1\}$, based on which a DAG can be plotted, where $\boldsymbol{\Gamma}_{\cdot,m_1}$ points to $\boldsymbol{\Gamma}_{\cdot,m_2}$ if $\boldsymbol{\Gamma}_{\cdot,m_1}\preceq \boldsymbol{\Gamma}_{\cdot,m_2}$.
One can see the partial order matrix $\bm{P}$ in fact is the adjacency matrix of the DAG.
An example of the indicator matrix, the partial order matrix and the corresponding DAG is shown in Figure \ref{fig:Gamma-P}.
In a DAG, we call a node at the start of an arrow as a parent node, and a node at the end of an arrow as a child node.
Note that since we always include the most basic attribute profile with all attributes being 0 and the most capable attribute profile with all attributes being 1, and any other latent attribute profile will lie between them, there is always a path  passing each latent attribute profile from the most basic one to the most capable one.

\begin{figure*}[h!]
	\centering
	\subfigure{\label{fig:Gamma}
	$\bm{\Gamma} = 
    \begin{pmatrix}
    0 & 1 & 1\\
    0 & 0 & 1\\
    0 & 0 & 1\\
    \end{pmatrix},$
	}
	\hspace{0.2in}
	\subfigure{\label{fig:P}
	$
	\bm{P} = 
	\begin{pmatrix}
    0 & 1 & 0\\
    0 & 0 & 1\\
    0 & 0 & 0
    \end{pmatrix},$
	}
	\hspace{0.2in}
	\subfigure{
	\begin{tikzpicture}[scale=0.9]
    \path
    node at (-1.5, -2.5) [place] (1) {$\Gamma_{\cdot, 1}$}
    node at (0, -2.5) [place] (2) {$\Gamma_{\cdot, 2}$}
    node at (1.5, -2.5) [place] (3) {$\Gamma_{\cdot, 3}$};
    \draw [->, thick] (1) to (2);
    \draw [->, thick] (2) to (3);
	\end{tikzpicture}
	}
	\caption{Indicator matrix, partial order matrix and corresponding DAG}
	\label{fig:Gamma-P}
\end{figure*}

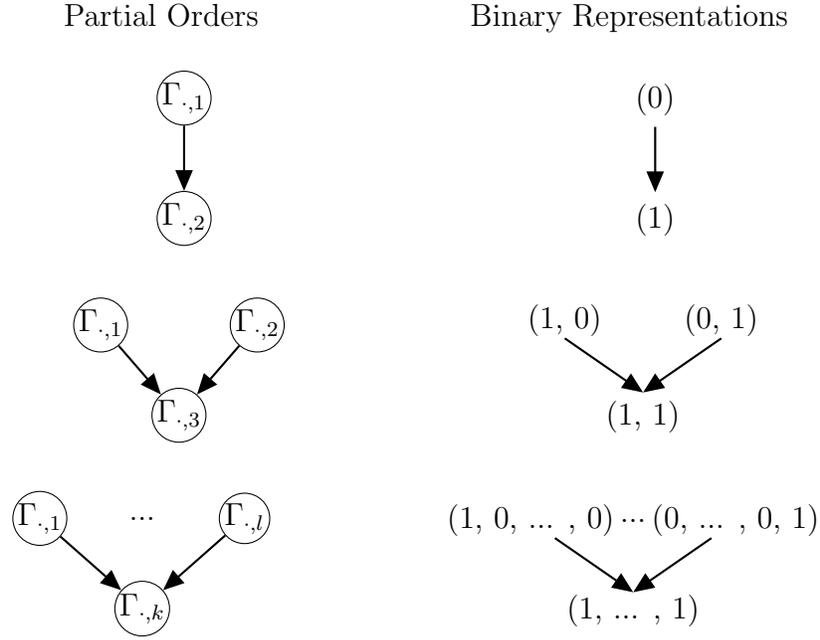
\begin{figure}[h!]

\centering
\subfigure{
	Partial Orders}\hspace{1in}
	\subfigure{
	Binary Representations}
\\\vspace{0.05in}
	\subfigure{
	\begin{tikzpicture}[scale=0.8]
    \path    node at (0, -2) [place] (1) {$\Gamma_{\cdot, 1}$}
    node at (0, -4) [place] (2) {$\Gamma_{\cdot, 2}$};
    \draw [->, thick] (1) to (2);
	\end{tikzpicture}
	}\hspace{2in}
	\subfigure{
	\begin{tikzpicture}[scale=0.8]
    \draw (0,-2) node {(0)}
  (0, -4) node {(1)};
  \draw [->, thick] (1) to (2);
	\end{tikzpicture}
	}
	\\
	\vspace{0.1in}
	\subfigure{
	\begin{tikzpicture}[scale=0.8]
    \path
    node at (-1.3, -2) [place] (1) {$\Gamma_{\cdot, 1}$}
    node at (0, -3.5) [place] (3) {$\Gamma_{\cdot, 3}$}
    node at (1.3, -2) [place] (2) {$\Gamma_{\cdot, 2}$};
    \draw [->, thick] (1) to (3);
    \draw [->, thick] (2) to (3);
	\end{tikzpicture}
	}
	\hspace{1in}
	\subfigure{
	\begin{tikzpicture}[scale=0.8]
    \draw (-1.3,-2) node {(1, 0)}
  (0, -3.6) node {(1, 1)}
  (1.3,-2) node {(0, 1)};
  \draw [->, thick] (-1.3,-2.3) to (0, -3.2);
  \draw [->, thick] (1.3,-2.3) to (0, -3.2);
	\end{tikzpicture}
	}
	\\\vspace{0.1in}
	\subfigure{
	\begin{tikzpicture}[scale=0.8]
    \path
    node at (-1.7, -2) [place] (1) {$\Gamma_{\cdot, 1}$}
    node at (0, -3.5) [place] (3) {$\Gamma_{\cdot, k}$}
    node at (1.7, -2) [place] (2) {$\Gamma_{\cdot, l}$}
    (0, -2) node {...};
    \draw [->, thick] (1) to (3);
    \draw [->, thick] (2) to (3);
	\end{tikzpicture}
	}
	\hspace{0.2in}
	\subfigure{
	\hspace{0.4in}
	\begin{tikzpicture}[scale=0.8]
    \draw (-1.7,-2) node {(1, 0, ... , 0)}
  (0, -3.5) node {(1, ... , 1)}
  (0, -2) node {...}
  (1.7,-2) node {(0, ... , 0, 1)};
  \draw [->, thick] (-1.3,-2.3) to (0, -3.2);
  \draw [->, thick] (1.3,-2.3) to (0, -3.2);
	\end{tikzpicture}
	}
	\\
		\caption{Examples of binary representations from partial orders}
	\label{fig-partial}
\end{figure}

\begin{figure}[h!]
\vspace{0.3in}
\centering
\subfigure{
	Partial Orders}\hspace{0.9in}
	\subfigure{
	Binary Representations}
\\
	\vspace{0.1in}
	\subfigure{
	\begin{tikzpicture}[baseline= -15ex][scale=0.9]
    \path
    node at (0, 1.5) [place] (1) {$\Gamma_{\cdot, 1}$}
    node at (0, 0) [place] (2) {$\Gamma_{\cdot, 2}$}
    node at (-1.3, -1) [place] (3) {$\Gamma_{\cdot, 3}$}
    node at (0, -2) [place] (5) {$\Gamma_{\cdot, 5}$}
    node at (1.3, -1) [place] (4) {$\Gamma_{\cdot, 4}$}
    node at (0, -3.5) [place] (6) {$\Gamma_{\cdot, 6}$};
    \draw [->, thick] (1) to (2);
    \draw [->, thick] (2) to (3);
    \draw [->, thick] (2) to (4);
    \draw [->, thick] (3) to (5);
    \draw [->, thick] (4) to (5);
    \draw [->, thick] (5) to (6);
	\end{tikzpicture}
	}
	\hspace{0.5in}
	\subfigure{
	\begin{tikzpicture}[baseline= -13ex][scale=0.8]
    \draw (0,2) node {(\textcolor{blue}{\bf{0}}, 0, 0, 0)}
  (0, 0.3) node {(\textcolor{blue}{\bf{1}}, 0, 0, 0)}
  (-1.7,-0.7) node {(\textcolor{blue}{\bf{1}}, \textcolor{blue}{\bf{1}}, \textcolor{blue}{\bf{0}}, 0)}
  (0,-2) node {(\textcolor{blue}{\bf{1}}, \textcolor{blue}{\bf{1}}, \textcolor{blue}{\bf{1}}, 0)}
  (1.7,-0.7) node {(\textcolor{blue}{\bf{1}}, \textcolor{blue}{\bf{0}}, \textcolor{blue}{\bf{1}}, 0)}
  (0,-3.5) node {(\textcolor{blue}{\bf{1}}, \textcolor{blue}{\bf{1}}, \textcolor{blue}{\bf{1}}, \textcolor{blue}{\bf{1}})};
  \draw [->, thick] (-1.9,-1) to (0, -1.7);
  \draw [->, thick] (0,0.1) to (-1.9, -0.5);
  \draw [->, thick] (1.9,-1) to (0, -1.7);
  \draw [->, thick] (0,0.1) to (1.9, -0.5);
  \draw [->, thick] (0,-2.3) to (0, -3.3);
  \draw [->, thick] (0,1.6) to (0, 0.6);
	\end{tikzpicture}
	}	\caption{A more complicated example of binary representations from partial orders}
\label{fig-recover}
\end{figure}
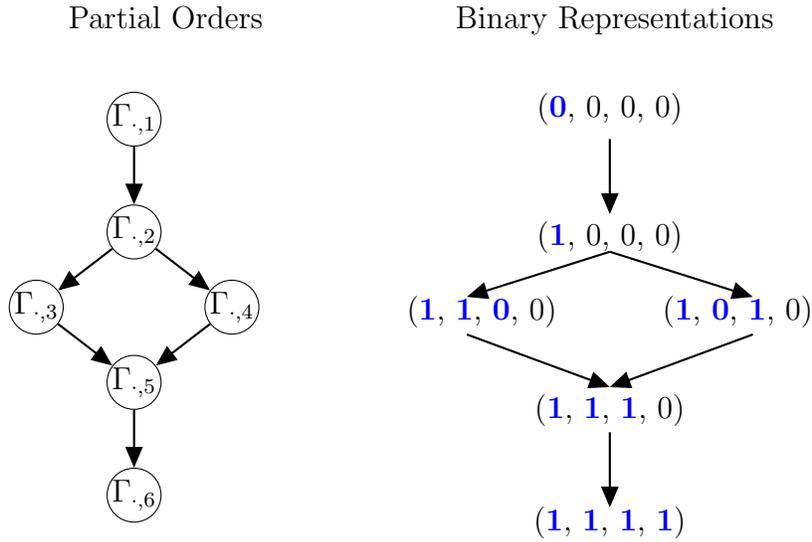

After we plot the DAG, we then recover the binary representations of the latent classes.
We start from the most basic one and move forward layer by layer.
Specifically, when we construct binary representations, we can find its parent nodes, and follow two rules below:
\begin{itemize}
	\item If the node has only one parent node, then we need to add a dimension in the binary representations.
	\item If the node has several parent nodes, then we set the binary representation of the node to be the union of all of its parent nodes.
\end{itemize}
We use examples in Figure \ref{fig-partial} and Figure \ref{fig-recover} to illustrate the procedures of recovering binary representations.
In the upper plot of Figure \ref{fig-partial}, $\Gamma_{\cdot,2}$ only has one parent node $\Gamma_{\cdot, 1}$, then we need to add a dimension in the binary representations.
In the middle plot of Figure \ref{fig-partial}, $\Gamma_{\cdot,3}$ has two parent nodes $\Gamma_{\cdot,1}$ and $\Gamma_{\cdot,2}$.  
Since there is no partial order between $\Gamma_{\cdot,1}$ and $\Gamma_{\cdot,2}$, then there are at least two dimensions in which $\Gamma_{\cdot,1}$ and $\Gamma_{\cdot,2}$ have different values.
Then we should set $\Gamma_{\cdot,3}$ to be the union of $\Gamma_{\cdot,1}$ and $\Gamma_{\cdot,2}$, which will be larger than $\Gamma_{\cdot,1}$ and $\Gamma_{\cdot,2}$.
A more general case is shown in the lower plot of Figure \ref{fig-partial}.
In Figure \ref{fig-recover}, we provide a more complicated example.
Since $\Gamma_{\cdot,2}$ only has one parent node $\Gamma_{\cdot,1}$, we need one binary digit for $\Gamma_{\cdot,2}$ and set $\Gamma_{\cdot,1} = (0)$ and $\Gamma_{\cdot,2} = (1)$.
Since $\Gamma_{\cdot,3}$ and $\Gamma_{\cdot,4}$ also have only one parent node, we need two additional dimensions, and set $\Gamma_{\cdot,3} = (1,1,0)$ and $\Gamma_{\cdot,4} = (1,0,1)$.
Next because $\Gamma_{\cdot,5}$ has two parent nodes $\Gamma_{\cdot,3}$ and $\Gamma_{\cdot,4}$, we set $\Gamma_{\cdot,5} = (1,1,1)$.
Lastly since $\Gamma_{\cdot,6}$ has one parent node $\Gamma_{\cdot,5}$, we need one more dimension and set $\Gamma_{\cdot,6} = (1,1,1,1)$.
Therefore, in total we have four latent attributes and the reconstruction process is highlighted in blue in Figure \ref{fig-recover}.
We want to point out that when we recover the latent structures using Algorithm \ref{algo-binary}, we choose the smallest $K$ such that the corresponding binary representations of the latent classes satisfy the partial orders. 
A larger value of $K$ is possible and may not be unique, but here we use the smallest one to make the latent structure concise.
Moreover, researchers can also use their domain knowledge to help specify these binary representations.

\begin{figure*}[h!]
	\centering	
	\subfigure{
	Latent Attribute Profile Matrix}\hspace{0.4in}
	\subfigure{
	Attribute Hierarchy}
\\
	\subfigure{\label{fig:A}
	$\mathcal{A} = 
    \begin{pmatrix}
    0 & 0 & 0 & 0\\
    1 & 0 & 0 & 0\\
    1 & 1 & 0 & 0\\
    1 & 0 & 1 & 0\\
    1 & 1 & 1 & 0\\
    1 & 1 & 1 & 1\\ 
    \end{pmatrix}$
	}
 	\hspace{1.1in}
	\subfigure{
	\begin{tikzpicture}[baseline= 3ex][scale=0.9]
	\centering
   \path   
    node at (0, 2) [place] (1) {$\alpha_1$}
    node at (-1.3, 0.7) [place] (2) {$\alpha_2$}
    node at (0, -0.6) [place] (4) {$\alpha_4$}
    node at (1.3, 0.7) [place] (3) {$\alpha_3$};
    \draw [->, thick] (1) to (2);
    \draw [->, thick] (1) to (3);
    \draw [->, thick] (2) to (4);
    \draw [->, thick] (3) to (4);
   	\end{tikzpicture}
	}
	\caption{Latent attribute profile matrix and attribute hierarchy; rows of $\mathcal{A}$ are the binary representations of the select latent classes in Figure \ref{fig-recover}.}
	\label{fig:A-H}
\end{figure*}
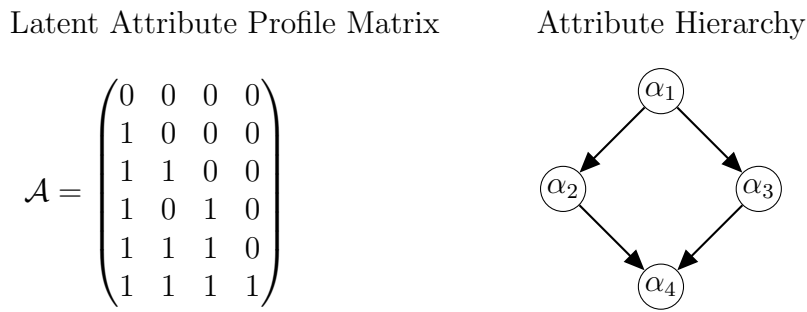

After we reconstruct binary representations of the latent classes, we can infer the attribute hierarchy accordingly.
Specifically, we can get partial orders among latent attributes.
For the example in Figure \ref{fig-recover}, our reconstructed latent attribute profile matrix $\mathcal{A}$ is shown in Figure \ref{fig:A-H}, where rows of $\mathcal{A}$ are the binary representations of the latent classes. 
We can see that $\mathcal{A}_{\cdot, 1} \succeq \mathcal{A}_{\cdot, k}$ for all $k\in [K]$, which indicates that the first latent attribute is the most basic one and the prerequisite for all the other latent attributes. 
Moreover, the fourth attribute is 1 only if all the other attributes are 1, indicating that the fourth attribute is the highest and requires all the other attributes as prerequisites.
Formally, we can use $\mathcal{E} = \{k \rightarrow l: \text{attribute } k \text{ is a prerequisite for attribute } l \}$ introduced in Section \ref{sec-hcdm} to denote the prerequisite relationship set, where $k \rightarrow l$ if $\mathcal{A}_{\cdot, k} \succeq \mathcal{A}_{\cdot, l}$. 
For latent attribute profile matrix $\mathcal{A}$ in Figure \ref{fig:A-H}, we have $\mathcal{E} = \{ 1\rightarrow 2, \ 1\rightarrow 3, \ 2\rightarrow 4, \ 3\rightarrow 4 \}$.
We can also plot a DAG according to the prerequisite relationship set $\mathcal{E}$ as shown in the right plot of Figure \ref{fig:A-H}.

Finally we need to reconstruct the $Q$-matrix, which can be done by comparing the indicator matrix $\boldsymbol{\Gamma}$ and the reconstructed latent attribute profile $\mathcal{A}$.
Specifically, since capable subjects have the same highest item parameters, for each item, the row in the $Q$-matrix will equal to the smallest latent attribute profile such that the corresponding indicator is 1.
To be more formal, let $\boldsymbol{q}_j$ be the $j$th row of the $Q$-matrix, we have
\[
\boldsymbol{q}_j = \mathcal{A}_{m,\cdot} \text{ such that } \Gamma_{j,m}=1 \text{ and for any } m' \text{ with } \Gamma_{j,m'}= 1, \mathcal{A}_{m,\cdot} \preceq  \mathcal{A}_{m',\cdot}, \text{ for } j = 1, \dots, J,
\] where $\mathcal{A}_{m,\cdot}$ denotes the $m$th row vector of the latent attribute profile matrix $\mathcal{A}$, i.e., the  binary representation of the $m$th latent class in $\Gamma$.
The procedures are summarized in Algorithm \ref{algo-binary}.

\begin{algorithm}[h!]
\caption{Recover Latent Attribute Profiles, Hierarchical Structure and $Q$-matrix}
\label{algo-binary}
\SetKwInOut{Input}{Input}
\SetKwInOut{Output}{Output}

\Input{Item parameter Matrix $\bm{\Theta}$}

\textbf{Step 1\ :} Construct the indicator matrix $\bm{\Gamma}=\big(\mathbb{I}\{\theta_{j,m}=\max_{l\in[M]}\theta_{j,l}\}\big)$.

\textbf{Step 2\ :} Construct $\boldsymbol{P}$ based on the partial orders among the columns of $\boldsymbol{\Gamma}$;\\ 
	\quad \quad \quad \quad plot a DAG based on $\boldsymbol{P}$.

\textbf{Step 3\ :} Reconstruct binary representations and get latent attribute profile set $\mathcal{A}$:

\quad \For{node from top to bottom}{
	\quad \If{the node has only one parent node}{add a dimension in the binary representations}
	\quad \If{the node has more than one parent node}{set the binary representation to be the union of all of its parent nodes}
}

\textbf{Step 4\ :} Construct prerequisite relationship set $\mathcal{E}$ and thus recover latent hierarchy.

\textbf{Step 5\ :} Reconstruct the $Q$-matrix $\boldsymbol{Q} = \big(\boldsymbol{q}_j\big)_{j=1}^J$: 
\[
\boldsymbol{q}_j = \mathcal{A}_{m,\cdot} \text{ such that } \Gamma_{j,m}=1 \text{ and for any } m' \text{ with } \Gamma_{j,m'}= 1, \mathcal{A}_{m,\cdot} \preceq  \mathcal{A}_{m',\cdot}, \text{ for } j = 1, \dots, J.
\]

\Output{Latent attribute profile set $\mathcal{A}$, prerequisite relationship set $\mathcal{E}$ and the $Q$-matrix $\boldsymbol{Q}$.}

\end{algorithm}{}

\section{Simulation Studies}
\label{sec-simu}

In this section, we conducted comprehensive simulation studies under various settings to evaluate the performance of the proposed method. 

For the underlying models, we considered three settings.
In the first setting, all the items conformed to the DINA model.
In the second setting, half of the items were from the DINA model and the others followed the DINO model.
In the third setting, we considered the GDINA model as the underlying data generating model.
To satisfy identifiability conditions \citep{xu2016identifiability,gu2019sufficient,gu2019learning}, in the DINA setting, the $Q$-matrix contained two identity sub-matrices and the remaining items were randomly generated.
In the DINA + DINO setting, the $Q$-matrix contained an identity sub-matrix for each type of the models and the remains were randomly generated.
For the GDINA setting, we had two identity sub-matrices, and the remaining items were randomly generated and required at most 3 latent attributes.

We considered four hierarchical structures shown in Figure \ref{fig:hier} with $K=4$.
The test length was set to 30 ($J=30$). 
For the DINA and DINA + DINO settings, we considered two sample sizes with $N = 500$ or $1000$.
Two different signal strengths for true item parameters were included:
$\{\theta_j^+=0.9,\ \theta_j^-=0.1;\ j\in[J]\}$ and $\{\theta_j^+=0.8,\ \theta_j^-=0.2;\ j\in[J]\}$.
For the GDINA setting, we considered two different sample sizes, $N=1000$ or 2000.
The sample sizes considered in the GDINA settings are relatively larger than those for the DINA and DINA + DINO settings, since in the GDINA model there are  more item parameters to be estimated.
As before, we set two different signal strengths, where the highest item parameter was $0.9$ or $0.8$, and the lowest item parameters was $0.1$ or $0.2$.
The other item parameters in between were equally spaced.
For each scenario, we performed 50 independent repetitions. 
All model parameters were randomly initialized and the implementations were done in Matlab.

To tune hyperparameters for the proposed method, we used a two-stage training strategy.
In the first training stage, we primarily tuned $\tilde{\lambda}_1$ and $\tilde{\lambda}_2$ to select significant latent classes, and used a fixed relatively large $\tau$.
In the second stage, we did not put penalty on the proportion parameters (i.e. $\tilde{\lambda}_1$ was set to 0), and fine-tuned $\tilde{\lambda}_2$ and $\tau$ for the TLP penalty term to further merge identical item parameters.
Specifically, in the first stage, the candidates for $\tilde{\lambda}_2$ were set to relatively small and the value of the threshold $\tau$ was set to relatively large.
In this work, we chose $\tau = 0.3$ and selected $\tilde{\lambda}_1 \in\{0.01,0.015,\dots,0.05\}$ and $\tilde{\lambda}_2 \in\{0.001,0.005,0.01,0.015\}$.
The reason to use a small penalty coefficient and a relatively large threshold for the TLP penalty during the first training stage is that we mainly aim to select the correct number of latent classes instead of learning identical item parameters.
A small TLP penalty would facilitate the shrinkage of the proportion parameters, while a large TLP penalty would merge the latent classes too fast.
After we selected the significant latent classes from the first stage, we next moved to the second stage where we used a larger $\tilde{\lambda}_2$ and a smaller threshold $\tau$ for the TLP penalty to further merge identical item parameters.
Specifically, the penalty for the proportion parameters $\tilde{\lambda}_1$ was set to 0, and we selected $\log(\tilde{\lambda}_2) \in\{-1,0,1,2,3\}$ and $\tau \in \{0.03, 0.05, 0.1\}$.
For the $\gamma$ parameter, similarly to \cite{wu2016new}, we used a fixed $\gamma$ for simplicity with $\gamma=0.02$.  
If computation allows, we could also tune for $\gamma$ or adaptively select it in each iteration \citep{wang2001decomposition}. The candidate sets of all the other tuning parameters were the same across the simulation settings. In total there were 480 possible combinations of tuning parameters, while using the two-stage training procedure, the number of combinations was reduced to around 50.  	On average, the computation time in our simulation study was less than 2.0 seconds per repetition per set of hyper-parameters.
We can also try larger candidate sets for these hyperparameters, but our simulation results below showed that the aforementioned candidate sets were enough to provide good results.

Following \cite{chen2017regularized} and  \cite{wang2021learning}, we also fitted the regularized LCMs under the same settings for comparison.
For the regularized LCM method, the number of latent classes   and the coefficient for the penalty term need to be selected according to some information criteria.
As suggested in \cite{chen2017regularized}, we used $\text{GIC}_2$ to select these tuning parameters in regularized LCMs.
In our simulation, for the number of latent classes, we chose $M\in\{M_0-2, M_0-1,M_0,M_0+1,M_0+2\}$, where $M_0$ is the true number of latent classes. 
We also conducted a sensitivity analysis to investigate the impacts of different specifications of the upper bound $M$ on our algorithm. 
The results show that our method is robust to the choice of different $M$.
The detailed results of the sensitivity analysis are included in Section~4 of Supplementary Material.
For the penalty term, we selected $\lambda \in \{0.01, 0.02,\dots,0.1\}$ as in \cite{wang2021learning}.

We inspect the results from different aspects. 
Firstly we examine the accuracy of selecting the number of latent classes $\hat{M}$, which is denoted as $\text{Acc}(\hat{M})$. 
Based on the learned item parameters, we reconstruct the indicator matrix 
$\hat{\bm{\Gamma}}= \big(\mathbb{I}\{\theta_{j,m}=\max_{l\in[\hat{M}]}\theta_{j,l}\}\big)\in\{0,1\}^{J\times\hat{M}}$ and the corresponding partial order matrix $\hat{\boldsymbol{P}}$.
It's worth noting that when we extract the partial orders among the latent classes, a single misspecification of the elements in the indicator matrix may lead to different ordering results, making the method of directly estimating   the partial orders not robust.
Based on this observation, we shall allow for certain tolerance on the estimation errors of the indicator matrix when reconstructing the partial orders. 
In particular, we relax the construction condition of the partial order such that we regard $\Gamma_{\cdot,k} \succeq \Gamma_{\cdot,j}$, if $\Gamma_{j,k} \geq \Gamma_{j,l}$ except for a small proportion $t$ of $j\in[J]$. 
In our simulation, we used $t = 5\%$ when the noise was small, and $t = 10\%$ when the noise was large.
Another issue to note here is that directly comparing two indicator matrices is not straightforward due to the label switching.
To address this issue, we apply the Hungarian algorithm \citep{kuhn1955hungarian} to find the best match of the columns of the estimated indicator matrix and the true indicator matrix, based on which the following comparisons can be made accordingly.
We use $\text{Acc}(\hat{\boldsymbol{P}})$ to denote the accuracy of reconstruction of the partial orders.
If all the partial orders among the columns of the indicator matrix are correctly recovered, then we will successfully reconstruct the binary latent pattern representations and accordingly the hierarchical structures among the latent attributes.
We use $\text{Acc}(\hat{\mathcal{E}})$ to denote the recovery rate of the hierarchical structure.
Here we count it a success only if the entire hierarchical structure is recovered.
If the number of latent classes is successfully selected, we also compute the mean squared error of the item parameters $\text{MSE}(\hat{\boldsymbol{\Theta}})$.
Finally, if the hierarchical structure is correctly recovered, we compute the accuracy of the estimated $Q$-matrix, denoted by $\text{Acc}(\hat{\boldsymbol{Q}})$.
In summary, we have five evaluation metrics : $\text{Acc}(\hat{M})$, $\text{Acc}(\hat{\boldsymbol{P}})$, $\text{Acc}(\hat{\mathcal{E}})$, $\text{MSE}(\hat{\boldsymbol{\Theta}})$ and $\text{Acc}(\hat{\boldsymbol{Q}})$.

\begin{table}[h!]
  \centering
    \begin{tabular}{c|c|c|c|c|c|c|c|c}
    \toprule
    Hierarchy & $N$     & $r$     & Method & Acc($\hat{M}$) & Acc($\hat{\boldsymbol{P}}$) & Acc($\hat{\mathcal{E}}$) & MSE($\hat{\boldsymbol{\Theta}}$) & Acc($\hat{\boldsymbol{Q}}$) \\
    \hline
    \multirow{8}[8]{*}{Linear} & \multirow{4}[4]{*}{500} & \multirow{2}[2]{*}{0.1} & Proposed & 1   & 1  & 1    & 0.0004 & 0.99\\
          &       &       & regularized LCM  & 0.72   & 0.71 & 0.60   & 0.0006 & 0.96 \\
\cline{3-9}          &       & \multirow{2}[2]{*}{0.2} & Proposed & 0.68   & 0.68 & 0.66 & 0.0012 & 0.95 \\
          &       &       & regularized LCM  & 0.42      & 0.42     & 0.42    & 0.0012 & 0.99 \\
\cline{2-9}          & \multirow{4}[4]{*}{1000} & \multirow{2}[2]{*}{0.1} & Proposed & 1    & 1     & 1     & 0.0002 & 1 \\
          &       &       & regularized LCM  & 0.80  & 0.80 & 0.80 & 0.0013 & 0.99 \\
\cline{3-9}          &       & \multirow{2}[2]{*}{0.2} & Proposed & 0.96   & 0.96     & 0.96     & 0.0004 & 0.99 \\
          &       &       & regularized LCM  & 0.54   & 0.53 & 0.52 & 0.0025 & 0.99\\
\hline    \multirow{8}[8]{*}{Convergent} & \multirow{4}[4]{*}{500} & \multirow{2}[2]{*}{0.1} & Proposed & 1   & 1     & 0.98    & 0.0005 & 0.99\\
          &       &       & regularized LCM  & 0.62   & 0.61 & 0.48   & 0.0013 & 0.96\\
\cline{3-9}          &       & \multirow{2}[2]{*}{0.2} & Proposed & 0.56  & 0.56 & 0.50 & 0.0014 & 0.93\\
          &       &       & regularized LCM  & 0.20      & 0.18     & 0.08     & 0.0245 & 0.93 \\
\cline{2-9}          & \multirow{4}[4]{*}{1000} & \multirow{2}[2]{*}{0.1} & Proposed & 1    & 1     & 1     & 0.0002 & 1\\
          &       &       & regularized LCM  & 0.48  & 0.48  & 0.40   & 0.0003 & 0.98 \\
\cline{3-9}          &       & \multirow{2}[2]{*}{0.2} & Proposed & 0.84   & 0.84 & 0.84 & 0.0005 & 0.98\\
          &       &       & regularized LCM  & 0.38   & 0.37 & 0.34 & 0.0062 & 0.98 \\
\hline    \multirow{8}[8]{*}{Divergent} & \multirow{4}[4]{*}{500} & \multirow{2}[2]{*}{0.1} & Proposed & 1   & 1     & 0.97    & 0.0005 & 0.97\\
          &       &       & regularized LCM  & 0.44   & 0.43 & 0.28   & 0.0047 & 0.93 \\
\cline{3-9}          &       & \multirow{2}[2]{*}{0.2} & Proposed & 0.48  & 0.47 & 0.34 & 0.0016 & 0.88\\
          &       &       & regularized LCM  & 0.22      & 0.20     & 0.08     & 0.0194  & 0.95 \\
\cline{2-9}          & \multirow{4}[4]{*}{1000} & \multirow{2}[2]{*}{0.1} & Proposed & 0.98   & 0.98     & 0.98     & 0.0002 & 1\\
          &       &       & regularized LCM  & 0.48  & 0.48  & 0.44   & 0.0003 & 0.97 \\
\cline{3-9}          &       & \multirow{2}[2]{*}{0.2} & Proposed & 0.86   & 0.86 & 0.80 & 0.0006 & 0.96\\
          &       &       & regularized LCM  & 0.26   & 0.25 & 0.20 & 0.0108 & 0.97 \\
\hline    \multirow{8}[8]{*}{Unstructured} & \multirow{4}[4]{*}{500} & \multirow{2}[2]{*}{0.1} & Proposed & 0.82   & 0.82     & 0.66    & 0.0006 & 0.93\\
          &       &       & regularized LCM  & 0.22   & 0.21 & 0.10   & 0.0103 & 0.90 \\
\cline{3-9}          &       & \multirow{2}[2]{*}{0.2} & Proposed & 0.06  & 0.06 & 0.02 & 0.0031 & 0.90\\
          &       &       & regularized LCM  & 0.14      & 0.13     & 0.04     & 0.0126 & 0.87 \\
\cline{2-9}          & \multirow{4}[4]{*}{1000} & \multirow{2}[2]{*}{0.1} & Proposed & 0.92   & 0.92     & 0.92     & 0.0002 & 0.99\\
          &       &       & regularized LCM  & 0.36  & 0.35  & 0.18   & 0.0074 & 0.98 \\
\cline{3-9}          &       & \multirow{2}[2]{*}{0.2} & Proposed & 0.48   & 0.48 & 0.48 & 0.0006 & 0.94\\
          &       &       & regularized LCM  & 0.28   & 0.26 & 0.14 & 0.0124 & 0.93 \\
    \bottomrule
    \end{tabular}%
    \caption{DINA Results; Acc($\hat{M}$), Acc($\hat{\boldsymbol{P}}$) and Acc($\hat{\mathcal{E}}$) are calculated for all the cases; MSE($\hat{\boldsymbol{\Theta}}$) is calculated for the cases when the number of latent classes are correctly selected; Acc($\hat{\boldsymbol{Q}}$) is calculated for the cases when the hierarchical structure is successfully recovered.}
  \label{table-DINA}%
\end{table}%

\begin{table}[h!]
  \centering
    \begin{tabular}{c|c|c|c|c|c|c|c|c}
    \toprule
    Hierarchy & $N$     & $r$     & Method & Acc($\hat{M}$) & Acc($\hat{\boldsymbol{P}}$) & Acc($\hat{\mathcal{E}}$) & MSE($\hat{\boldsymbol{\Theta}}$) & Acc($\hat{\boldsymbol{Q}}$) \\
    \toprule
    \multirow{8}[8]{*}{Linear} & \multirow{4}[4]{*}{500} & \multirow{2}[2]{*}{0.1} & Proposed & 1     & 1     & 1     & 0.0004 & 0.99 \\
          &       &       & regularized LCM  & 0.96   & 0.93 & 0.68 & 0.0006 & 0.96 \\
\cline{3-9}          &       & \multirow{2}[2]{*}{0.2} & Proposed & 0.98  & 0.98 & 0.96 & 0.0010 & 0.94\\
          &       &       & regularized LCM  & 0.72  & 0.72 & 0.70 & 0.0013 & 0.97\\
\cline{2-9}          & \multirow{4}[4]{*}{1000} & \multirow{2}[2]{*}{0.1} & Proposed & 1    & 1     & 1     & 0.0002 & 1\\
          &       &       & regularized LCM  & 0.96  & 0.96 & 0.94 & 0.0002 & 0.98\\
\cline{4-9}          &       & \multirow{2}[2]{*}{0.2} & Proposed & 1   & 1     & 1     & 0.0004 & 0.99\\
          &       &       & regularized LCM  & 0.78  & 0.78 & 0.78 & 0.0004 & 0.99\\
    \hline
    \multirow{8}[8]{*}{Convergent} & \multirow{4}[4]{*}{500} & \multirow{2}[2]{*}{0.1} & Proposed & 1     &  1 & 0.86  & 0.0004 & 0.98 \\
          &       &       & regularized LCM  & 0.88  & 0.84 & 0.52     & 0.0012 & 0.93\\
\cline{4-9}          &       & \multirow{2}[2]{*}{0.2} & Proposed & 0.96  & 0.94 & 0.76 & 0.0013 & 0.89\\
          &       &       & regularized LCM  & 0.60  & 0.60  & 0.54     & 0.0017 & 0.92 \\
\cline{2-9}          & \multirow{4}[4]{*}{1000} & \multirow{2}[2]{*}{0.1} & Proposed & 1    & 1    & 1     & 0.0002 & 1\\
          &       &       & regularized LCM  & 0.88  & 0.87 & 0.76 & 0.0003 & 0.97 \\
\cline{4-9}          &       & \multirow{2}[2]{*}{0.2} & Proposed & 1  & 0.99 & 0.82 & 0.0004 & 0.99 \\
          &       &       & regularized LCM  & 0.64  & 0.64 & 0.58     & 0.0006 & 0.98\\
\hline
    \multirow{8}[8]{*}{Divergent} & \multirow{4}[4]{*}{500} & \multirow{2}[2]{*}{0.1} & Proposed & 0.98     &  0.98 & 0.96 & 0.0005 & 0.97 \\
          &       &       & regularized LCM  & 0.80  & 0.77 & 0.40     & 0.0009 & 0.92\\
\cline{4-9}          &       & \multirow{2}[2]{*}{0.2} & Proposed & 0.86  & 0.84 & 0.46 & 0.0016 & 0.86\\
          &       &       & regularized LCM  & 0.40  & 0.39  & 0.26     & 0.0023 & 0.88 \\
\cline{2-9}          & \multirow{4}[4]{*}{1000} & \multirow{2}[2]{*}{0.1} & Proposed & 1    & 1    & 1     & 0.0002 & 1 \\
          &       &       & regularized LCM  & 0.82  & 0.81 & 0.56 & 0.0003 & 0.96 \\
\cline{4-9}          &       & \multirow{2}[2]{*}{0.2} & Proposed & 1  & 0.99 & 0.78 & 0.0005 & 0.97\\
          &       &       & regularized LCM  & 0.48  & 0.48 & 0.40     & 0.0009 & 0.95\\
    \hline
    \multirow{8}[8]{*}{Unstructured} & \multirow{4}[4]{*}{500} & \multirow{2}[2]{*}{0.1} & Proposed & 0.92     &  0.91 & 0.70  & 0.0006 & 0.94\\
          &       &       & regularized LCM  & 0.54  & 0.51 & 0.14     & 0.0039 & 0.88 \\
\cline{4-9}          &       & \multirow{2}[2]{*}{0.2} & Proposed & 0.28  & 0.27 & 0 & 0.0010 & 0.75 \\
          &       &       & regularized LCM  & 0.28  & 0.27  & 0.08     & 0.0112 & 0.88 \\
\cline{2-9}          & \multirow{4}[4]{*}{1000} & \multirow{2}[2]{*}{0.1} & Proposed & 0.98    & 0.98    & 0.96     & 0.0002 & 1 \\
          &       &       & regularized LCM  & 0.58  & 0.57 & 0.34 & 0.0005 & 0.94 \\
\cline{4-9}          &       & \multirow{2}[2]{*}{0.2} & Proposed & 0.82  & 0.81 & 0.48 & 0.0007 & 0.92 \\
          &       &       & regularized LCM  & 0.18  & 0.17 & 0.06    & 0.0066 & 0.88\\
    \bottomrule
    \end{tabular}%
    \caption{DINA+DINO Results; Acc($\hat{M}$), Acc($\hat{\boldsymbol{P}}$) and Acc($\hat{\mathcal{E}}$) are calculated for all the cases; MSE($\hat{\boldsymbol{\Theta}}$) is calculated for the cases when the number of latent classes are correctly selected; Acc($\hat{\boldsymbol{Q}}$) is calculated for the cases when the hierarchical structure is successfully recovered.}
  \label{table-MIX}%
\end{table}%

\begin{table}[h!]
  \centering
    \begin{tabular}{c|c|c|c|c|c|c|c|c}
    \toprule
    Hierarchy & $N$     & $r$     & Method & Acc($\hat{M}$) & Acc($\hat{\boldsymbol{P}}$) & Acc($\hat{\mathcal{E}}$) & MSE($\hat{\boldsymbol{\Theta}}$) & Acc($\hat{\boldsymbol{Q}}$) \\
    \hline    \multirow{8}[8]{*}{Linear} & \multirow{4}[4]{*}{1000} & \multirow{2}[2]{*}{0.1} & Proposed & 0.98       & 0.98     & 0.98     & 0.0005 & 1 \\
          &       &       & regularized LCM  & 0.76        & 0.76     & 0.76     & 0.0005 & 0.99 \\
\cline{3-9}          &       & \multirow{2}[2]{*}{0.2} & Proposed & 0.96    & 0.96 & 0.96  & 0.0010 & 0.99\\
          &       &       & regularized LCM  & 0.52     & 0.51     & 0.48     & 0.0036 & 0.97 \\
\cline{2-9}          & \multirow{4}[4]{*}{2000} & \multirow{2}[2]{*}{0.1} & Proposed & 0.94   &  0.94     & 0.94     & 0.0003 & 1 \\
          &       &       & regularized LCM  & 0.92      & 0.92     & 0.92     & 0.0002 & 1\\
\cline{3-9}          &       & \multirow{2}[2]{*}{0.2} & Proposed & 0.96   & 0.96     & 0.96     & 0.0005 & 1\\
          &       &       & regularized LCM  & 0.62      & 0.62    & 0.62     & 0.0009 & 1\\
\hline
    \multirow{8}[8]{*}{Convergent} & \multirow{4}[4]{*}{1000} & \multirow{2}[2]{*}{0.1} & Proposed & 0.98   & 0.98     & 0.98     & 0.0006 & 1 \\
          &       &       & regularized LCM  & 0.68     & 0.68     & 0.66     & 0.0008 & 0.97 \\
\cline{3-9}          &       & \multirow{2}[2]{*}{0.2} & Proposed & 0.90   & 0.90     & 0.86     & 0.0013 & 0.98 \\
          &       &       & regularized LCM  & 0.36         & 0.35     & 0.30     & 0.0161 & 0.96 \\
\cline{2-9}          & \multirow{4}[4]{*}{2000} & \multirow{2}[2]{*}{0.1} & Proposed & 0.98    & 0.98     & 0.98     & 0.0003 & 1 \\
          &       &       & regularized LCM  & 0.82  & 0.82     & 0.80    & 0.0003 & 0.99 \\
\cline{3-9}          &       & \multirow{2}[2]{*}{0.2} & Proposed & 1    & 1     & 1     & 0.0005 & 1 \\
          &       &       & regularized LCM  & 0.38  & 0.38 & 0.36     & 0.0029 & 0.99 \\
\hline
    \multirow{8}[8]{*}{Divergent} & \multirow{4}[4]{*}{1000} & \multirow{2}[2]{*}{0.1} & Proposed & 0.98   & 0.98     & 0.98     & 0.0006 & 1 \\
          &       &       & regularized LCM  & 0.84     & 0.83     & 0.66     & 0.0061 & 0.94 \\
\cline{3-9}          &       & \multirow{2}[2]{*}{0.2} & Proposed & 0.86   & 0.86     & 0.82     & 0.0014 & 0.96\\
          &       &       & regularized LCM  & 0.38         & 0.36     & 0.26     & 0.0148 & 0.89 \\
\cline{2-9}          & \multirow{4}[4]{*}{2000} & \multirow{2}[2]{*}{0.1} & Proposed & 1    & 1     & 1     & 0.0003 & 1 \\
          &       &       & regularized LCM  & 0.76  & 0.76     & 0.74    & 0.0003 &  0.99\\
\cline{3-9}          &       & \multirow{2}[2]{*}{0.2} & Proposed & 0.98    & 0.98     & 0.92     & 0.0006 & 0.99 \\
          &       &       & regularized LCM  & 0.52  & 0.51 & 0.48     & 0.0018 & 0.97 \\
          \hline
    \multirow{8}[8]{*}{Unstructured} & \multirow{4}[4]{*}{1000} & \multirow{2}[2]{*}{0.1} & Proposed & 1   & 1     & 0.98     & 0.0007 & 0.99 \\
          &       &       & regularized LCM  & 0.62     & 0.61     & 0.44     & 0.0013 & 0.92 \\
\cline{3-9}          &       & \multirow{2}[2]{*}{0.2} & Proposed & 0.48   & 0.47     & 0.36     & 0.0021  & 0.89\\
          &       &       & regularized LCM  & 0.36         & 0.34     & 0.20     & 0.0220 & 0.88 \\
\cline{2-9}          & \multirow{4}[4]{*}{2000} & \multirow{2}[2]{*}{0.1} & Proposed &  1    & 1     & 1     & 0.0003 & 1 \\
          &       &       & regularized LCM  & 0.66  & 0.66     & 0.60    & 0.0005 & 0.96 \\
\cline{3-9}          &       & \multirow{2}[2]{*}{0.2} & Proposed & 0.86  & 0.85 & 0.78     & 0.0008 & 0.99 \\
          &       &       & regularized LCM  & 0.38  & 0.37 & 0.24     & 0.0056 & 0.94\\
\bottomrule
    \end{tabular}%
    \caption{GDINA Results; Acc($\hat{M}$), Acc($\hat{\boldsymbol{P}}$) and Acc($\hat{\mathcal{E}}$) are calculated for all the cases; MSE($\hat{\boldsymbol{\Theta}}$) is calculated for the cases when the number of latent classes are correctly selected; Acc($\hat{\boldsymbol{Q}}$) is calculated for the cases when the hierarchical structure is successfully recovered.}
  \label{table-GDINA}%
\end{table}%

The simulation results of the DINA, and DINA + DINO settings are presented in Table \ref{table-DINA} and Table \ref{table-MIX}.
The results of the GDINA model are shown in Table \ref{table-GDINA}.
The simulations show that compared with the regularized LCM approach, our method provided much better results in almost all the settings and from all the evaluation aspects. 
In many settings, the proposed method could achieve nearly perfect selection of the number of latent classes, reconstruction of the partial orders and hierarchies, and the estimation of the $Q$-matrix, especially when the noise was small or there was sufficiently large data size.
Among the four hierarchical structures, the unstructured hierarchy was the most difficult one, especially when the noise was large but the sample size was relatively small.
This is expected since under the unstructured hierarchy, there are 9 latent classes, and the hierarchical structure is more complicated compared with the others.
However, with increasing sample sizes, the proposed method also provided satisfactory results, while the regularized LCM approach did not.
In terms of the underlying data generating model, the DINA setting was the most difficult one to learn.
This is because the DINA models the conjunctive ``AND" relationship among the latent attributes, which makes it hard to distinguish the latent classes under hierarchical structures.
For example, consider latent classes $\boldsymbol{\alpha} = (1, 0, 0, 0)$ and $\boldsymbol{\alpha}' = (1, 1, 0, 0)$.
Under the DINA model, only the items with the q-vector $\boldsymbol{q}_j = (0, 1, 0, 0)$ or $(1, 1, 0, 0)$ can distinguish these two latent classes.
By contrast, under the DINO model, the items with $\boldsymbol{q}_j = (0, 1, *, *)$ where $``*"$ can be either 0 or 1, will distinguish them.
And under the GDINA model, the two latent classes can be differentiated by the items with $\boldsymbol{q}_j = (*, 1, *, *)$.
Therefore, if the underlying data generating model is the DINA model, it requires larger sample size to achieve good performance.
It is also noted that for the $Q$-matrix estimation, $\text{Acc}(\hat{\bm{Q}})$ for both methods are similar from the tables.
However, since we calculate the accuracy of the $Q$-matrix only if the hierarchical structure is correctly recovered, given the worse performance on hierarchical structure recovery of the regularized LCM method, the proposed method in fact provided much better  overall $Q$-matrix estimation.

\section{Real Data Analysis}
\label{sec-data}

\subsection{Analysis of ECPE Data}

We applied the proposed approach to the Examination for the Certificate of Proficiency in English (ECPE) data to learn the latent hierarchical structure.
The ECPE data was collected by the English Language Institute of the University of Michigan, and we used the data from R package \verb!CDM!.
The dataset includes 2,922 examinees and 28 ECPE items with three target attributes including lexical rules, cohesive rules and morphosyntactic rules.
In the literature of the analysis of the ECPE data, 
\cite{templin2014hierarchical} fitted an HCDM with the corresponding $Q$-matrix  pre-specified by exam designers and tested the presence of the linear hierarchy through bootstrap, which supports the linear hierarchy among the three attributes under the CDM framework.
In \cite{wang2021learning}, the authors also studied this ECPE data using the latent variable selection approach and regularized LCM approach respectively.
In the latent variable selection approach, they used three ``anchor'' items which formed a known identity sub-matrix in the $Q$-matrix.
The latent variable selection approach selected 5 significant latent classes, and the learned model implied a convergent structure, that is, two latent attributes were prerequisites of the third one.
Though estimations of the ECPE data have been widely studied under the CDM setting, \cite{von2014hier} pointed out that ECPE data appear to have mainly a unidimensional structure, which may not be suitable for CDM modeling.

\begin{figure}[h!]
    \centering
    \subfigure[]{     \includegraphics[width=3in]{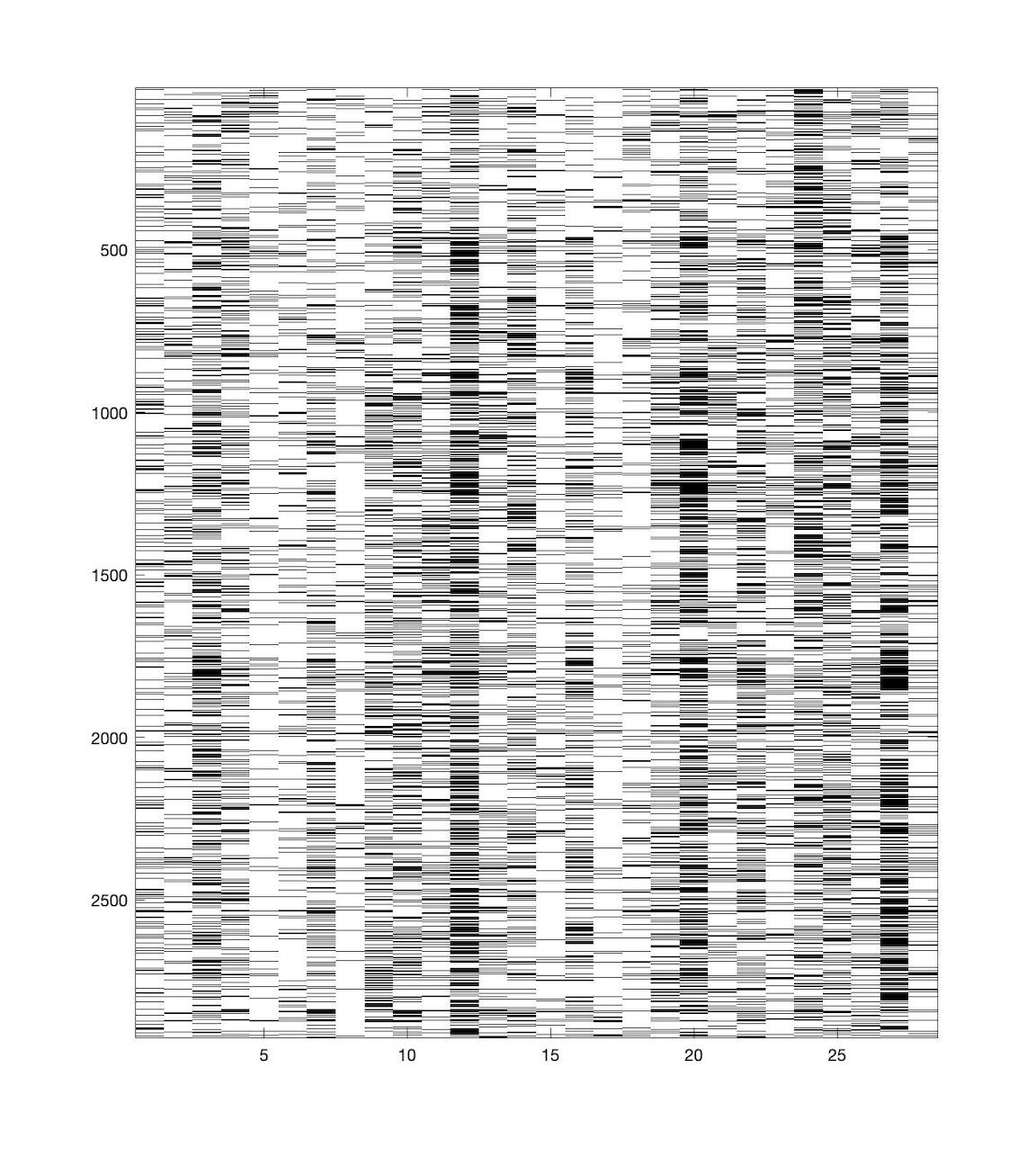} \label{fig:original}}
    \hspace{0in}
    \subfigure[]{
    \includegraphics[width=3in]{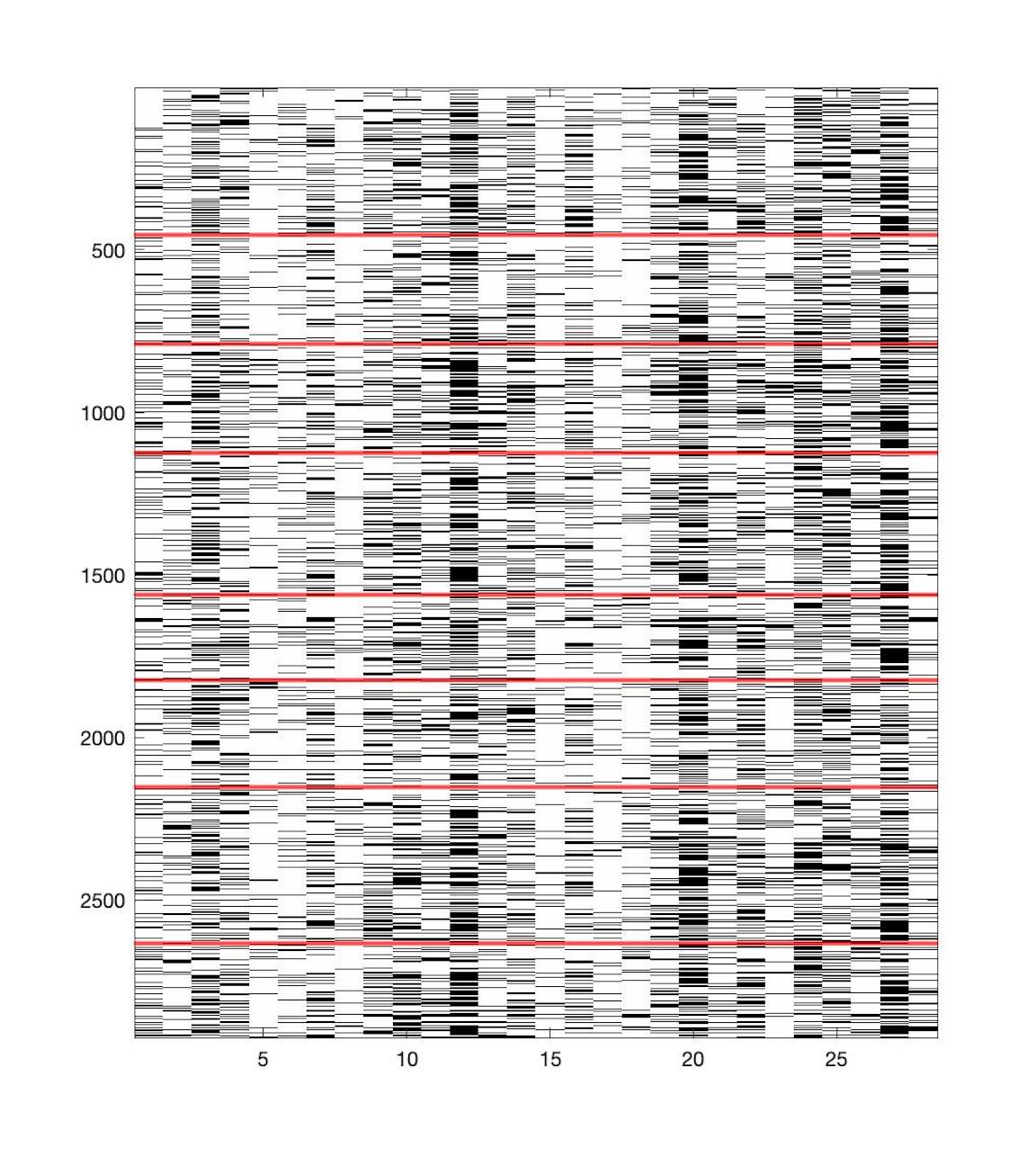} \label{fig:spectral}}
	\\
    \subfigure[]{
    \includegraphics[width=3in]{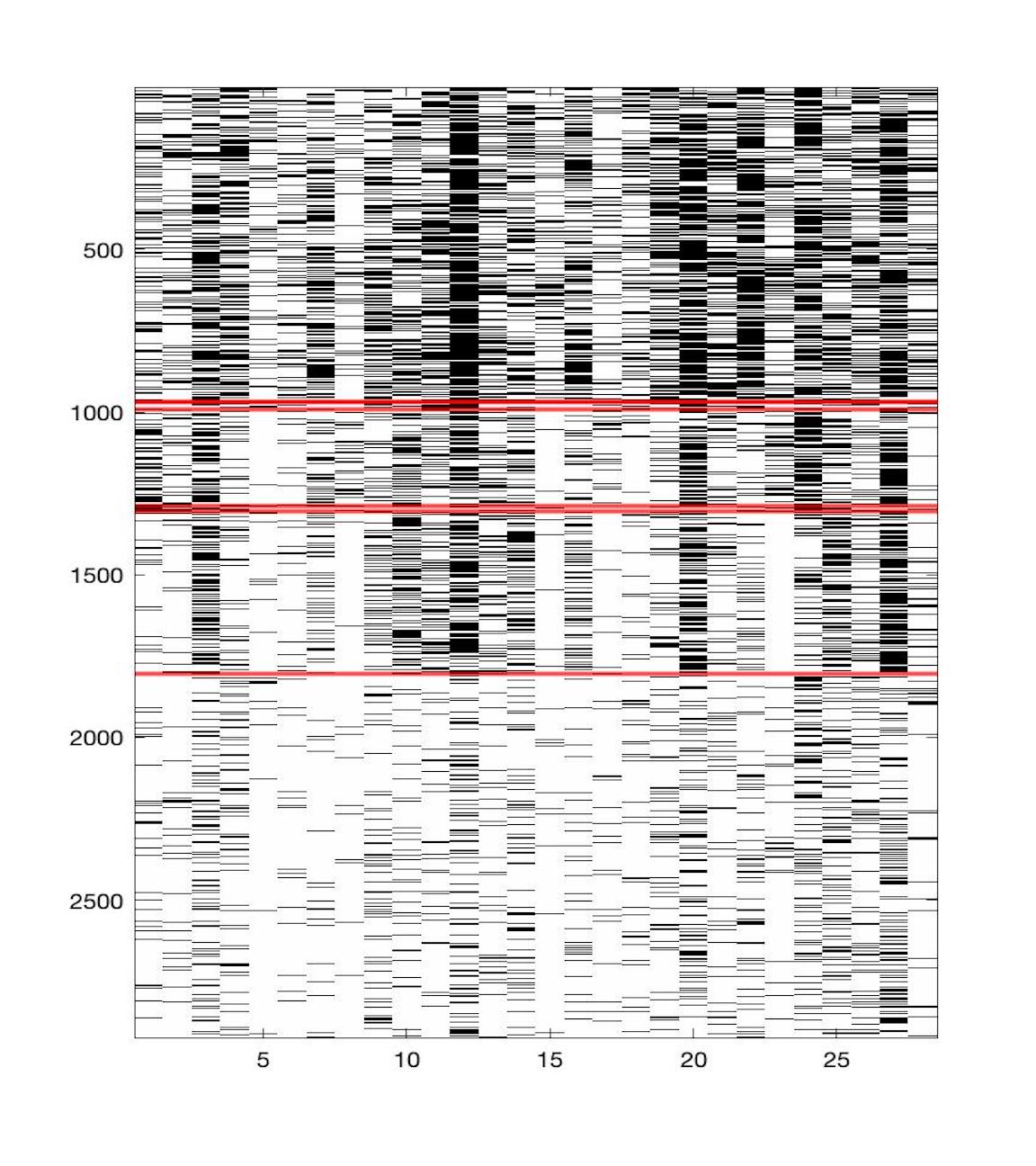} \label{fig:gdina}}
    \hspace{0in}
    \subfigure[]{
    \includegraphics[width=3in]{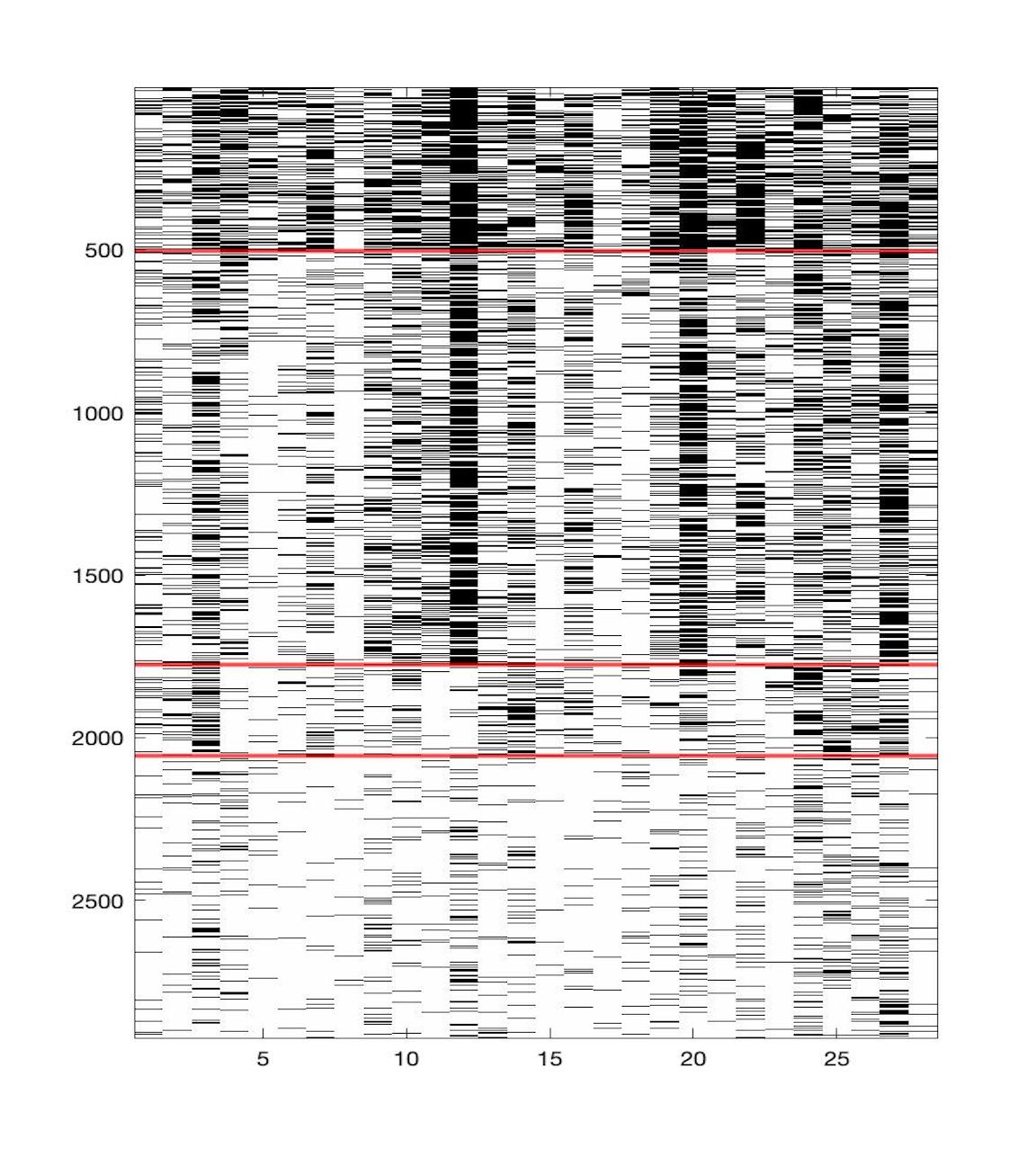} \label{fig:pcdm}}
    \caption{(a): the original data; (b): clustered data from spectral initialization; (c): clustered data from GDINA initialization with known $Q$-matrix; (d): clustered data from the proposed method. Note that the rows of the data matrices in (b), (c), and (d) are permuted differently to better show the clustering structures. The black points stand for response value 1, and the white ones stand for response value 0.}
    \label{fig:ecpe-data}  
\end{figure}

Our proposed method uses a penalized  exploratory latent class analysis approach, which does not  depend on the CDM settings such as the Q-matrix structure and multi-dimensionality of the attributes.
The proposed method does not require any prior information except for an upper bound of number of latent classes $M$.
Here we took $M = 8$, and used spectral clustering to initialize the model parameters.
Specifically, given the data matrix $\mathcal{R} \in \{0,1\}^{N\times J}$, we calculated the symmetric normalized Laplacian matrix $L^{\text{norm}} := I - D^{-1/2}\mathcal{R} D^{-1/2}$, where $D = \text{diag}\{\sum_j R_{1j},\sum_j R_{2j},\dots,\sum_j R_{Nj} \}$.
Then we took the first $M$ eigenvectors of $L$ and performed $k$-means clustering on the eigenvectors.
Based on the clustering results, we had an initialization of the partition of the subjects to $M$ classes and then used class proportions and mean responses to the items as the model initializations. 
The clustered data from spectral initialization is shown in Figure \ref{fig:spectral} and the final estimation results with spectral initialization is  in Figure \ref{fig:pcdm}.
In the plots, each row represents the response vector from a subject and each column represents an item, with dark cells standing for ``1"'s and white cells standing for ``0"'s.
The resulting clusters are separated by red lines.
For ease of visualization, we have rearranged the rows of data to form clusters.

\begin{figure}[h]
    \centering
    \subfigure[]{     \includegraphics[height=2.3in]{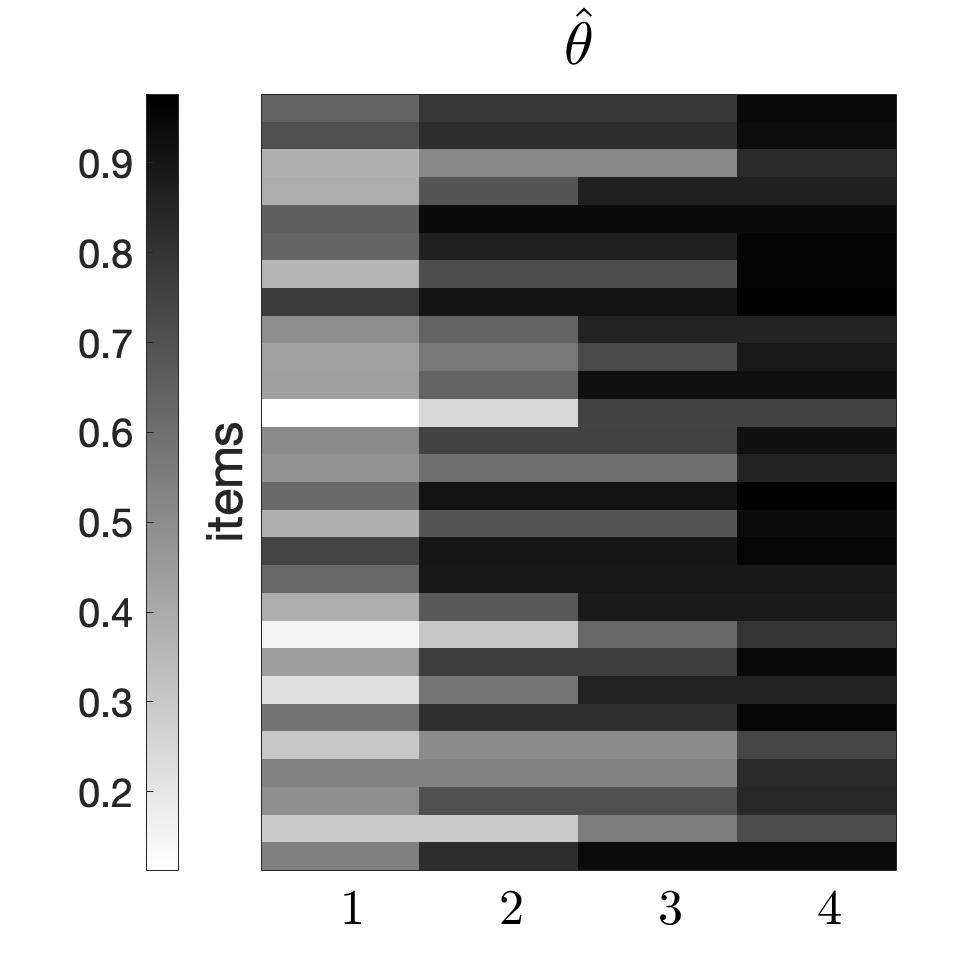}\label{fig:theta}}
    \hspace{0in}
    \subfigure[]{
    \includegraphics[height=2.25in]{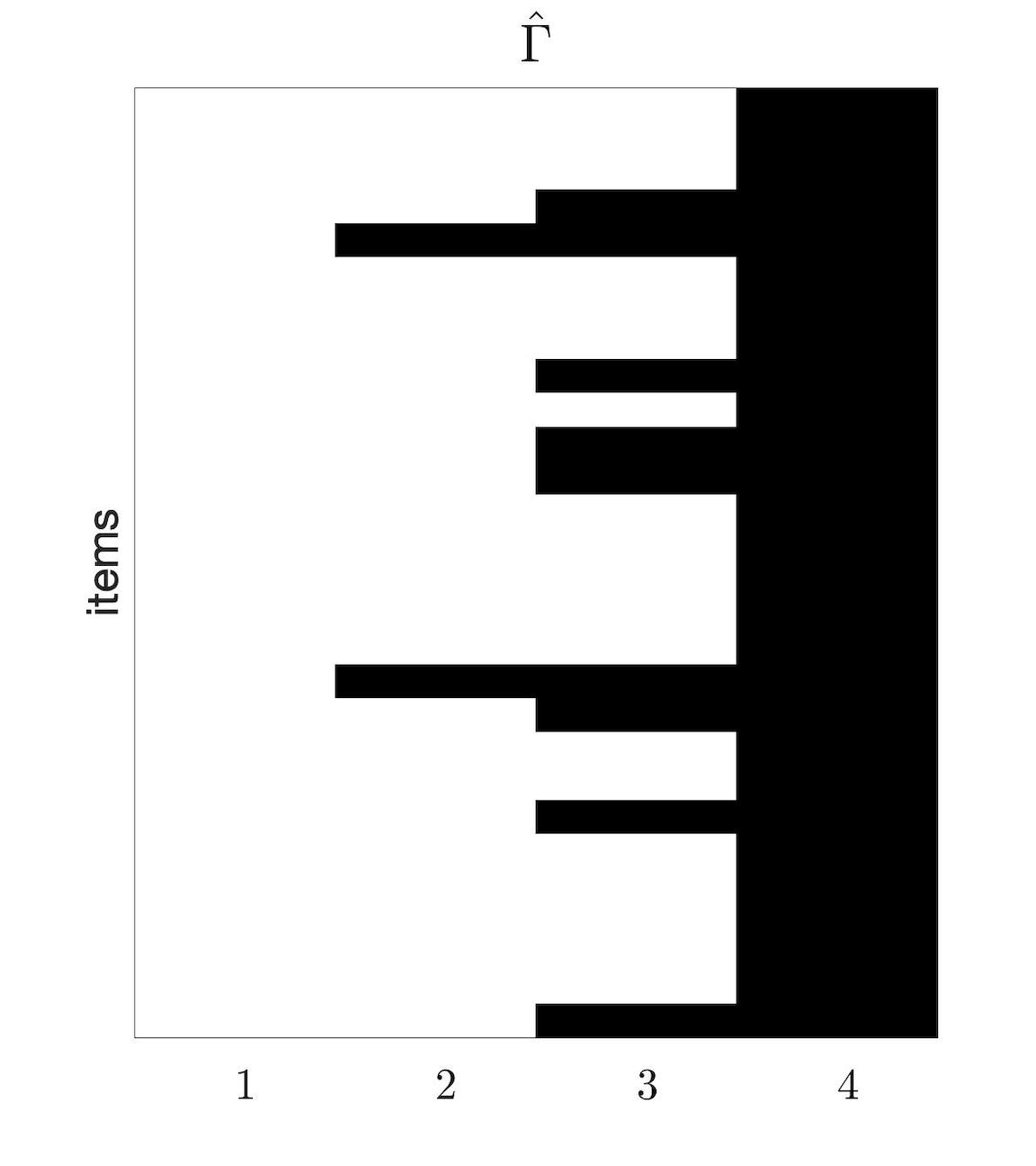}\label{fig:gamma_hat}}
    \hspace{0in}
    \subfigure[]{
    \includegraphics[height=2.25in]{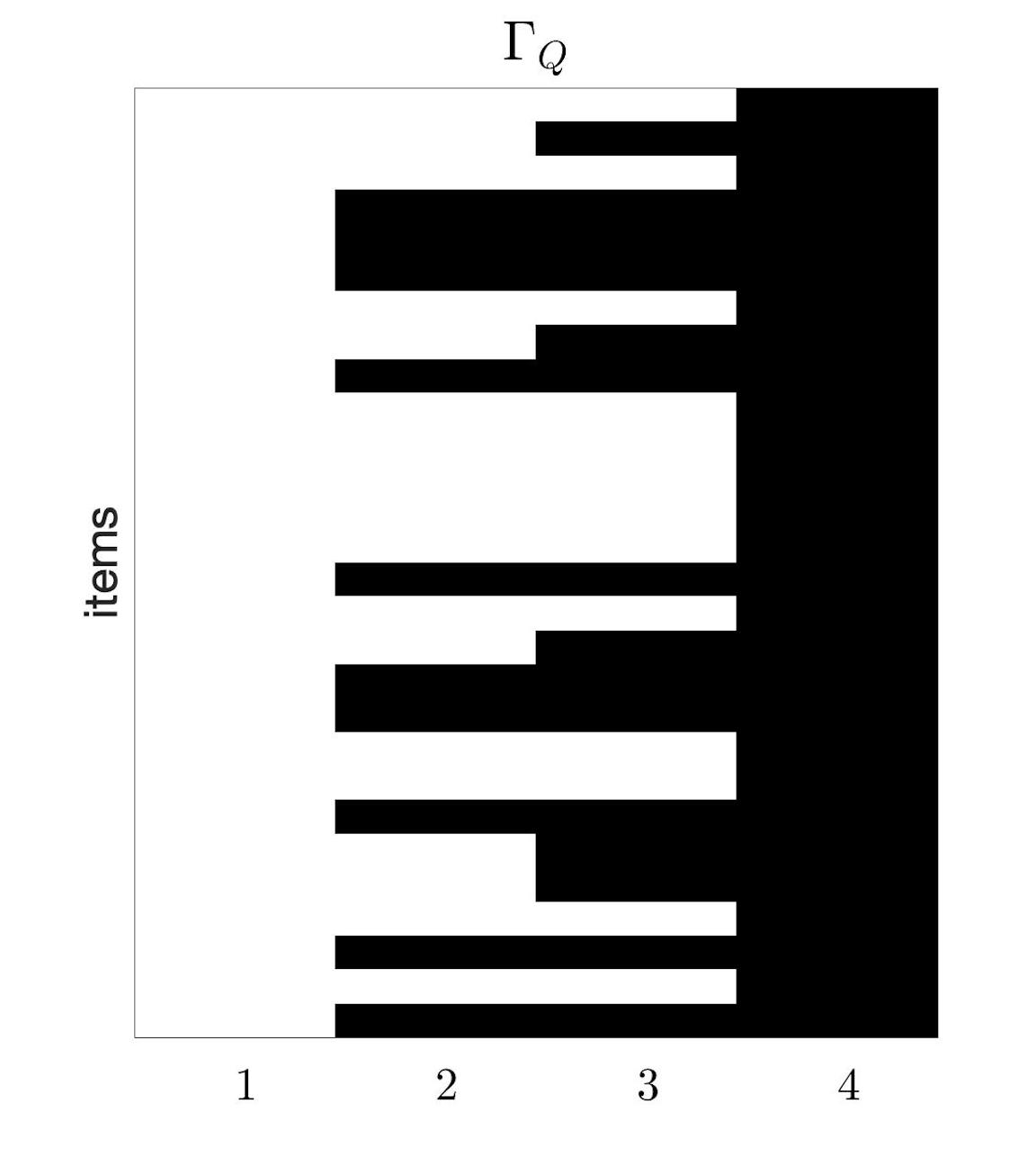}\label{fig:gamma_Q}}
    \caption{(a): estimated $\hat{\bm{\Theta}}$ matrix; (b): reconstructed indicator matrix $\hat{\bm{\Gamma}}$; (c): the indicator matrix based on the pre-specified $Q$-matrix. Black blocks indicate value 1, and white blocks indicate value 0.}
    \label{fig:ecpe}  
\end{figure}

For a comparison purpose, we also used the pre-specified $Q$-matrix to fit a GDINA model with $2^K$ latent classes,
 and then used the learned GDINA estimation results as initialization, which is shown in Figure \ref{fig:gdina}.
We found that the GDINA model initialization using the pre-specified $Q$-matrix  resulted in the same learned model as the spectral initialization, which does not require the pre-specified $Q$-matrix. 
It is also noted that by directly fitting a GDINA model with the pre-designed $Q$-matrix and $2^K$ latent classes, it learned four latent groups with large proportions and all the other proportion parameters were very small, but not exactly zeros. 
Comparing Figure \ref{fig:gdina} and Figure \ref{fig:pcdm}, we can also see that the clustered data based on our method showed a much clearer ordered structure among the latent classes.
Specifically, using the proposed method, we obtained four significant latent classes, as shown in Figure \ref{fig:pcdm}.
From the clustered results, there seemed to be an ordered structure:  the subjects in the first cluster were more likely to give positive responses than those in the second clusters, and the second cluster tended to have more positive responses than the third cluster, and the same for the results in the third and the fourth clusters.
To better identify the hierarchical structure, we further calculated the indicator matrix.
The estimated item parameter matrix $\hat{\bm{\Theta}}$ and the reconstructed indicator matrix $\hat{\bm{\Gamma}}$ are shown in Figure \ref{fig:ecpe}. 
It is easy to see $\hat{\Gamma}_{\cdot,1} \prec \hat{\Gamma}_{\cdot,2} \prec \hat{\Gamma}_{\cdot,3} \prec \hat{\Gamma}_{\cdot,4}$, which indicated a unidimensional located latent class model structure, or in other words, a model structure with strictly ordered latent classes \citep{von2014hier}. 
This finding is consistent with the observation in \cite{von2014hier}.

To present the latent class structure under the HCDM framework, we can apply the proposed Algorithm~\ref{algo-binary}.
Since there are four latent classes and $\hat{\Gamma}_{\cdot,1} \prec \hat{\Gamma}_{\cdot,2} \prec \hat{\Gamma}_{\cdot,3} \prec \hat{\Gamma}_{\cdot,4}$, the smallest $K$ will be 3 and the corresponding binary representations of the latent classes will be $(0,0,0), (1,0,0), (1,1,0), (1,1,1)$, which is consistent with the analysis in \cite{templin2014hierarchical} under the CDM framework. 
Moreover, we also fitted the GDINA models with three latent attributes and linear hierarchy based on the inferred $Q$-matrix from our model and the original designed $Q$-matrix, respectively.
The corresponding indicator matrices obtained from our method and the original $Q$-matrix are shown in Figure \ref{fig:gamma_hat} and \ref{fig:gamma_Q}.
From the fitted GDINA models, we found that the BIC for the original $Q$-matrix was 86,117, while the BIC for our learned $Q$-matrix was 86,000, indicating that our learned $Q$ fits the data better in terms of BIC.

\subsection{Analysis of PISA Data}

To test on a more complex and realistic setting, we also applied the proposed approach to a dataset from Programme for International
Student Assessment (PISA), an international reading assessment for 15-year-old students.
In particular, we used a PISA 2000 dataset from R package \verb!CDM!, which was previously studied in \cite{chen2014pisa}. 
This dataset contains $J = 26$ items from six independent articles assessing 1096 examinee's reading abilities.
Most of 26 items are dichotomous items except for some trichotomous items.
We converted the trichotomous items to dichotomous by combining all non-zero response values as one category, where we regarded any partial or full credit case as a success  and no credit as a failure. 
In \cite{chen2014pisa}, the authors specified six latent attributes for the PISA 2000 data: $(1)$ locating information; $(2)$ forming a broad general understanding; $(3)$ developing a logical interpretation; $(4)$ evaluating a number-rich text with number sense; $(5)$ evaluating the quality or appropriateness of a text; $(6)$ test speededness.

To apply the proposed method, we set the initial number of latent classes $M = 2^6=64$ 
and initialized the models parameters using the pre-specified $Q$-matrix in \cite{chen2014pisa}.
Specifically, we first fitted a GDINA model using the pre-specified $Q$-matrix, and then used the estimated item and mixture proportion parameters as the initial values for the item parameter matrix and the proportion parameter vector $\bm{\pi}$.

\begin{figure}[h!]
    \centering
    \subfigure[]{
    \includegraphics[height=2.3in]{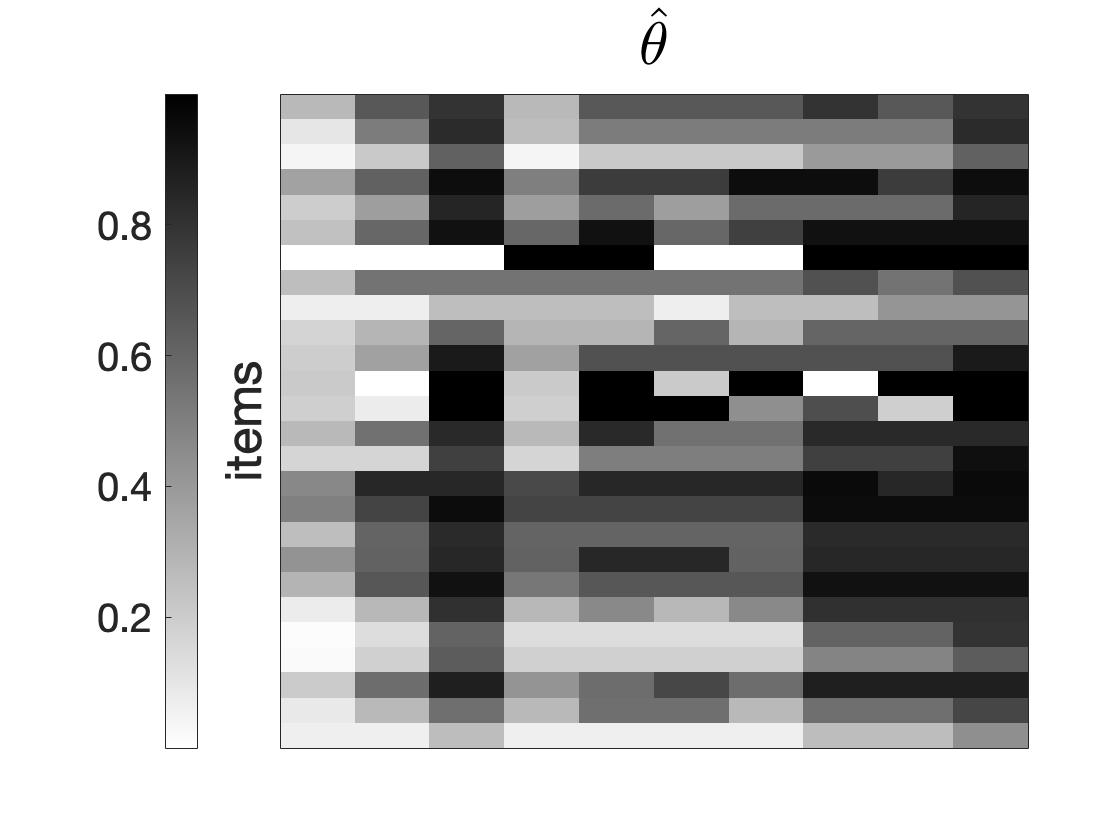}\label{fig:theta_p}}~\subfigure[]{
    \includegraphics[height=2.25in]{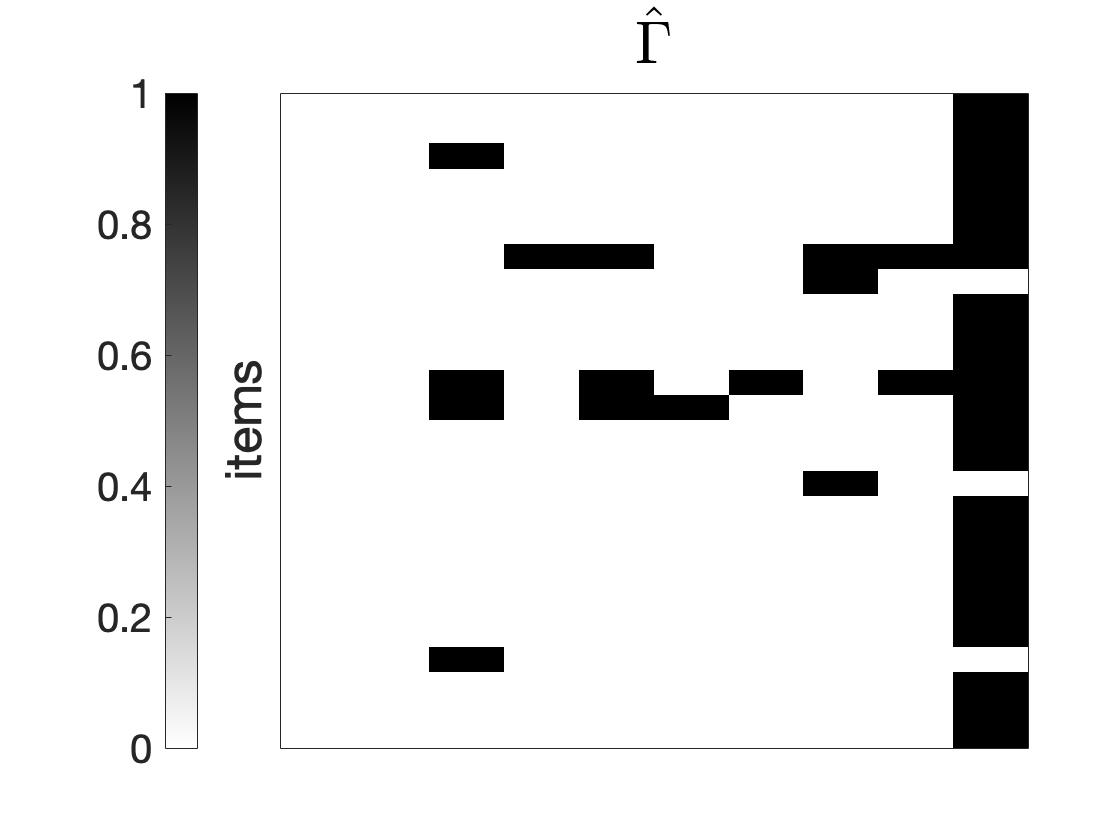}\label{fig:gamma_pisa}}
    \hspace{0in} 
	\vspace{0.3in}
    \centering
    \subfigure[]{%
	\begin{tikzpicture}[baseline= -0ex][scale=0.2]
	\centering 
    \path
    node at (0, 1.5) [place] (1) {$\Gamma_{\cdot, 1}$}
    	node at (0, 0) [place] (2) {$\Gamma_{\cdot, 2}$}
    	node at (1, -2.5) [place] (3) {$\Gamma_{\cdot, 3}$}
    	
    	node at (-1.3, -1.3) [place] (4) {$\Gamma_{\cdot, 4}$}
    	node at (0, -3.7) [place] (5) {$\Gamma_{\cdot, 5}$}
    	node at (0, -1.3) [place] (6) {$\Gamma_{\cdot, 6}$}    
    	node at (1.3, -1.3) [place] (7) {$\Gamma_{\cdot, 7}$}
    	
    	node at (-2, -2.5) [place] (8) {$\Gamma_{\cdot, 8}$}
    	node at (-0.5, -2.5) [place] (9) {$\Gamma_{\cdot, 9}$}
    	node at (0, -5) [place] (10) {$\Gamma_{\cdot, 10}$};

    \draw [->, thick] (1) to (2);
    
    \draw [->, thick] (2) to (4);
    \draw [->, thick] (2) to (6);
    \draw [->, thick] (2) to (7);

    \draw [->, thick] (4) to (8);
    \draw [->, thick] (4) to (9);
    
    \draw [->, thick] (7) to (9);
    \draw [->, thick] (7) to (3);
    \draw [->, thick] (9) to (5);
    \draw [->, thick] (6) to (5);
    \draw [->, thick] (5) to (10);
	\end{tikzpicture}%
	}\qquad\qquad\qquad\qquad\qquad \subfigure[]{%
    \begin{tikzpicture}[baseline= -10ex][scale=1]
	\centering
   \path 
    node at (10.5, -1.5) [place] (1) {$\alpha_1$}
    node at (9, -3) [place] (2) {$\alpha_2$}
    node at (10, -3) [place] (3) {$\alpha_3$}
        node at (11, -3) [place] (4) {$\alpha_4$}
    node at (12, -3) [place] (5) {$\alpha_5$}
    node at (10.5, -4.5) [place] (6) {$\alpha_6$};
    \draw [->, thick] (1) to (2);
    \draw [->, thick] (1) to (3);
    \draw [->, thick] (1) to (4);
    \draw [->, thick] (1) to (5);
    \draw [->, thick] (2) to (6);
        \draw [->, thick] (3) to (6);
    \draw [->, thick] (4) to (6);

    \end{tikzpicture}%
       \hspace{0.2in}
} 
    \caption{(a): Estimated $\hat{\bm{\Theta}}$ matrix; (b): Reconstructed indicator matrix $\hat{\bm{\Gamma}}$. (c): Partial orders; (d): Constructed hierarchical structure of latent attributes.}
    \label{fig:pisa}  
\end{figure}
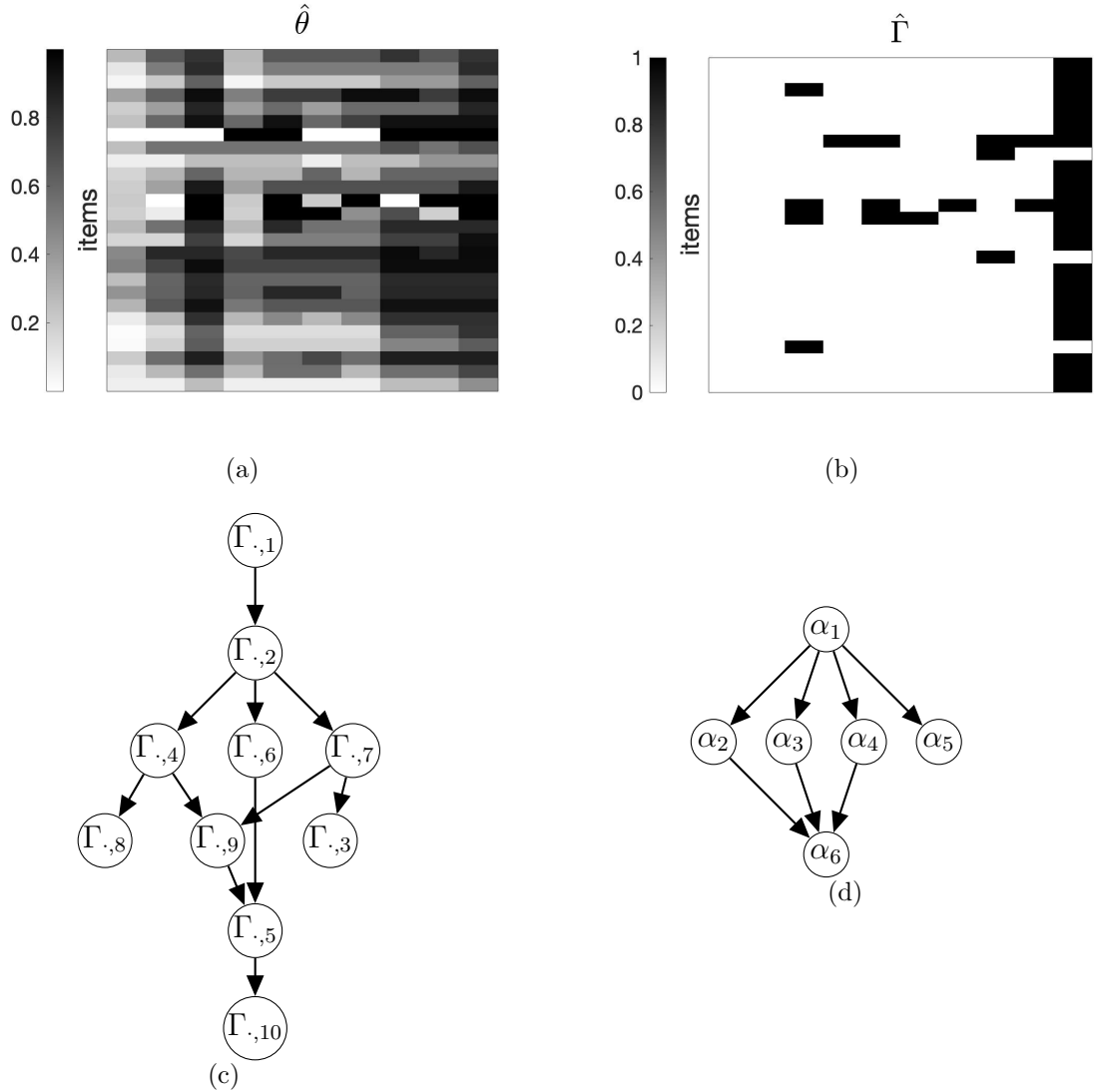

After applying the proposed method to the PISA 2000 data, we learned 10 significant latent classes. 
On average, for each set of hyperparameters, the computation time  was 14.87 seconds. 
The estimated item parameter matrix $\hat{\Theta}$ and  the reconstructed indicator matrix $\hat{\Gamma}$ are shown in Figure~\ref{fig:pisa}(a) and Figure~\ref{fig:pisa}(b), respectively.  
Based on the indicator matrix, we recovered the partial orders of these 10 latent classes in Figure~\ref{fig:pisa}(c), which suggests a multi-dimensional latent structure. 
With partial orders recovered, we applied Algorithm~\ref{algo-binary} and recovered six latent attributes with a hierarchical structure as shown in Figure~\ref{fig:pisa}(d).

The six latent attributes and their hierarchical structures learned from the data may match the prior study in \cite{chen2014pisa} as follows.
The recovered attributes $\alpha_1$ to $\alpha_6$ may correspond to ``locating information", ``forming a broad general understanding", ``evaluating a number-rich text with number sense", ``evaluating the quality or appropriateness of a text", ``test speededness", and ``developing a logical interpretation", respectively.
From the hierarchical structure in Figure~\ref{fig:pisa}(d), $\alpha_1$ can be viewed as a basic prerequisite for other attributes, 
which makes sense because examinees need to first understand the item and correctly identify the key information in the article before forming understanding or evaluating the text,
while developing a logical interpretation ($\alpha_6$) can be interpreted as a more advanced skill.

Besides the interpretable hierarchical structure, we also assessed the model fit using the BIC.
Specifically, the BIC of our estimation was 7045, while the BIC of the full GDINA model with the pre-specified $Q$-matrix was 31495, indicating that the proposed method improved the model fitting in terms of BIC.

\section{Discussion}
\label{sec-discuss}

In this paper we propose a penalized likelihood approach to simultaneously learn the number of latent attributes, the hierarchical structure, the item-attribute $Q$-matrix and item-level diagnostic models in hierarchical CDMs.
We achieve these goals by imposing two regularization terms on an exploratory latent class model: one is a log-type penalty on proportion parameters and the other is truncated Lasso penalty on the differences among item parameters.
The nice form of the penalty terms facilitates the computation and an efficient EM-type algorithm is developed.
A latent structure recovery algorithm is also provided based on the learned model parameters.
The simulation study and real data analysis demonstrate good performance of the proposed method.

In most existing works of learning CDMs, the hierarchical structures of latent attributes are either not considered or pre-specified by domain experts.
Moreover, related works using exploratory approaches to learn latent hierarchies also require additional pre-specifications such as the number of latent attributes.
By contrast, in this work, we develop an exploratory regularized likelihood approach with minimal model specifications.
In particular, we estimate the number of latent attributes, recover the hierarchical structure, the $Q$-matrix, and the item-level diagnostic models simultaneously.
The price we have to pay for the minimal model assumptions is that a set of hyperparameters need to be tuned, while our simulation results show that we can achieve it computationally efficiently.

In addition to computational efficiency, our proposed method also has theoretical guarantees.	Specifically, we show that the number of latent classes, the model parameters and the constraint structure of the item parameter matrix can be consistently estimated. 
Moreover, based on the assumption that more capable subjects have higher item parameters, we develop estimation procedures to recover the number of latent attributes, hierarchical structures and $Q$-matrix from the introduced indicator matrix.
Due to the consistency of the item parameter matrix and its constraint structure, under the identifiability conditions, the indicator matrix is also consistently estimated, which leads to the consistency of these latent specifications as well.
Although the method is purely data-driven, our analysis also provides sound theoretical support. 
With the theoretical foundation established, our method is consistent and robust in learning the hierarchical structure and other CDM characteristics.

A natural follow-up question would be how we conduct hypothesis testing for the learned hierarchies.
Since the existence of hierarchical structures would result in the sparsity structure of the proportion parameter vector, it is equivalent to testing the zero elements in the proportion vector.
However, due to the irregularity of the problem since the true parameter now is lying on the boundary of the parameter space, the limiting distribution of the likelihood ratio statistic would be   complicated.
 As noted in the literature  \citep{ma2021hypothesis}, such nonstandard tests need to be further investigated theoretically.

Currently the proposed model is applied  to a static setting where we only have a data set for a fixed time point.
It would be also interesting to extend it to the dynamic setting, where multiple measurement data sets for a sequence of time points are available.
We can also learn such hierarchical structures by inferring learning trajectories of the subjects.
Moreover, taking the hierarchical structures into consideration, we can generate recommendations for learning materials or test items by formulating a sequential decision problem.
We leave these interesting directions for future work.

\section*{Acknowledgments}
The authors are grateful to the Editor-in-Chief Professor Matthias von Davier, an Associate Editor, and three referees for their valuable comments and suggestions. This research is partially supported by NSF CAREER SES-1846747 and Institute of Education Sciences R305D200015.

\bibliographystyle{apalike}
\bibliography{refs}

\end{document}


\maketitle

\pagestyle{plain}
\setcounter{page}{1}
\pagenumbering{arabic}

In the supplementary material, we provide the proof of the main theorem, the derivations for the penalized EM algorithm and a sensitivity analysis of our algorithm with varying upper bounds for the number of latent classes.

\section{Proof for Theorem 3}

\label{sec:S1}

In this section, we provide the proof of Theorem 3.

\begin{proof}
We first introduce some notations. 
For two sequences $\{a_N\}$ and $\{b_N\}$, we denote $a_N \lesssim b_N$ if $a_N = O(b_N)$, and $a_N \asymp b_N$ if $a_N \lesssim b_N$ and $b_N \lesssim a_N$. 
We use $(\bm{\pi}^0, \bm{\Theta}^0)$ to denote the true model parameter and use $(\hat{\bm{\pi}}^0, \hat{\bm{\Theta}}^0)$ to denote the oracle MLE obtained by assuming the number of latent attributes, the hierarchical structure, the $Q$-matrix and the item-level diagnostic models are known.
Let $(\hat{\bm{\pi}}^*, \hat{\bm{\Theta}}^*)$ be the MLE obtained by directly optimizing log-likelihood \eqref{eq:log-like} and $(\hat{\bm{\pi}}, \hat{\bm{\Theta}})$ be the estimator obtained by optimizing the regularized log-likelihood \eqref{eq-objective}.
We define $\hat{\bm{\pi}}_{\rho_N} := \{\hat{\pi}_m: \hat{\pi}_m > \rho_N, \ m\in [M]\}$ and $\hat{\bm{\Theta}}_{\rho_N} := \{\hat{\theta}_{j,m}: \hat{\pi}_m > \rho_N, \ j\in [J], \ m\in [M]\}$, the model parameters corresponding to the selected latent classes.
Let $M$ be the upper bound for the number of latent classes, $M_0$ be the true number of latent classes, and $\hat{M} = \big|\{m: \hat{\pi}_m > \rho_N, \ m \in [M]\}\big|$ be the estimated number of latent classes.
Without loss of generality, let $\hat{\bm{\pi}}^0_{\text{full}} = (\hat{\bm{\pi}}^0, \bm{0}_{M - M_0})$.
For the true item parameter matrix $\bm{\Theta}^0$, we defined the set of identical item parameter pairs $S^0 = \big\{(j,k_1,k_2):\theta_{j,k_1}^0=\theta_{j,k_2}^0,1\leq k_1<k_2\leq M_0\big\}$.
Similarly, for $(\hat{\bm{\pi}}, \hat{\bm{\Theta}})$ 
we define $\hat{S} = \big\{(j,k_1,k_2): \hat{\theta}_{j,k_1}=\hat{\theta}_{j,k_2},\ 1\leq k_1<k_2\leq M, \ \hat{\pi}_{k_1}>\rho_N, \ \hat{\pi}_{k_2}>\rho_N\big\}$.
We say $\hat{S} \sim S^0$ if there exists a column permutation $\sigma$ of $\hat{\bm{\Theta}}$ such that $\hat{S}_{\sigma} = S^0$.

The probability $\mathbb{P} \big( \hat{M} \neq M_0 \big)$ can be decomposed into two parts:
\begin{equation}
	\mathbb{P} \big(\hat{M} \neq M_0 \big) = \mathbb{P} \big(\hat{M} < M_0 \big) + \big(\hat{M} > M_0 \big).
	\label{eq:P-M}
\end{equation}
Similarly, the probability $\mathbb{P} \big( \hat{S} \neq S^0 \big)$ can be decomposed into three parts:
\begin{equation}
	\mathbb{P} \big(\hat{S} \nsim S^0 \big) = \mathbb{P} \big(\hat{M} < M_0 \big) + \big(\hat{M} > M_0 \big) + \mathbb{P}\big( \hat{S} \nsim S^0, \hat{M} = M_0\big).
	\label{eq:P-S}
\end{equation}
In the following ,we will bound each part in \eqref{eq:P-M} and \eqref{eq:P-S} respectively.
Therefore, we will consider three cases below:
\begin{enumerate}
	\item overfitted case: $\hat{M} > M_0$,  
	\item underfitted case: $\hat{M} < M_0$, 
	\item $\hat{M} = M_0$ but $\hat{S} \nsim S^0$.
\end{enumerate}

The objective function is
\begin{equation}
    G_N(\bm{\pi}, \bm{\Theta}) = \frac{l_N(\bm{\pi}, \bm{\Theta};\mathcal{R})}{N} - \frac{\lambda_N^{(1)}}{N}\sum_{k = 1}^M \log_{[\rho_N]}\pi_k - \frac{\lambda_N^{(2)}}{N}\sum_{j=1}^J \mathcal{J}_{\tau,\rho_N}(\bm{\theta}_j),
\end{equation}
where $\log_{[\rho_N]}\pi_k = \log\pi_k\cdot\mathbb{I}\big(\pi_k>\rho_N\big) + \log{\rho_N}\cdot\mathbb{I}\big(\pi_k\leq \rho_N\big)$. 
Let $\log_{[\rho_N]}(\bm{\pi}) = \sum_{k = 1}^M \log_{[\rho_N]}\pi_k$. 

First consider the overfitted case where $\hat{M} > M_0$. 
The event $\big\{ G_N \big( \hat{\bm{\pi}}, \hat{\bm{\Theta}} \big) > G_N \big(\hat{\bm{\pi}}^0, \hat{\bm{\Theta}}^0 \big) \big\}$ implies that
\begin{align}\nonumber
    & \frac{1}{N}\sum_{i=1}^N \log \Big[\frac{\sum_{k=1}^M \hat{\pi}_k \prod_{j=1}^J \hat{\theta}_{j,k}^{R_{ij}}(1-\hat{\theta}_{j,k})^{1-R_{ij}}}
    {\sum_{k=1}^M \hat{\pi}_k^0 \prod_{j=1}^J (\hat{\theta}_{j,k}^0)^{R_{ij}}(1-\hat{\theta}_{j,k}^0)^{1-R_{ij}}}\Big]
    \\ \label{eq-overfit}
    > & \ \frac{\lambda_N^{(1)}}{N}\big\{
    \log_{[\rho_N]}(\bm{\hat{\pi}})-\log_{[\rho_N]}(\hat{\bm{\pi}}_{full}^0)
    \big\}+ \frac{\lambda_N^{(2)}}{N}\Big\{\sum_{j=1}^J \mathcal{J}_{\tau,\rho_N}(\hat{\bm{\theta}}_j) - \sum_{j=1}^J \mathcal{J}_{\tau,\rho_N}(\hat{\bm{\theta}}_j^0)\Big\}\\ \nonumber
    := & \ J_1+J_2.
\end{align}
\noindent For the RHS of \eqref{eq-overfit}, we have $J_1\gtrsim N^{-1} \lambda_N^{(1)}|\log \rho_N|$ and $J_2\gtrsim - N^{-1} \lambda_N^{(2)} \tau J M^2$. 
Since $\lambda_N^{(2)}\tau = o(\lambda_N^{(1)}|\log \rho_N|)$, we have $\text{RHS}\gtrsim N^{-1}\lambda_N^{(1)}|\log \rho_N|$.

For the LHS of \eqref{eq-overfit}, we have 
\begin{align*}
	\text{LHS of }\eqref{eq-overfit} = & \frac{1}{N} \log \big[ \sum_{k=1}^M \hat{\pi}_k \prod_{j=1}^J \hat{\theta}_{j,k}^{R_{ij}}(1-\hat{\theta}_{j,k})^{1-R_{ij}} \big] - \frac{1}{N} \log \big[ \sum_{k=1}^M \hat{\pi}_k^0 \prod_{j=1}^J (\hat{\theta}_{j,k}^0)^{R_{ij}}(1-\hat{\theta}_{j,k}^0)^{1-R_{ij}} \big]\\
		& \leq \frac{1}{N} \log \big[ \sum_{k=1}^M \hat{\pi}_k^* \prod_{j=1}^J (\hat{\theta}_{j,k}^*)^{R_{ij}}(1-(\hat{\theta}_{j,k}^*))^{1-R_{ij}} \big] - \frac{1}{N} \log \big[ \sum_{k=1}^M \hat{\pi}_k^0 \prod_{j=1}^J (\hat{\theta}_{j,k}^0)^{R_{ij}}(1-\hat{\theta}_{j,k}^0)^{1-R_{ij}} \big]\\
		& \lesssim N^{-\delta},
\end{align*}
where the last inequality follows from Assumption \ref{assum-mle}. 
When $N^{1-\delta}/|\log(\rho_N)| = o\big(\lambda_N^{(1)}\big)$, we have $N^{-\delta} = o\big(N^{-1}\lambda_N^{(1)}|\log \rho_N|\big)$, which implies that the event described in (\ref{eq-overfit}) will happen with probability tending to zero.
Therefore we have $\mathbb{P}\big(\hat{M} > M_0 \big) \xrightarrow{}0$ as $N\xrightarrow{}\infty$.
That is to say, with the appropriate choice of tuning parameters, the extent that the log-penalty part favors a smaller model would dominate the extent that the likelihood part favors a larger model in the overfitted case.

Now consider the under-fitted case where $\hat{M} < M_0$.
We need to bound
\begin{equation}
    \mathbb{P}\Big(\sup_{\hat{M}<M_0}\big[G_N(\hat{\bm{\pi}}, \hat{\bm{\Theta}})-G_N(\hat{\bm{\pi}}^0, \hat{\bm{\Theta}}^0)\big]>0\Big).
\end{equation}
\noindent We follow a similar argument to \cite{shen2012likelihood}.
More specifically, since
\begin{equation}
    \begin{split}
        \mathbb{P} \Big(    \sup_{\hat{M} < M_0} \big[
        G_N(\hat{\bm{\pi}}, \hat{\bm{\Theta}}) - G_N(\hat{\bm{\pi}}^0, \hat{\bm{\Theta}}^0)
        \big]>0
        \Big)
        &\leq \sum_{m=1}^{M_0 - 1} 
        \mathbb{P}\Big(
        \sup_{\hat{M}=m } \big[
        G_N(\hat{\bm{\pi}}, \hat{\bm{\Theta}}) - G_N(\hat{\bm{\pi}}^0, \hat{\bm{\Theta}}^0)
        \big] > 0
        \Big),
    \end{split}
    \label{eq-decom}
\end{equation}
we will bound each term in the RHS of \eqref{eq-decom}.
By the large deviation inequality in Theorem 1 of \cite{wong1995probability}, we have
\begin{equation}
    \begin{split}
    &\mathbb{P}\Big(
    \sup_{h^2\big((\hat{\bm{\pi}}, \hat{\bm{\Theta}}), (\bm{\pi}^0, \bm{\Theta}^0)\big)\geq \epsilon_N^2} \big[
    \frac{1}{N}l_N\big(\hat{\bm{\pi}}, \hat{\bm{\Theta}}\big) - \frac{1}{N}l_N\big(\bm{\pi}^0, \bm{\Theta}^0\big)
    \big]
    > -\epsilon_N^2
    \Big)\\
    \leq & \ \mathbb{P}\Big(
    \sup_{h^2\big((\hat{\bm{\pi}}, \hat{\bm{\Theta}}), (\bm{\pi}^0, \bm{\Theta}^0)\big)\geq \epsilon_N^2} \big[
    \frac{1}{N}l_N\big(\hat{\bm{\pi}}, \hat{\bm{\Theta}}\big) - \frac{1}{N}l_N\big(\bm{\pi}^0, \bm{\Theta}^0\big)
    \big]
    > -\epsilon_N^2
    \Big) \leq \exp (-N\epsilon_N^2),
    \end{split}{}
    \label{inequal}
\end{equation}
where $h^2\big((\hat{\bm{\pi}}, \hat{\bm{\Theta}}), (\bm{\pi}^0, \bm{\Theta}^0)\big) = \sum_{\bm{R}\in\{0,1\}^J} \big[\mathbb{P}(\bm{R} \mid \hat{\bm{\pi}}, \hat{\bm{\Theta}})^{1/2} -  \mathbb{P}(\bm{R} \mid \bm{\pi}^0, \bm{\Theta}^0)^{1/2}\big]$ is the Hellinger distance.
From the remark in \cite{wong1995probability}, the inequality (\ref{inequal}) holds for any $t>\epsilon_N$. 

To use this large deviation inequality, we need to introduce the notion of bracketing Hellinger metric entropy $H(t,\mathcal{B}_m)$, which characterizes the size of the local parameter space. 
Consider the local parameter space $\mathcal{B}_m=\big\{
\big(\hat{\bm{\pi}}, \hat{\bm{\Theta}}\big): \hat{M} = m \leq  M_0, \ h^2\big((\hat{\bm{\pi}}, \hat{\bm{\Theta}}),\big(\bm{\pi}^0, \bm{\Theta}^0)\big)\leq 2\epsilon_N^2
\big\}$,
then $H(t,\mathcal{B}_m)$ is defined as the logarithm of the cardinality of the t-bracketing of $\mathcal{B}_m$ of the smallest size.
Specifically, following the definition in \cite{shen2012likelihood}, consider a bracket covering $S(t,m)=\{f_1^l,f_1^u,\cdots,f_m^l,f_m^u\}$ such that $\max_{1\leq j\leq m}||f_j^u-f_j^l||_2\leq t$ 
and for any $f\in\mathcal{B}_m$, there is some $j$ such that $f_j^l\leq f \leq f_j^u$ almost surely.
Then $H(t,\mathcal{B}_m)$ is defined as $\log\big(\min \{m:S(t,m)\}\big)$.
Following Lemma 3 in \cite{gu2019learning}, for any $2^{-4}\epsilon < t < \epsilon$, there is
\begin{equation}
    H(t,\mathcal{B}_m)\lesssim M_0\log M \log (2\epsilon/t).
    \label{eq-H}
\end{equation}{}

Next we need to verify the conditions in \cite{wong1995probability}.
Let's take 
$\epsilon_N=\sqrt{M_0\log M /N}$ and verify the entropy integral condition in Theorem 1 of \cite{wong1995probability} for such $\epsilon_N$.
The integral of bracketing Hellinger metric entropy on the interval $[2^{-8}\epsilon_N^2, \sqrt{2}\epsilon_N]$ satisfies the following inequality
\begin{align*}
	\int_{2^{-8}\epsilon_N^2}^{\sqrt{2}\epsilon_N}H^{1/2}(t,\mathcal{B}_m)dt &\leq \int_{2^{-8}\epsilon_N^2}^{\sqrt{2}\epsilon_N} \sqrt{M_0\log M \log (2\epsilon_N/t)}dt\\
    &=\sqrt{M_0\log M}\int_{\sqrt{\log \sqrt{2}}}^{\sqrt{\log \frac{2^9}{\epsilon_N}}}
    4\epsilon_N u^2 e^{-u^2}du\\
    &=\sqrt{M_0\log M}\cdot 2 \epsilon_N \int_{\log \sqrt{2}}^{\log \frac{2^9}{\epsilon_N}}\sqrt{u}e^{-u}du\\
    &\lesssim\sqrt{N}\epsilon_N^2.
\end{align*}
Note that $\epsilon_N = o(1)$ as $N\rightarrow \infty$.  



Following the proof in \cite{gu2019learning}, there exists a constant $c_0$, for some small constant $t > \epsilon_N$, we have 
\[
C_{\min}(\bm{\pi}^0, \bm{\Theta}^0):= \inf_{(\hat{\bm{\pi}}, \hat{\bm{\Theta}}):\hat{M} \leq M_0}\Big\{
    \frac{h^2\big((\hat{\bm{\pi}}, \hat{\bm{\Theta}}),(\bm{\pi}^0, \bm{\Theta}^0)\big)}{\max\big(M_0-\hat{M}, 1 \big)}\Big\}  \geq c_0  \gtrsim t^2 > \epsilon_N^2.
\]
Moreover, for $\hat{M} = m < M_0$, there is $h^2((\hat{\bm{\pi}}, \hat{\bm{\Theta}}),(\bm{\pi}^0, \bm{\Theta}^0)) \geq \big(M_0 - m \big)C_{\min}(\bm{\pi}^0, \bm{\Theta}^0)$.
In order to have the probability of the event (\ref{eq-overfit}) go to zero in the under-fitted case, the log-penalty term should not be too large such that the likelihood part is dominated by the log-penalty term that favors a smaller model. 
Here we take $\lambda_N^{(1)}=o(N \log \rho_N|^{-1})$. 
Then for (\ref{eq-decom}) we have
\begin{align*}
    &\quad \text{RHS of (\ref{eq-decom})}\\
    &\leq \sum_{m=1}^{M_0 - 1} 
    \mathbb{P} \Big(
    \sup_{ h^2((\hat{\bm{\pi}}, \hat{\bm{\Theta}}),(\bm{\pi}^0, \bm{\Theta}^0))\geq (M_0- m)C_{\min}(\bm{\pi}^0, \bm{\Theta}^0),\hat{M}=m} \big[
    G_N(\hat{\bm{\pi}}, \hat{\bm{\Theta}}) - G_N(\hat{\bm{\pi}}^0, \hat{\bm{\Theta}}^0)
    \big]>0\Big)
    \\   
    &\leq \sum_{m=1}^{M_0 - 1} 
    \mathbb{P} \Big(
    \sup_{h^2((\hat{\bm{\pi}}, \hat{\bm{\Theta}}),(\bm{\pi}^0, \bm{\Theta}^0))\geq (M_0- m)C_{\min}(\bm{\pi}^0, \bm{\Theta}^0),\hat{M}=m} \big[
    l_N(\hat{\bm{\pi}}, \hat{\bm{\Theta}}) - l_N(\hat{\bm{\pi}}^0, \hat{\bm{\Theta}}^0)
    \big]> -\frac{\lambda_N^{(1)}M_0|\log \rho_N|}{N}
    \Big)\\
    & \leq \sum_{m=1}^{M_0 - 1} 
    \mathbb{P} \Big(
    \sup_{ h^2((\hat{\bm{\pi}}, \hat{\bm{\Theta}}),(\bm{\pi}^0, \bm{\Theta}^0))\geq (M_0- m)C_{\min}(\bm{\pi}^0, \bm{\Theta}^0),\hat{M}=m} \big[
    l_N(\hat{\bm{\pi}}, \hat{\bm{\Theta}}) - l_N(\bm{\pi}^0, \bm{\Theta}^0)
    \big]> -\frac{\lambda_N^{(1)}M_0|\log \rho_N|}{N}
    \Big)
    \\
    & \leq \sum_{m=1}^{M_0 - 1} 
    \mathbb{P} \Big(
    \sup_{ h^2((\hat{\bm{\pi}}, \hat{\bm{\Theta}}),(\bm{\pi}^0, \bm{\Theta}^0)) \geq (M_0- m)C_{\min}(\bm{\pi}^0, \bm{\Theta}^0),\hat{M}=m} \big[
    l_N(\hat{\bm{\pi}}, \hat{\bm{\Theta}}) - l_N(\bm{\pi}^0, \bm{\Theta}^0)
    \big]> - (M_0- m)C_{\min}(\bm{\pi}^0, \bm{\Theta}^0)
    \Big)\\
    & \leq \sum_{m=1}^{M_0-1} \exp \big(
    -c_2 N (M_0 - m) C_{\min}(\bm{\pi}^0, \bm{\Theta}^0)
    \big)\\
    &\leq c_3 \exp\big(-c_2 N C_{\text{min}}(\bm{\pi}^0, \bm{\Theta}^0)\big).
\end{align*}
Therefore we have $\mathbb{P}\big(\hat{M}<M_0\big) \xrightarrow{}0$ as $N\xrightarrow{}\infty$.
So far we have proved \eqref{eq:consistency-pi} in Theorem \ref{thm:consistency},
\[
\mathbb{P}\big(\hat{M}\neq M_0\big) = \mathbb{P}\big(\hat{M}<M_0\big)  + \mathbb{P}\big(\hat{M}>M_0\big) \longrightarrow{}0.
\]

Finally we consider the third case where $\hat{M}=M_0$ but $\hat{S}\nsim S^0$. 
The argument is similar to the proof of Proposition 2 in \cite{xu2018identifying}.
We first show $(\hat{\bm{\pi}}_{\rho_N}, \hat{\bm{\Theta}}_{\rho_N})$ converge to $(\bm{\pi}^0, \bm{\Theta}^0)$ with rate $N^{-1/2}$.
For $(\bm{\pi}, \bm{\Theta})$ with $(\bm{\pi}_{\rho_N}, \bm{\Theta}_{\rho_N})$ in a small neighborhood of $(\bm{\pi}^0, \bm{\Theta}^0)$,
\begin{align*}
	G'_N(\bm{\pi}_{\rho_N}, \bm{\Theta}_{\rho_N}) := & \frac{l_N(\bm{\pi}_{\rho_N}, \bm{\Theta}_{\rho_N};\mathcal{R})}{N} - \frac{\lambda_N^{(1)}}{N}\sum_{k: \pi_k > \rho_N} \log\pi_k - \frac{\lambda_N^{(2)}}{N}\sum_{j=1}^J \mathcal{J}_{\tau,\rho_N}(\bm{\theta}_j) \\
	= & \frac{l_N(\bm{\pi}_{\rho_N}, \bm{\Theta}_{\rho_N};\mathcal{R})}{N} - O(\lambda_N^{(1)}N^{-1}|\log \rho_N|) - O(\lambda_N^{(2)}\tau N^{-1}),
\end{align*}
converges uniformly to the same limit of $l_N(\bm{\pi}_{\rho_N}, \bm{\Theta}_{\rho_N};\mathcal{R})/N$ by the uniform law of large number, since $\lambda_N^{(1)}N^{-1}|\log \rho_N| \rightarrow 0$ and $\lambda_N^{(2)}\tau N^{-1} \rightarrow 0$.
We use $G_0(\bm{\pi}_{\rho_N}, \bm{\Theta}_{\rho_N})$  to denote the limit process, which is the expectation of the negative log-likelihood of a single observation.
By Taylor’s expansion, we have $G_0(\bm{\pi}_{\rho_N}, \bm{\Theta}_{\rho_N}) - G_0(\bm{\pi}^0, \bm{\Theta}^0) = O(\big|\big|(\bm{\pi}_{\rho_N}, \bm{\Theta}_{\rho_N})) - (\bm{\pi}^0, \bm{\Theta}^0) \big|\big|^2)$.

For the log-likelihood function $l_N (\hat{\bm{\pi}}, \hat{\bm{\Theta}}; \mathcal{R}) =\sum_{i=1}^N \log \big( \sum_{k=1}^M \hat{\pi}_k \prod_{j=1}^J \hat{\theta}_{j,k}^{R_{ij}}(1-\hat{\theta}_{j,k}^{1 - R_{ij}}) \big)$,
we have
\begin{align}\nonumber
	& \ \frac{1}{N} \big|l_N (\hat{\bm{\pi}}, \hat{\bm{\Theta}}; \mathcal{R}) - l_N (\hat{\bm{\pi}}_{\rho_N}, \hat{\bm{\Theta}}_{\rho_N}; \mathcal{R}) \big|  \\\nonumber
	\leq & \frac{1}{N}\sum_{i=1}^N \Big|\log \big( \sum_{k=1}^M \hat{\pi}_k \prod_{j=1}^J \hat{\theta}_{j,k}^{R_{ij}}(1-\hat{\theta}_{j,k}^{1 - R_{ij}}) \big) - \log \big( \sum_{k: \hat{\pi}_k > \rho_N} \hat{\pi}_k \prod_{j=1}^J \hat{\theta}_{j,k}^{R_{ij}}(1-\hat{\theta}_{j,k}^{1 - R_{ij}}) \big) \Big|\\\label{eq:log}
	\leq & \frac{1}{N}\sum_{i=1}^N\frac{\big|\big( \sum_{k=1}^M \hat{\pi}_k \prod_{j=1}^J \hat{\theta}_{j,k}^{R_{ij}}(1-\hat{\theta}_{j,k}^{1 - R_{ij}}) \big) - \big( \sum_{k: \hat{\pi}_k > \rho_N} \hat{\pi}_k \prod_{j=1}^J \hat{\theta}_{j,k}^{R_{ij}}(1-\hat{\theta}_{j,k}^{1 - R_{ij}}) \big) \big|}{ \sqrt{ \big( \sum_{k=1}^M \hat{\pi}_k \prod_{j=1}^J \hat{\theta}_{j,k}^{R_{ij}}(1-\hat{\theta}_{j,k}^{1 - R_{ij}}) \big) \times  \big( \sum_{k: \hat{\pi}_k > \rho_N} \hat{\pi}_k \prod_{j=1}^J \hat{\theta}_{j,k}^{R_{ij}}(1-\hat{\theta}_{j,k}^{1 - R_{ij}}) \big)}}\\\nonumber
	\leq & \frac{1}{N}\sum_{i=1}^N\frac{(M - \hat{M})\rho_N}{  \sum_{k: \hat{\pi}_k > \rho_N} \hat{\pi}_k \prod_{j=1}^J \hat{\theta}_{j,k}^{R_{ij}}(1-\hat{\theta}_{j,k}^{1 - R_{ij}})}\\ \label{eq:log-diff}
	 = & O(\rho_N) = O(N^{-d}),\ d \geq 1, 
\end{align}
where inequality \eqref{eq:log} follows from an upper bound for log function.
Specifically, for $x\geq 1$, we know $\log x \leq (x-1) / \sqrt{x}$, and thus for $0 < x \leq y$, we have $\log y - \log x \leq (y - x) / \sqrt{xy}$.
From \eqref{eq:log-diff}, $G'_N(\hat{\bm{\pi}}, \hat{\bm{\Theta}}) = G'_N(\hat{\bm{\pi}}_{\rho_N}, \hat{\bm{\Theta}}_{\rho_N}) + O(N^{-d}) \geq G'_N (\bm{\pi}^0, \bm{\Theta}^0)$ and thus $G'_N(\hat{\bm{\pi}}_{\rho_N}, \hat{\bm{\Theta}}_{\rho_N}) > G'_N (\bm{\pi}^0, \bm{\Theta}^0) - O(N^{-d}) \geq  G'_N (\bm{\pi}^0, \bm{\Theta}^0) - O(N^{-1})$.
Since $N^{-1/2}\lambda_N^{(1)} \rightarrow 0$ and $N^{-1/2}\lambda_N^{(2)}\tau \rightarrow 0$, then for sufficiently small $\zeta$, by Taylor's expansion,
\[
\mathbb{E} \Big( \underset{|| (\bm{\pi}_{\rho_N}, \bm{\Theta}_{\rho_N}) - (\bm{\pi}^0, \bm{\Theta}^0) || \leq \zeta }{\sup} G'_N(\bm{\pi}_{\rho_N}, \bm{\Theta}_{\rho_N}; \mathcal{R}) - G_0(\bm{\pi}_{\rho_N}, \bm{\Theta}_{\rho_N}) - G'_N(\bm{\pi}^0, \bm{\Theta}^0; \mathcal{R}) + G_0(\bm{\pi}^0, \bm{\Theta}^0) \Big) = O(\zeta N^{-1/2}).
\]
By Theorem 3.2.5 in \cite{van1996weak}, we have $(\hat{\bm{\pi}}_{\rho_N}, \hat{\bm{\Theta}}_{\rho_N}) - (\bm{\pi}^0, \bm{\Theta}^0) = O_p(N^{-1/2})$.

We next show selection consistency of $S^0$.
If true item parameters $\theta_{j,k_1}^0 \neq \theta_{j,k_2}^0$, then from the above convergence result, we know $\hat{\theta}_{j,k_1} \rightarrow \theta_{j,k_1}^0$ and $\hat{\theta}_{j,k_2} \rightarrow \theta_{j,k_2}^0$, and thus $\hat{\theta}_{j,k_1} \neq \hat{\theta}_{j,k_2}$ in probability.
If true item parameters $\theta_{j,k_1}^0 = \theta_{j,k_2}^0$ but $\hat{\theta}_{j,k_1} \neq \hat{\theta}_{j,k_2}$, by the Karush-Kuhn-Tucker (KKT) conditions, we have 
$N^{-1/2} \partial l_N(\bm{\pi}, \bm{\Theta};\mathcal{R})/\partial \theta_{j,k_1} |_{(\bm{\pi}, \bm{\Theta}) = (\hat{\bm{\pi}}, \hat{\bm{\Theta}})} = N^{-1/2} \lambda_N^{(2)}\rightarrow \infty$ in probability. 
However $N^{-1/2} \partial l_N(\bm{\pi}, \bm{\Theta};\mathcal{R})/\partial \theta_{j,k_1} |_{(\bm{\pi}, \bm{\Theta}) = (\hat{\bm{\pi}}, \hat{\bm{\Theta}})} = O_p(1)$.
Therefore, if $\theta_{j,k_1}^0 = \theta_{j,k_2}^0$, we have $\hat{\theta}_{j,k_1} = \hat{\theta}_{j,k_2}$ in probability, which proved the selection consistency that $\mathbb{P}(\hat{S}\nsim S^0)\rightarrow 0$ as $N\rightarrow \infty$.
\end{proof}{}

\section{Derivations of PEM Algorithm}
\label{sec-S2}

In this section, we give detailed derivations of the penalized EM algorithm in Section \ref{em}.
First let's introduce a new variable $\bm{d} = (d_{jkl},j=1,\dots,J, 1\leq k < l \leq M)$ to be the differences of the item parameters for each item. Then our problem becomes

\begin{equation}
\begin{aligned}
\min_{\bm{\pi},\bm{\Theta},\bm{d}} \quad & G(\bm{\pi},\bm{\Theta},\bm{d})\\
\textrm{s.t.} \quad &d_{jkl} = \theta_{jk} - \theta_{jl} \\
  & j=1,\dots,J,\ 1\leq k < l \leq M. \\
\end{aligned}
\end{equation}

By using the difference convex property of the truncated Lasso penalty, we can decompose the objective function into two parts:
\begin{equation}
G(\bm{\pi},\bm{\Theta},\bm{d}) = G_1(\bm{\pi},\bm{\Theta},\bm{d}) - G_2(\bm{d}),        
\end{equation}
where 
\begin{equation}
    G_1(\bm{\pi},\bm{\Theta},\bm{d}) = - \frac{1}{N} Q(\bm{\pi},\bm{\Theta}|\bm{\pi}^{(c)},\bm{\Theta}^{(c)}) + \tilde{\lambda}_1 \sum_{k=1}^M \log \pi_k + \tilde{\lambda}_2 \sum_{j=1}^J \sum_{1\leq k < l \leq M} |d_{jkl}|,
\end{equation}
\begin{equation}
    G_2(\bm{d}) = \tilde{\lambda}_2\sum_{j=1}^J\sum_{1\leq k < l \leq M}  \big(|d_{jkl}-\tau|\big)_+.
\end{equation}

Then we construct a sequence of upper approximation of $G(\bm{\pi},\bm{\Theta},\bm{d})$ iteratively by replacing $G_2(\bm{d})$ at iteration $c+1$ with its piecewise affine minorization:
\begin{equation}
    G_2^{(c)}(\bm{d})=G_2(\hat{\bm{d}}^{(c)})+\tilde{\lambda}_2 \sum_{j=1}^J \sum_{1\leq k<l \leq M} \big(|d_{jkl}|-|\hat{d}_{jkl}^{(c)}|\big)\cdot \mathbb{I}\big(|\hat{d}_{jkl}^{(c)}|\geq\tau\big),
\end{equation}
at the current estimate $\hat{\bm{d}}^{(c)}$, which lead to an upper convex approximation:
\begin{align*}
        G^{(c+1)}(\bm{\pi},\bm{\Theta},\bm{d})=&-\frac{1}{N} Q(\bm{\pi},\bm{\Theta}|\bm{\pi}^{(c)},\bm{\Theta}^{(c)}) + \tilde{\lambda}_1 \sum_{k=1}^M \log \pi_k \\
        &+\tilde{\lambda}_2\sum_{j=1}^J\sum_{1\leq k < l \leq M}|d_{jkl}|\cdot \mathbb{I}\big(|\hat{d}_{jkl}^{(c)}|<\tau\big) \\
        &+ \tilde{\lambda}_2 \tau \sum_{j=1}^J\sum_{1\leq k<l \leq M}\mathbb{I}\big(|\hat{d}_{jkl}^{(c)}|\geq \tau \big).
\end{align*}
Now we can apply ADMM. At iteration $c+1$, the augmented Lagrangian is 
\begin{equation}
     L_{\gamma}(\bm{\pi},\bm{\Theta},\bm{d}, \bm{y}) = G^{(c+1)}(\bm{\pi},\bm{\Theta},\bm{d}) + \sum_{j=1}^J\sum_{1\leq k <l \leq M} y_{jkl}\cdot \big(d_{jkl}-(\theta_{jk}-\theta_{jl})\big) + \frac{\gamma}{2}\sum_{j=1}^J\sum_{1\leq k < l \leq M}\big|d_{jkl}-(\theta_{jk}-\theta_{jl})\big|^2, 
\end{equation}
where $y_{jkl}$'s are the dual variables and $\gamma$ is a nonnegative penalty parameter.
Then ADMM \citep{boyd2011distributed} consists of the following iterations:
\begin{align*}
	\bm{\pi}^{(c+1)} &= \underset{\bm{\pi}}{\mathrm{argmin}}\ L_{\gamma}(\bm{\pi},\bm{\Theta}^{(c)},\bm{d}^{(c)}, \bm{y}^{(c)}),\\
	\bm{\Theta}^{(c+1)} &= \underset{\bm{\Theta}}{\mathrm{argmin}}\ L_{\gamma}(\bm{\pi}^{(c+1)},\bm{\Theta},\bm{d}^{(c)}, \bm{y}^{(c)}),\\
	\bm{d}^{(c+1)} &= \underset{\bm{d}}{\mathrm{argmin}}\ L_{\gamma}(\bm{\pi}^{(c+1)},\bm{\Theta}^{(c+1)},\bm{d}, \bm{y}^{(c)}),\\
	y^{(c+1)}_{jkl} &= y^{(c)}_{jkl} + \gamma (d^{(c+1)}_{jkl} - \big(\theta_{jk}^{(c+1)} - \theta_{jl}^{(c+1)})\big), \ j = 1, ..., J, 1 \leq k < l \leq M.
\end{align*}

Using the scaled Lagrangian multiplier $\mu_{jkl} = y_{jkl}/\gamma$ and defining the residual $r_{jkl} = d_{jkl} - (\theta_{jk} - \theta_{jl})$, we have:
\begin{align*}
	& y_{jkl}\cdot \big(d_{jkl}-(\theta_{jk}-\theta_{jl})\big) + \frac{\gamma}{2} \big|d_{jkl}-(\theta_{jk}-\theta_{jl})\big|^2\\
	= & y_{jkl} \cdot r_{jkl} + \frac{\gamma}{2} r_{jkl}^2 \\
	= & \frac{\gamma}{2}\big(r_{jkl} + (1/\gamma) y_{jkl} \big)^2 - \frac{1}{2\gamma} \mu_{jkl}^2 \\
	= & \frac{\gamma}{2}\big(r_{jkl} + \mu_{jkl} \big)^2 - \frac{1}{2\gamma} \mu_{jkl}^2.
\end{align*}
Then using the scaled dual variable, we can express ADMM as:
\begin{align*}
	\bm{\pi}^{(c+1)} &= \underset{\bm{\pi}}{\mathrm{argmin}}\ G^{(c+1)}(\bm{\pi},\bm{\Theta}^{(c)},\bm{d}^{(c)}),\\
	\bm{\Theta}^{(c+1)} &= \underset{\bm{\Theta}}{\mathrm{argmin}}\ G^{(c+1)}(\bm{\pi}^{(c+1)},\bm{\Theta},\bm{d}^{(c)}) + \frac{\gamma}{2}\sum_{j=1}^J\sum_{1\leq k <l \leq M}\big(d_{jkl}^{(c)} - (\theta_{jk}^{(c)} - \theta_{jl}^{(c)}) + \mu_{jkl}^{(c)}\big),\\
	\bm{d}^{(c+1)} &= \underset{\bm{d}}{\mathrm{argmin}}\ G^{(c+1)}(\bm{\pi}^{(c+1)},\bm{\Theta}^{(c+1)},\bm{d}) + \frac{\gamma}{2}\sum_{j=1}^J\sum_{1\leq k <l \leq M}\big(d_{jkl} - (\theta_{jk}^{(c+1)} - \theta_{jl}^{(c+1)}) + \mu_{jkl}^{(c)}\big),\\
	\mu^{(c+1)}_{jkl} &= \mu^{(c)}_{jkl} + d^{(c+1)}_{jkl} - \big(\theta_{jk}^{(c+1)} - \theta_{jl}^{(c+1)}), \ j = 1, ..., J, 1 \leq k < l \leq M.
\end{align*}
Specifically, we get the following updates:
\renewcommand\labelenumi{(\theenumi)}
\begin{enumerate}
    \item \begin{equation*}
       \pi_k^{(c+1)} = \frac{\sum_{i=1}^N s_{ik}^{(c+1)}/N - \tilde{\lambda}_1}{1-M\tilde{\lambda}_1}, \quad \text{where } s_{ik}^{(c+1)}=\frac{\pi_k^{(c)}\varphi_k(\bm{R}_i;\bm{\Theta}_k^{(c)})}{\sum_{k'}^{(c)}\pi_{k'}^{(c)}\varphi_{k'}^{(c)}(\bm{R}_i;\bm{\theta}_{k'}^{(c)})},
    \end{equation*}
    \item
    \begin{align*}
        \hat{\theta}_{jk}^{(c+1)} = \underset{\theta_{jk}}{\mathrm{argmin}}&
        \Big\{-\frac{\sum_{i=1}^N s_{ik}^{(c)}R_{ij}}{N}\log\theta_{jk}
        - \frac{\sum_{i=1}^N s_{ik}^{(c)}(1-R_{ij})}{N} \log (1-\theta_{jk})\\
        &+ \frac{\gamma}{2}\sum_{l>k}\big(\hat{d}_{jkl}^{(c)}-(
        \theta_{jk}-\hat{\theta}_{jl}^{
        (c)})+\hat{\mu}_{jkl}^{(c)}\big)^2
        \\
        &+\frac{\gamma}{2}\sum_{l<k}\big(\hat{d}_{jlk}^{(c)}-(
        \hat{\theta_{jl}}^{(c+1)}-\theta_{jk})+\hat{\mu}_{jlk}^{(c)}\big)^2
        \Big\}
    \end{align*}
    \item \begin{equation*}
        \hat{d}_{jkl}^{(c+1)}=\left\{
        \begin{aligned}
        &\hat{\theta}_{jk}^{(c+1)} - \hat{\theta}_{jl}^{(c+1)} -\hat{\mu}_{jkl}^{(c)},\quad \quad \text{if }  |\hat{d}_{jkl}^{(c)}| \geq \tau\\
        &\text{ST}\big(\hat{\theta}_{jk}^{(c+1)} - \hat{\theta}_{jl}^{(c+1)} -\hat{\mu}_{jkl}^{(c)};\tilde{\lambda}_2/\gamma \big),\quad \text{if } |\hat{d}_{jkl}^{(c)}| < \tau, \text{where } \text{ST}(x;\gamma) = (|x|-\gamma)_+ x/|x|
        \end{aligned},
    \right.
    \end{equation*}
    \item 
    \begin{equation*}
        \hat{\mu}_{jkl}^{(c+1)} = \hat{\mu}_{jkl}^{(c)} + \hat{d}_{jkl}^{(c+1)} - \big(
    \hat{\theta}_{jk}^{(c+1)} - \hat{\theta}_{jl}^{(c+1)}\big).
    \end{equation*}
\end{enumerate}
Note that the objective in step (2) is convex in $\theta_{jk}$, therefore we use gradient descent to perform the minimization.

\section{PEM Algorithm with Missing Values}
\label{sec:S3}

In this section, we present the penalized EM algorithm with missing values. 
Here we use a mask matrix $M\in\{0,1\}^{N\times J}$ to indicate the locations of the missing values, where $M_{i,j} = 0$ means the $i$th subject's response to the $j$th item is missing.
The details of the algorithm is summarized in Algorithm \ref{algo-missing}.
\begin{algorithm}[h!]
\caption{Penalized EM with missing data}
\label{algo-missing}
\SetKwInOut{Input}{Input}
\SetKwInOut{Output}{Output}

\KwData{Binary response matrix $\mathcal{R}=(R_{i,j})_{N\times J}$ and the mask matrix $\bm{M}=(M_{ij})_{N\times J}$ indicating missing values.}
Set hyperparameters $\tilde{\lambda}_1,\ \tilde{\lambda}_2, \ \tau, \ \gamma$ and $\rho$.

Set an upper bound of the number of latent classes $L$.

Initialize parameters $\bm{\pi}$, $\bm{\Theta}$, and the conditional expectations $\bm{s}$.

\While{not converged}{

In the $(c+1)$th iteration,


\For{$(i,k)\in [N]\times[L]$}{
$s_{ik}^{(c+1)}=\frac{\pi_k^{(c)}\varphi_k(\bm{R}_i;\bm{\theta}_k^{(c)})}{\sum_{k'}^{(c)}\pi_{k'}^{(c)}\varphi_{k'}^{(c)}(\bm{R}_i;\bm{\theta}_{k'}^{(c)})},\quad \varphi(\bm{r}_i;\bm{\theta}_k) = \prod_{j=1}^J \big(\theta_{jk}^{R_{ij}}(1-\theta_{kj})^{1-R_{ij}}\big)^{m_{ij}}$
}

\For{$k\in[L]$ and $\pi_k^{(c)} > \rho$}{
$\pi_k^{(c+1)} = \frac{\sum_{i=1}^N s_{ik}^{(c+1)}/N - \tilde{\lambda}_1}{1-L\tilde{\lambda}_1}.$
}

\For{$(j,k)\in[J]\times [L]$ and $\pi^{(c+1)}_k > \rho$}{
\begin{align*}
        \theta_{jk}^{(c+1)} = \underset{\theta_{jk}}{\mathrm{argmin}}&
        \Big\{-\frac{\sum_{i=1}^N s_{ik}^{(c)}R_{ij} m_{ij}}{\sum_{i=1}^N m_{ij}}\log \theta_{jk}
        - \frac{\sum_{i=1}^N s_{ik}^{(c)}(1-_{ij})m_{ij}}{\sum_{i=1}^N m_{ij}} \log (1-\theta_{jk})\\
        &+ \frac{\gamma}{2}\sum_{l>k}\big(\hat{d}_{jkl}^{(c)}-(
        \theta_{jk}-\hat{\theta}_{jl}^{
        (c)})+\hat{\mu}_{jkl}^{(c)}\big)^2
        \\
        &+\frac{\gamma}{2}\sum_{l<k}\big(\hat{d}_{jlk}^{(c)}-(
        \hat{\theta_{jl}}^{(c+1)}-\theta_{jk})+\hat{\mu}_{jlk}^{(c)}\big)^2
        \Big\}
    \end{align*}
}

\For{$j\in [J],k,l \in [L],\  k<l$ and $\pi^{(c+1)}_k > \rho$, $\pi^{(c+1)}_l > \rho$}{
\begin{align*}
    &\hat{d}_{jkl}^{(c+1)}=\left\{
        \begin{aligned}
        &\hat{\theta}_{jk}^{(c+1)} - \hat{\theta}_{jl}^{(c+1)} -\hat{\mu}_{jkl}^{(c)},\quad \quad \text{if }  |\hat{d}_{jkl}^{(c)}| \geq \tau\\
        &\text{ST}\big(\hat{\theta}_{jk}^{(c+1)} - \hat{\theta}_{jl}^{(c+1)} -\hat{\mu}_{jkl}^{(c)};\tilde{\lambda}_2/\gamma \big),\quad \text{if } |\hat{d}_{jkl}^{(c)}| < \tau
        \end{aligned},
    \right.\\
    & \hat{\mu}_{jkl}^{(c+1)} = \hat{\mu}_{jkl}^{(c)} + \hat{d}_{jkl}^{(c+1)} - \big(
    \hat{\theta}_{jk}^{(c+1)} - \hat{\theta}_{jl}^{(c+1)}\big).
\end{align*}{}
}
}
\Output{$\big\{\bm{\hat{\pi}},\ \bm{\hat{\Theta}}, \ \bm{\hat{s}}\big\}$}

\end{algorithm}{}

\section{Sensitivity Analysis}
\label{chapter:sa}
\begin{figure}[htbp]
    \centering
    \subfigure[]{
    \includegraphics[width=0.8\textwidth]{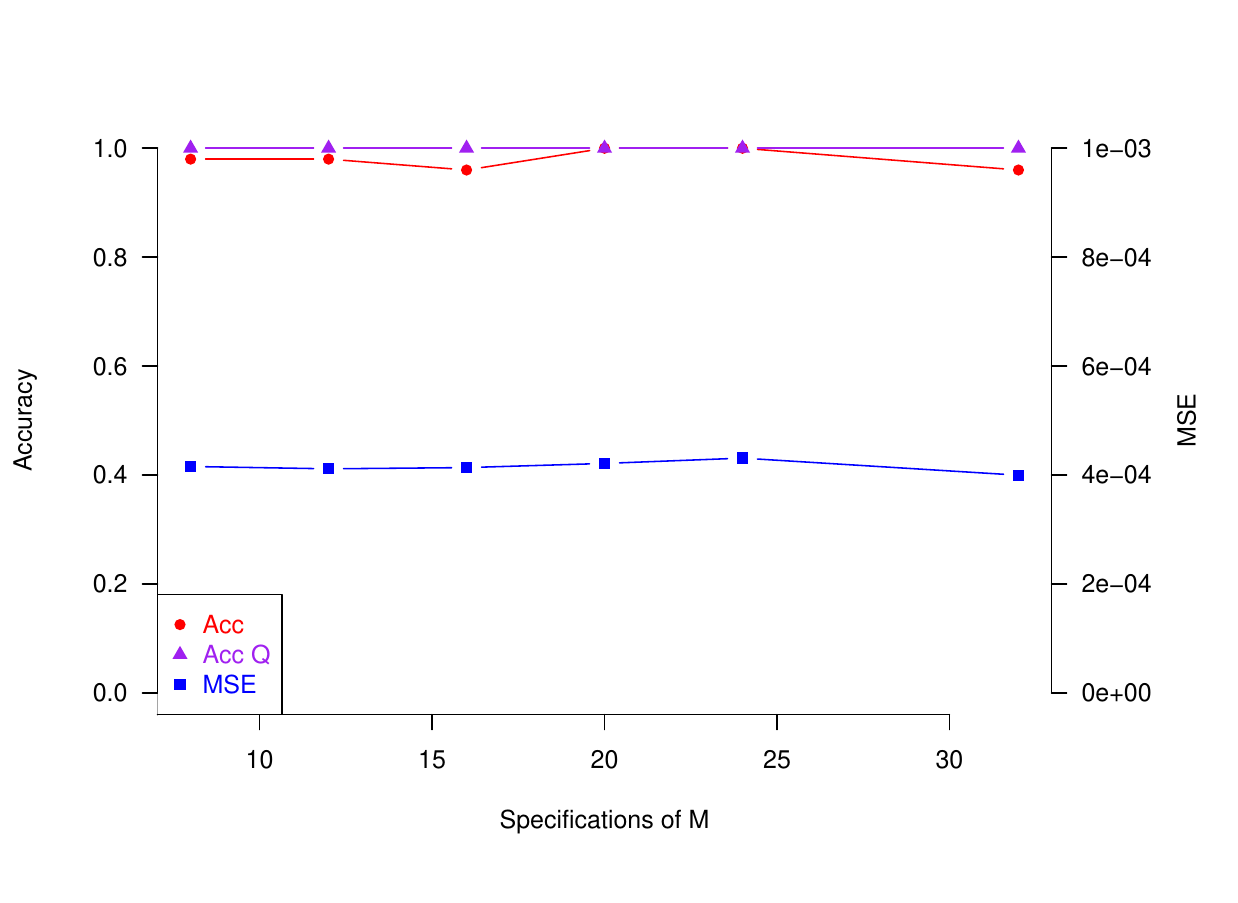}}
    \hspace{0in}
    \subfigure[]{
    \includegraphics[width=0.8\textwidth]{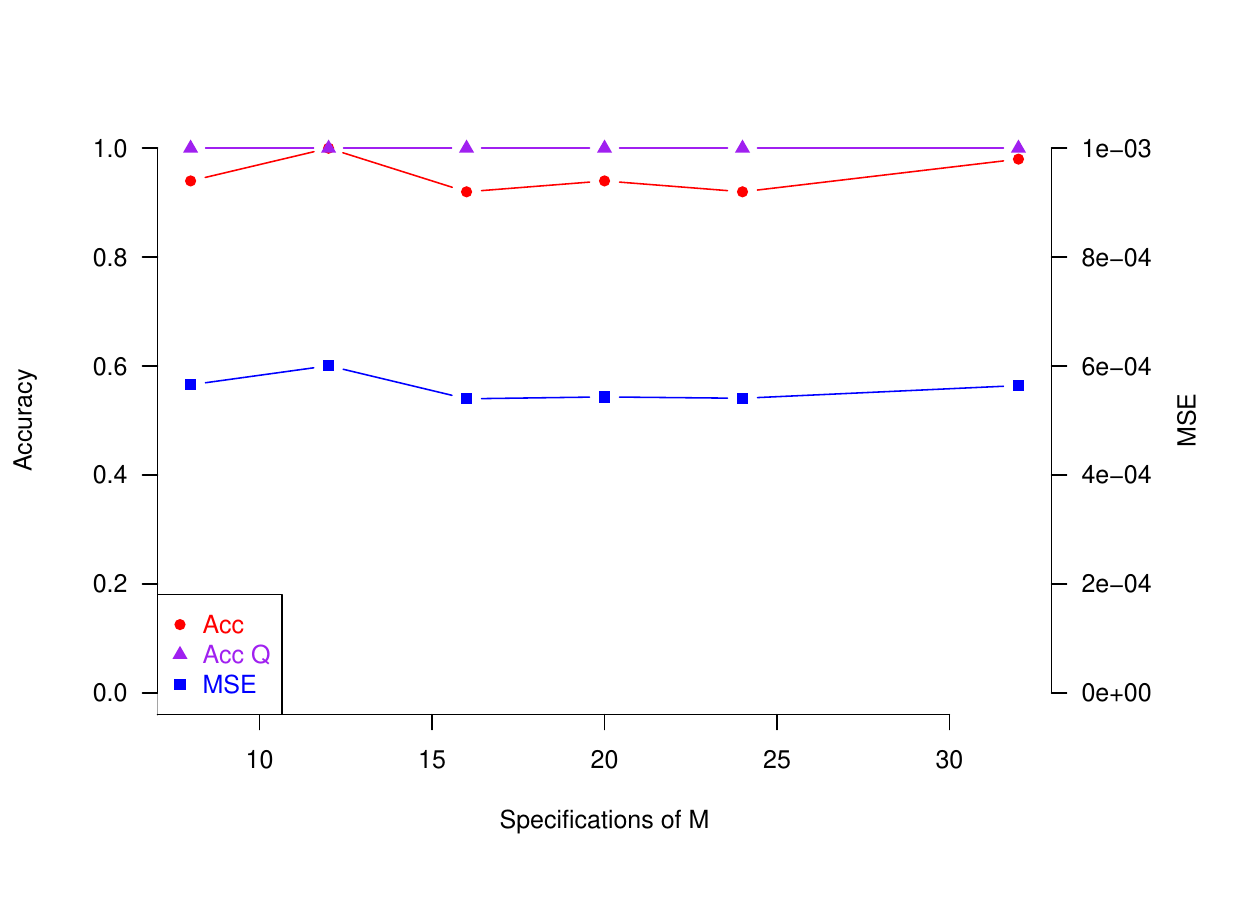}}
    \caption{Sensitivity analysis results. (a) DINA results; (b) GDINA results. The red curve captures the Acc($\hat{M}$), Acc($\hat{\boldsymbol{P}}$), Acc($\hat{\mathcal{E}}$), the blue curve captures MSE($\hat{\boldsymbol{\Theta}}$) and the purple curve captures the Acc($\hat{\boldsymbol{Q}}$) for various $M$.  }
    \label{fig:sa evaluation}
\end{figure}

In this section, we conduct the sensitivity analysis of our algorithm by investigating the effects of different inputs of $M$, the upper bound of the number of latent classes, on the simulation results. In particular, we focus on two simulation settings: (1) DINA model with linear hierarchical structure, $N = 500$ and $r = 0.1$; (2) GDINA model with linear hierarchical structure, $N = 1000$ and $r = 0.1$. Both two settings have $K = 4$ latent attributes and $J = 30$ test items, and run for 50 repetitions. We keep the parameter generation process and the hyperparameter tuning strategy consistent with the simulation studies in the main article. In this sensitivity analysis, we fit our proposed method with various $M = \{8,12, 16, 20, 24, 32\}$ in the two simulations settings. The evaluation results in DINA and GDINA settings are based on metrics Acc($\hat{M}$), Acc($\hat{\boldsymbol{P}}$), Acc($\hat{\mathcal{E}}$), MSE($\hat{\boldsymbol{\Theta}}$) and Acc($\hat{\boldsymbol{Q}}$). Consistent with the simulation studies in the main article, the Acc($\hat{M}$), Acc($\hat{\boldsymbol{P}}$) and Acc($\hat{\mathcal{E}}$) are calculated for all the cases; MSE($\hat{\boldsymbol{\Theta}}$) is calculated for the cases when the number of latent classes is correctly selected; Acc($\hat{\boldsymbol{Q}}$) is calculated for the cases when the hierarchical structure is successfully recovered. The results are plotted in Figure~\ref{fig:sa evaluation}. 

From the simulation results in Figure~\ref{fig:sa evaluation}, we see our proposed method is robust to the different specifications of $M$, in terms of all metrics. Among cases with different $M$, our method achieves a high accuracy in estimating the number of latent classes, and in recovering the partial orders, the hierarchical structures, the item parameter matrix, and the $Q$-matrix. In terms of computation time, the average running time is 0.36 seconds and 1.12 seconds for DINA and GDINA, respectively, per repetition per set of tuning hyperparameters.

\newpage 
\bibliographystyle{apalike}
\bibliography{refs}